\documentclass[10.5pt,letterpaper]{article}
\pdfoutput=1 

\usepackage[utf8]{inputenc} 
\usepackage{lipsum} 

\usepackage[top=2.5cm, bottom=3cm, left=3.5cm, right=3.5cm,
           heightrounded,marginparwidth=1.5cm, marginparsep=1cm]{geometry} 

\usepackage{changepage} 

\usepackage[shortlabels]{enumitem} 

\usepackage[square,numbers,merge,comma,sort&compress]{natbib} 
\makeatletter
\def\NAT@spacechar{\,}  
\makeatother

\usepackage{amsmath,amssymb,amsfonts,amsthm}
\usepackage{mathtools} 
\usepackage{old-arrows}
\usepackage[scale=0.7]{miama} 
\usepackage{slashed,cancel}
\usepackage{comment}   
\usepackage{relsize}   
\usepackage{setspace}  
\usepackage{moresize}  
\usepackage{epsfig}
\usepackage{latexsym}
\usepackage{mathrsfs} 
\usepackage{calligra,aurical} 
\usepackage{calc}
\usepackage{float}
\usepackage{appendix}
\usepackage{xargs}
\usepackage{extarrows}
\usepackage{empheq}

\usepackage[blocks]{authblk}  

\usepackage[fulladjust]{marginnote}%

\usepackage{graphicx} 

\usepackage[labelsep=colon]{caption}  
\usepackage[labelsep=colon,aboveskip=10pt,belowskip=10pt]{subcaption}
\usepackage{array,multirow,makecell,booktabs}  
\newcolumntype{X}[2]{>{\centering\arraybackslash$}#1{#2\linewidth}<{$}}
\newcolumntype{R}[1]{>{\raggedleft\arraybackslash$}m{#1\linewidth}<{$}}
\newcolumntype{L}[1]{>{\raggedright\arraybackslash}m{#1\linewidth}}

\makeatletter
\renewcommand\mcell@classz{\@classx
   \@tempcnta \count@
   \prepnext@tok
   \@addtopreamble{
      \ifcase\@chnum
         \hfil
         \mcell@agape{\d@llarbegin\insert@column\d@llarend}\hfil \or
         \hskip1sp
         \mcell@agape{\d@llarbegin\insert@column\d@llarend}\hfil \or
         \hfil\hskip1sp
         \mcell@agape{\d@llarbegin \insert@column\d@llarend}\or
         \mcell@agape{$\vcenter
         \@startpbox{\@nextchar}\insert@column\@endpbox$}\or
         \mcell@agape{\vtop
         \@startpbox{\@nextchar}\insert@column\@endpbox}\or
         \mcell@agape{\vbox
         \@startpbox{\@nextchar}\insert@column\@endpbox}%
      \fi
      \global\let\mcell@left\relax\global\let\mcell@right\relax
    }\prepnext@tok}
\makeatother

\usepackage{titlesec}

\titleformat{\section}{\normalfont\large\bfseries}{\thesection}{1em}{}
\titleformat{\subsection}{\normalfont\normalsize\bfseries}{\thesubsection}{0.75em}{}
\titleformat{\subsubsection}{\normalfont\normalsize\bfseries}{\thesubsubsection}{0.75em}{}

\titlespacing*{\section}{0pt}%
                {4ex plus 1ex minus .5ex}{1.75ex plus .25ex minus .25ex}
\titlespacing*{\subsection}{0pt}%
                {3.5ex plus 1ex minus .5ex}{1.25ex plus .2ex minus .2ex}
\titlespacing*{\subsubsection}{0pt}%
                {2.5ex plus 0.75ex minus .2ex}{0.75ex plus .15ex minus .15ex}
\titlespacing*{\paragraph}{0pt}%
                {1.85ex plus 0.5ex minus .15ex}{1em}

\usepackage{titletoc}

\addtocontents{toc}{\addvspace{-0.75em}}  

\titlecontents{section}
  [1.25em] {\addvspace{0.7em plus 0pt}\small}
  {\thecontentslabel\hspace{0.75em}}{}
  {\hspace{0.5em}\titlerule*[0.5em]{.}\contentspage}
  [\addvspace{0.0em plus 0pt}]

\titlecontents{subsection}
  [2.75em] {\addvspace{0.075em plus0pt}\fns}
  {\thecontentslabel\hspace{0.75em}}{\thecontentslabel\hspace{0.75em}}
  {\hspace{0.5em}\titlerule*[0.5em]{.}\small\contentspage}
  [\addvspace{0.075em plus 0pt}]

\usepackage{environ}
\makeatletter
\NewEnviron{subalign}[1]{
\begin{subequations}\label{#1}
\begin{align} \BODY \end{align}
\end{subequations}      }
\makeatother
\makeatletter
\newenvironment{subeqs}%
{\begingroup%
\setlength{\abovedisplayskip}{10pt plus 4pt minus 9pt}%
\setlength{\abovedisplayshortskip}{0pt plus 2pt minus 2pt}%
\setlength{\belowdisplayskip}{12pt plus 3pt minus 9pt}%
\setlength{\belowdisplayshortskip}{7pt plus 3pt minus 4pt}%
\begin{subequations}%
}%
{\end{subequations}\ignorespacesafterend%
\endgroup}%
\makeatother

\makeatletter
{\begingroup%
\setlength{\abovedisplayskip}{{#1}pt plus 2pt minus 9pt}%
\setlength{\abovedisplayshortskip}{0pt plus 0pt minus 2pt}%
\setlength{\belowdisplayskip}{{#2}pt plus 3pt minus 9pt}%
\setlength{\belowdisplayshortskip}{7pt plus 3pt minus 4pt}%
\begin{subequations}%
}%
{\end{subequations}\ignorespacesafterend%
\endgroup}%
\makeatother

\makeatletter
{\begingroup%
\setlength{\abovedisplayskip}{{#1}pt plus 3pt minus 9pt}%
\setlength{\abovedisplayshortskip}{0pt plus 3pt}%
\setlength{\belowdisplayskip}{{#2}pt plus 3pt minus 9pt}%
\setlength{\belowdisplayshortskip}{7pt plus 3pt minus 4pt}%
\begin{equation}%
}%
{\end{equation}\ignorespacesafterend%
\endgroup}%
\makeatother

\makeatletter
{%
\begin{equation}%
\begin{split}{#1}%
}%
{\end{split}%
\end{equation}\ignorespacesafterend%
}%
\makeatother

\usepackage{xcolor}
\definecolor{Green}{rgb}{0.05, 0.45, 0.25}
\definecolor{dogwoodrose}{rgb}{0.8, 0.1, 0.55}
\definecolor{RRed}{rgb}{0.7, 0.1, 0.525}

\usepackage{bm}  
\usepackage{dsfont}  

\DeclareMathAlphabet{\mathpzc}{OT1}{pzc}{m}{it}
\DeclareMathAlphabet{\mathcal}{OMS}{cmsy}{m}{n}
\DeclareSymbolFontAlphabet{\Scr}{rsfs}
\DeclareMathAlphabet{\mathbold}{U}{BOONDOX-ds}{m}{n}
\SetMathAlphabet{\mathbold}{bold}{U}{BOONDOX-ds}{b}{n}
\DeclareMathAlphabet{\mathcalboondox}{U}{BOONDOX-calo}{m}{n}
\SetMathAlphabet{\mathcalboondox}{bold}{U}{BOONDOX-calo}{b}{n}
\DeclareMathAlphabet{\mathbcalboondox}{U}{BOONDOX-calo}{b}{n}

\newcommand\eqlinkcol{RRed}



\usepackage[breaklinks=true,backref=page]{hyperref}
\hypersetup{
    bookmarks=true,         
    pdfmenubar=true,        
    pdffitwindow=false,     
    pdfpagemode={UseNone},
    pdfstartview={FitH},
    pdfauthor={},     
    pdfsubject={},   
    pdfcreator={},   
    pdfproducer={},  
    colorlinks=true,
    bookmarks=true,
    bookmarksnumbered=true,
    plainpages,
    a4paper,
    linktoc=page,
    citecolor=blue,
    filecolor=black,
    linkcolor=\eqlinkcol,
    urlcolor=Green,
}
\renewcommand*{\backref}[1]{}
\renewcommand*{\backrefalt}[4]{%
\ifcase #1 %
\relax
\or
~{\small [\textsc{p.~\fns{\!#2}}]}
\else
~{\small [\textsc{p.~\fns{\!#2}}]}%
\fi}

\usepackage{footnotebackref}
\usepackage{hypernat} 

\def\+{~+~}
\def\-{~-~}
\def\={~=~}
\def\'{``}
\def\*{{}^*}
\newcommand\fns{\footnotesize}

\newcommand\qqquad{\quad\quad\quad}
\newcommand\qqqquad{\quad\quad\quad\quad}
\newcommand\qRq{\quad\Rightarrow\quad}

\newcommand\qlq{\quad\longrightarrow\quad}
\newcommand\qLq{\quad\Longrightarrow\quad}
\newcommand\qqLqq{\qquad\Longrightarrow\qquad}
\newcommand\Real{\operatorname{Re}}
\newcommand\Img{\operatorname{Im}}
\newcommand{\ms}{\mathsmaller}

\newcommand{\ml}{\mathlarger}

\newcommand{\dd}{\partial}

\newcommand{\und}[1]{{\underline{#1}}}

\newcommand\eps{\epsilon}
\newcommand\veps{\varepsilon}
\newcommand\w{\omega}
\newcommand\Id{\mathds{1}}
\newcommand\Zero{\mathbold{0}}
\newcommand\N{\mathcal{N}}
\newcommand\NN{\mathfrak{N}}
\newcommand\eD{e_{\textsc{d}}}
\newcommand\nv{n_\text{v}}
\newcommand\ns{n_\text{s}}
\newcommand\II{\mathcal{I}}
\newcommand\RR{\mathcal{R}}
\newcommand\Ms{\mathscr{M}}
\newcommand\Mscal{\Ms_\text{scal}}
\newcommand\LL{\textsl{L}}
\newcommand\LLb{\mathbb{L}}
\newcommand\OO{\mathcal{O}}
\newcommand\D{\mathcal{D}}

\newcommand\Lagr{\mathscr{L}}
\newcommand\Lgaug{\mathscr{L}_\text{gaug}}
\newcommand\Rs{\mathscr{R}}
\newcommand\pp{\wp}
\newcommand\ww{\mathpzc{w}}

\newcommand\Tr{\operatorname{Tr}}
\newcommand\Span{\operatorname{Span}}
\newcommand\Det{\operatorname{Det}}
\newcommand\Adj{\operatorname{Adj}}
\newcommand\rank{\operatorname{rank}}
\newcommand\Gdual{\mathcal{G}}
\newcommand\M{\mathcal{M}}
\newcommand\FF{\mathbb{F}}
\newcommand\Cc{\mathbb{C}}
\newcommand\Sp{\mathrm{Sp}}
\newcommand\U{\mathrm{U}}
\newcommand\SL{\mathrm{SL}}
\newcommand\SU{\mathrm{SU}}
\newcommand\SO{\mathrm{SO}}
\newcommand\Rsv{\mathscr{R}_\text{v}}
\newcommand\Rsvu{\underline{\mathscr{R}}_\text{\,v}}
\newcommand\Rsvst{\mathscr{R}_\text{v*}}
\newcommand\VBH{V_\textsc{bh}}
\newcommand\AdS{\text{AdS}}
\newcommand\cex{c_\tts{ex}}
\newcommand\Ham{\mathscr{H}}
\newcommand\MADM{M_{\textsc{adm}}}
\newcommand\GN{\text{G}_{\textsc{n}}}
\newcommand\Sentr{\mathcal{S}}
\newcommand\Z{\mathcal{Z}}
\newcommand\Zch{\mathscr{Z}}
\newcommand\Q{\mathcal{Q}}
\newcommand\Qpsi{\mathcal{Q_\psi}}
\newcommand\QpsiZ{\mathcal{Q}_\psi^{(0)}}
\newcommand\J{\mathcal{J}_\psi}
\newcommand\Jextr{\J^{(\text{ex})}}
\newcommand\Sextr{\Sentr^{(\text{ex})}}

\newcommand\Jb{\mathbb{J}}
\newcommand\kalgstar{\mathfrak{K}_3^{^\mathlarger{*}}}
\newcommand\halgstar{\mathfrak{H}_3^{^\mathlarger{*}}}
\newcommand\Nnut{N_\text{\textsc{nut}}}

\newcommand\Gg{G_{\!\textbf{g}}}
\newcommand\Gel{G_\text{\!el}}

\newcommand\V{\mathcal{V}}
\newcommand\W{\mathcal{W}}
\newcommand\F{\mathcal{F}}
\newcommand\Fbo{\mathcalboondox{F}}
\newcommand\X{\mathcal{X}}
\newcommand\Ucal{\mathcal{U}}

\newcommand\T{\textsl{T}}
\newcommand\E{\textsl{E}}
\newcommand\K{\mathcal{K}}
\newcommand\zb{{\bar{z}}}
\newcommand\ib{{\bar{\imath}}}
\newcommand\jb{{\bar{\jmath}}}

\newcommand\A{\mathcal{A}}
\newcommand\B{\mathcal{B}}

\newcommandx{\tts}[1]{\text{\textsmaller{#1}}}
\newcommandx{\dm}[1][1=\mu,usedefault]{\partial_{#1}}
\newcommandx{\dmup}[1][1=\mu,usedefault]{\partial^{#1}}
\newcommandx{\subm}[2][1=p,2=A,usedefault]{{#1}_{\!\mathsmaller{#2}}}
\newcommandx{\subt}[2][1=p,2=A,usedefault]{{#1}_\text{\textsmaller{#2}}}
\newcommandx{\supm}[2][1=p,2=A,usedefault]{{#1}^{\!\mathsmaller{#2}}}
\newcommandx{\supt}[2][1=p,2=A,usedefault]{{#1}^\text{\textsmaller{#2}}}
\newcommandx{\subpt}[3][1=p,2=A,3=B,usedefault]{{#1}^\text{\textsmaller{#3}}_\text{\textsmaller{#2}}}
\newcommandx{\subpm}[3][1=p,2=A,3=B,usedefault]{{#1}^{\mathsmaller{#3}}_{\mathsmaller{#2}}}
\newcommandx{\sh}[1][1=\alpha,usedefault]{\sinh\left(#1\right)}
\newcommandx{\ch}[1][1=\alpha,usedefault]{\cosh\left(#1\right)}
\newcommandx{\sech}[1][1=\alpha,usedefault]{\mathrm{sech}\left(#1\right)}
\newcommandx{\cosech}[1][1=\alpha,usedefault]{\mathrm{cosech}\left(#1\right)} \newcommandx{\LCTd}[4][1=\mu,2=\nu,3=\rho,4=\sigma,usedefault]{\veps_{#1#2#3#4}}
\newcommandx{\LCTu}[4][1=\mu,2=\nu,3=\rho,4=\sigma,usedefault]{\veps^{#1#2#3#4}}
\newcommandx{\gmetr}[2][1=\mu,2=\nu,usedefault]{g_{{#1}{#2}}}
\newcommandx{\invgmetr}[2][1=\mu,2=\nu,usedefault]{g^{{#1}{#2}}}
\newcommandx{\spc}[3][1=\mu,2=a,3=b,usedefault]{{\w_{#1}}^{\!\!{#2}{#3}}}
\newcommandx{\Conn}[3][1=\mu,2=\nu,3=\lambda,usedefault]{{\Gamma_{{#1}{#2}}}^{\!\!#3}}
\newcommandx{\viel}[2][1=\mu,2=a,usedefault]{{e_{#1}}^{\!#2}}
\newcommandx{\inviel}[2][1=a,2=\mu,usedefault]{{e_{#1}}^{#2}}
\newcommandx{\vieluu}[2][1=\mu,2=a,usedefault]{e^{#1#2}}
\newcommandx{\Rdduu}[4][1=\mu,2=\nu,3=a,4=b,usedefault]{{R_{{#1}{#2}}}^{{#3}{#4}}}
\newcommandx{\gamui}[1][1=0,usedefault]{\gamma^{\mathsmaller{#1}}}
\newcommandx{\gamdi}[1][1=0,usedefault]{\gamma_{{}_{#1}}}
\newcommandx{\Gm}[2][1=i,2=j,usedefault]{\mathpzc{G}_{{#1}{#2}}}
\newcommandx{\Gmu}[2][1=i,2=j,usedefault]{\mathpzc{G}^{{#1}{#2}}}
\newcommandx{\G}[1][1=3,usedefault]{G_{(#1)}}
\newcommandx{\HH}[1][1=3,usedefault]{H_{(#1)}}
\newcommandx{\Hc}[1][1=3,usedefault]{H_{(#1)}^{^\text{\textlarger{c}}}}
\newcommandx{\HHstar}[1][1=3,usedefault]{H_{(#1)}^{^\mathlarger{*}}}
\newcommandx\AL[2][1=\Lambda,2=\mu,usedefault]{A^{#1}_{#2}}
\newcommandx\ALd[2][1=\Lambda,2=\mu,usedefault]{A_{{#1}\,{#2}}}
\newcommandx\ALe[2][1=\Lambdae,2=\mu,usedefault]{A^{#1}_{#2}}
\newcommandx\ALed[2][1=\Lambdae,2=\mu,usedefault]{A_{{#1}\,{#2}}}
\newcommandx\AM[2][1=M,2=\mu,usedefault]{\mathbb{A}^{#1}_{#2}}
\newcommandx\XLe[1][1=\Lambdae,usedefault]{X_{#1}}
\newcommandx{\Th}[2][1=M,2=\alpha,usedefault]{\Theta_{#1}{}^{#2}}
\newcommandx\LgaugO[1][1=0,usedefault]{\mathscr{L}_\text{gaug}^{{}^{(#1)}}}

\newcommandx{\overbar}[1]{\mkern                   1.5mu\overline{\mkern-2.0mu#1\mkern-2.0mu}\mkern 1.5mu}
\newcommandx{\overbarcal}[1]{\mkern                   6.0mu\overline{\mkern-5.5mu#1\mkern-1.0mu}\mkern 1.5mu}  

\makeatletter
\normalsize
\makeatother

\DeclareFixedFont\trfont{OT1}{phv}{b}{sc}{11}

\title{%
       \vspace{-1.5cm}
       \centering\boldmath\LARGE\bfseries%
       Constructing black hole solutions in supergravity theories
       \bigskip
       }

\medskip

\author{\textsc{Antonio Gallerati}
\vspace{0.15em}
}
\affil{%
\makebox[\textwidth][c]{Politecnico di Torino, Dipartimento di Scienza Applicata e Tecnologia, corso Duca degli Abruzzi 24, 10129 Torino, Italy}
}
\affil{Istituto Nazionale di Fisica Nucleare, Sezione di Torino, via Pietro Giuria 1, 10125 Torino, Italy
       }
\affil{\href{mailto:antonio.gallerati@polito.it}{\texttt{antonio.gallerati@polito.it}}
      }

\date{}

\begin{document}

\maketitle


\begin{abstract}
{
\parskip=0pt
We perform a detailed analysis of black hole solutions in supergravity models. 
After a general introduction on black holes in general relativity and supersymmetric theories, we provide a detailed description of ungauged extended supergravities and their dualities. Therefore, we analyze the general form of black hole configurations for these models, their near-horizon behavior and characteristic of the solution. An explicit construction of a black hole solution with its physical implications is given for the STU-model.\par
The second part of this review is dedicated to gauged supergravity theories. We describe a step-by-step gauging procedure involving the embedding tensor formalism, to be used to obtain a gauged model starting from an ungauged one. Finally, we analyze general black hole solutions in gauged models, providing an explicit example for the \,$\N=2$, \,$D=4$ case. A brief review on special geometry is also provided, with explicit results and relations for supersymmetric black hole solutions.
}
\end{abstract}

\bigskip

\tableofcontents



\pagebreak


\section{Introduction} \label{sec:intro}
A long-standing problem in theoretical physics is the definition of a quantum theory of gravity, due to the unique and particular features of this interaction. This kind of theory is called for when studying phenomena in which the gravitational field is so intense as to affect the dynamics of elementary particles: this can occur, for example, in the vicinity of a black hole or, presumably, in the early stages of the evolution of the Universe.\par
In general, gravity becomes important at energy scales comparable to the Planck mass.
Superstring theory in ten dimensions and M-theory in eleven, seem to provide a promising theoretical framework where this unification could be achieved and a consistent quantum theory of gravitation could be formulated. However, many shortcomings originate from their formulation. In particular, our mathematical tools seem not to be adequate to describe superstring theory in all its aspects, including non-perturbative ones. This makes difficult to obtain phenomenological predictions from it.\par
A valuable approach to the study of superstring theory is provided by the formulation of a \emph{supergravity theory}. Supergravity (SUGRA) is primarily a field theory, therefore it has a well-established mathematical framework. Moreover, a supergravity theory can describe a consistent low-energy approximation to some fundamental quantum theory of gravity, like superstring (or M-theory) in the chosen background. In this regard, SUGRA can provide a precise descriptions of physical systems even in non-perturbative regimes, where a superstring formulation is not known. Finally, supergravity encodes \emph{supersymmetry} (SUSY), a spontaneously broken symmetry relating bosons and fermions of the theory, that imposes an additional structure and makes a quantum gravity theory more consistent and easier to analyze.\par
%
Einstein's standard theory of gravity is based on the symmetry principle of invariance under general coordinate transformations, seen as local space-time transformations generated by the local translation generators $P_a$, whose gauge boson is the \emph{graviton}. In a supersymmetric theory of gravity, this invariance is realized as a natural consequence of a more fundamental symmetry principle, the invariance of the theory under space-time (local) dependent supersymmetry transformations.

\paragraph{Minimal and extended models.}
Supersymmetric theories differ in the amount of supersymmetry -- namely in the number \,$\N$ of the supersymmetry generators $Q$ -- and in the field content, which should correspond to multiplets of the super-Poincaré group $G_\textsc{sp}$\,. A number \,$\N$ of supersymmetry generators defines an $\N$-extended supersymmetry. The larger $\N$, the stronger the constraints on the interactions, the larger the maximum spin $j_\text{max}$ of the fields in the supermultiplets. In general, the least value of the maximum spin in the supermultiplets is related to $\N$; in four-dimensional space-time we have $j_\text{max}\geq\N/4$.\par
The construction of extended theories \cite{Cremmer:1982wb} can be performed by coupling the supergravity multiplet to a number $n_\text{c}$ of chiral (or Wess-Zumino) multiplets, each consisting of a chiral fermion and two scalar fields, and a number $\nv$ of vector multiplets, each consisting of a vector field and a chiral fermion. The vector multiplets define the gauge sector, the vector fields possibly gauging a suitable local internal symmetry group, while the chiral multiplets define the matter sector. The gauge sector consists of one vector field and one Majorana fermion. The matter sector has one chiral fermion and two scalar fields (one scalar, one pseudo-scalar). This couple of scalars $\{a,b\}$ in each chiral multiplet enter the Lagrangian and the SUSY transformation laws in the complex combination $z=a+i\,b$.\par
If one considers an extended theory describing the supergravity multiplet ($g_{\mu\nu},\,\psi^i$) coupled to a number of vector and matter multiplets, the consistent definition of a number $\N$ of massless gravitino fields $\psi^i$ ($i=1,\dotsc,\N$) on a curved space-time requires, for each of them, the decoupling of the spin-$1/2$ longitudinal modes. This, in turn, follows from  the invariance of the theory under $\N$-independent supersymmetry transformations. A consistent theory containing \,$\N$ massless gravitinos is an \,$\N$-extended supersymmetric theory of gravity.\par

\subsection{Supergravity}
Supergravity is an extension of Einstein's general relativity that includes supersymmetry \cite{Freedman:1976xh,Deser:1976eh}. General relativity demands extensions since it has many shortcomings. It is incompatible with quantum mechanics and, from a mathematical point of view, pure quantum gravity is not renormalizable and hence has little predictive power.
If we include supersymmetry in a theory of gravity the situation improves, as we can appreciate analysing the simplest example of divergences: zero point energy of the vacuum, can be potentially cancelled by superpartners of ordinary particles.\par
Since supergravity field theories are invariant under local supersymmetry, the underlying superalgebra states that invariance under local supersymmetry implies the invariance under spacetime diffeomorphisms (i.e.\ invariance under general coordinate transformation).

\paragraph{Ungauged SUGRA.}
Supersymmetry constrains the form of the Lagrangian, that is the structure of its kinetic terms, mass terms, couplings and scalar potential. The larger the amount $\N$ of supersymmetry, the more stringent these constraints. The theory is characterized by a bosonic sector and a fermionic one. Once the former is given, the latter is completely fixed by supersymmetry.\par
\sloppy
Let us consider the case of \emph{ungauged supergravities}, namely models where vector fields of the theory are not minimally coupled to other fields. The bosonic sector consists of the graviton field $g_{\mu\nu}$, \,$\nv$ vector fields $A^\Lambda_\mu$ \,({\small$\Lambda$}\;$=1,\,\dotsc,\,\nv)$, \,$\ns$ scalar fields $\phi^s$ \,${(s=1,\,\dotsc,\,\ns)}$.
The possible couplings are constrained by the request of gauge invariance and diffeomorphism invariance of the theory, which allow for the following terms
in the bosonic Lagrangian%
\footnote{%
in the ``mostly minus'' convention
}:
\begin{enumerate}[label=${\odot}$,topsep=1.5ex,itemsep=1.15em,labelsep=1.75ex,after=\vspace{0.75ex}]
\item   the Einstein-Hilbert term,
        \begin{equation}
        \Lagr_{\textsc{eh}}~=\,-\dfrac12\,|\eD|R\;,
        \end{equation}
        describing the gravitational sector of the theory;
\item   the scalar term,
        \begingroup%
        \setlength{\abovedisplayshortskip}{2pt plus 3pt}%
        \setlength{\abovedisplayskip}{2pt plus 3pt minus 4pt}
        \setlength{\belowdisplayshortskip}{5pt plus 3pt}%
        \setlength{\belowdisplayskip}{6pt plus 3pt minus 4pt}%
        \begin{equation}
        \Lagr_{\text{scal}}\=\dfrac12\,|\eD|\,\Gm[s][u](\phi)\,\dd_\mu\phi^s\,\dd_\nu\phi^u\,g^{\mu\nu}\-|\eD|\,V(\phi)\;,\quad \label{Lagrscal}
        \end{equation}
        \endgroup%
        $V(\phi)$ being the scalar potential and $\Gm[s][u](\phi)$ the positive definite metric of a \emph{scalar manifold} $\Mscal$\,;
\item   the vector term,
        \begingroup%
        \setlength{\abovedisplayshortskip}{4pt plus 3pt}%
        \setlength{\abovedisplayskip}{4pt plus 3pt minus 4pt}
        \setlength{\belowdisplayshortskip}{5pt plus 3pt}%
        \setlength{\belowdisplayskip}{6pt plus 3pt minus 4pt}%
        \begin{equation}
        \Lagr_{\text{vect}}\=\dfrac14\,|\eD|\,g^{\mu\rho}g^{\nu\sigma}\,F_{\mu\nu}^\Lambda\;\II_{\Lambda\Sigma}(\phi)\,F_{\rho\sigma}^\Sigma \+\dfrac{1}{8}\,\LCTu\,F_{\mu\nu}^\Lambda\;\RR_{\Lambda\Sigma}(\phi)\,F_{\rho\sigma}^\Sigma\;, \label{Lagrvect}
        \end{equation}
        \endgroup%
        where the first is a Maxwell term, and the latter is a topological term.
\end{enumerate}
Let us discuss more in detail the previous expressions.\par\smallskip
The scalar fields $\phi^s$ in the Lagrangian are described by a \emph{non-linear $\sigma$-model}, that is they are coordinates of a non-compact, Riemannian $\ns$-dimensional differentiable manifold, the target space $\Mscal$. If $G$ is the isometry group of $\Mscal$, a generic element of it will map the scalar fields $\phi^s$ in new ones by the action of $g \in G$ as non-linear functions of the original ones:
\begin{equation}
\phi^{\prime s}\=g\star\phi\=\phi^{\prime s}(\phi^u)\;,
\end{equation}
and the $\sigma$-model action turns out to be invariant under this action of global isometries of the manifold. \par\smallskip
The two terms containing the vector field strengths are called \emph{vector kinetic terms}. A general feature of supergravity theories is that the scalar fields are non-minimally coupled to the vector fields, as they enter these terms through symmetric matrices \,$\II_{\Lambda\Sigma}(\phi)$\, and \,$\RR_{\Lambda\Sigma}(\phi)$\, which contract the vector field strengths. The negative definite matrix \,$\II_{\Lambda\Sigma}(\phi)$ in the Maxwell term generalizes the standard $-1/g^2$ factor in Yang-Mills theories, while the \,$\RR_{\Lambda\Sigma}(\phi)$\, matrix in the topological term plays the role of the so-called $\theta$-angle. \par\smallskip
There is a $\U(1)^{\nv}$ gauge invariance associated with the vector fields:
\begin{equation}
A_\mu^\Lambda\;\;\rightarrow\;\; A_\mu^\Lambda+\partial_\mu\zeta^\Lambda\;;
\end{equation}
and all the fields are neutral with respect to this symmetry group.\par\smallskip
\sloppy
In an ungauged supergravity theory, a scalar potential is allowed only for ${\N=1}$ (\emph{F-term potential}). In extended supergravities, a non-trivial scalar potential can be introduced without explicitly breaking supersymmetry only through the \emph{gauging procedure}, which implies the introduction of a local symmetry group to be gauged by the vector fields of the theory.\par\smallskip
Finally, the isometry group $G$ would alter the vector field equations, due to the non-minimal coupling between scalar and vector fields. It was proven by Gaillard and Zumino that the group $G$ can be promoted to a global symmetry group of the field equations and Bianchi identities (i.e.\ \emph{on-shell global symmetry group}), provided its non-linear action on the scalar fields is combined with an electric-magnetic duality transformation on the vector field strengths and their magnetic duals \cite{Gaillard:1981rj}.\par

\paragraph{Bosonic Lagrangian.}
Summarizing, to fix the bosonic Lagrangian, one has to specify $\Gm(\phi)$, $\II_{\Lambda\Sigma}(\phi)$, $\RR_{\Lambda\Sigma}(\phi)$ and the potential $V(\phi)$. Once the bosonic action is given, the fermionic couplings of the Lagrangian are entirely fixed by supersymmetry, without any freedom left.\par
The bosonic data are not arbitrary, but constrained by supersymmetry, requiring the bosonic Lagrangian
\begin{equation}
\Lagr_{\textsc{bos}}\=\Lagr_{\textsc{eh}}+\Lagr_{\text{scal}}+\Lagr_{\text{vect}}
\end{equation}
to be completed by fermionic couplings to a Lagrangian invariant under local SUSY transformations.\par

\paragraph{Extended supergravities.}
In \,$\N>1$ supergravities, multiplets start becoming large enough to accommodate both scalar and vector fields. As we increase $\N$, the first instance of scalar and vector fields connected by supersymmetry is in the  $\N=2$ vector multiplet. This feature has profound implications on the mathematical structure of the models, since it poses strong constraints on the (non-minimal) scalar-vector couplings in the Lagrangian, that is on the matrices \,$\II_{\Lambda\Sigma}(\phi)$, $\RR_{\Lambda\Sigma}(\phi)$. Given the scalar manifold $\Mscal$, supersymmetry fixes these matrices up to a choice of the \emph{symplectic frame}. The latter is related to the geometric structure of the scalar manifold and associates with each point $\phi$ on the manifold a symmetric symplectic matrix \,$\M(\phi)_{MN}$, and to each isometry transformation $g\in G$ on the same manifold a corresponding constant symplectic matrix.\par
Global isometry transformations on the scalar fields induce, by supersymmetry, global transformations on the vector fields. These act as electric-magnetic transformations on the vector field strengths and their magnetic duals and define the on-shell global symmetries of the theory.\par


\subsection{Black holes}
As a theory of gravity, supergravity has black hole solutions \cite{Wald:1984rg}. Indeed, being invariant under local super-Poincaré transformations, supergravity includes general relativity and describes gravitation coupled to other fields in a supersymmetric framework.\par
A supergravity black hole can be seen as a solitonic solution, that is a time-independent, non-singular, localized solution of the classical equations of motion of a field theory, with finite energy density. It is associated to an additional particle-like (non-perturbative) quantum state that completes the spectrum of a fundamental field theory. This quantum states originate from regular solution of the classical field equation (Einstein equation of general relativity) with the new ingredient of supersymmetry, which requires the presence of vector and scalar fields in appropriate proportion. \par
Supergravity provides a macroscopic (large scale) description of the black hole solutions, analogous to the macroscopic thermodynamic description of gases. In the case of black holes, the microscopic description of the solution is provided by some higher dimensional superstring or M-theory. Following this analogy, the laws of black holes thermodynamics should be explained, at the fundamental level, by a superstring/M-theory just as the standard laws of thermodynamics can be derived from a molecular description of a gas.

\subsubsection{Black holes in General Relativity}
\label{subsubsec:GenRel}
In classical general relativity, the first exact solution to the Einstein equation in the vacuum was found in 1916 by Schwarzschild. It describes space-time around a point-particle of mass $M$, and it is the most general spherically-symmetric solution of Einstein equations in the vacuum.\par
A spherical symmetric solution describing a particle of mass $M$ and charge $Q$ was found in 1918 by Reissner and Nordstr\"om, with metric:
\begin{equation}
ds^2\=\left(1-\frac{2\,r_\textsc{m}}{r}+\frac{r_\textsc{q}^2}{r^2}\right)\,dt^2
    \-{\left(1-\dfrac{2\,r_\textsc{m}}{r}+\dfrac{r_\textsc{q}^2}{r^2}\right)}^{\!-1}dr^2 \-r^2\left(d\theta^2+\sin^2(\theta)\,d\varphi^2\right)\,,
\label{RN}
\end{equation}
where
\begingroup
\belowdisplayskip=12pt
\belowdisplayshortskip=12pt
\begin{equation}
r_\textsc{m}\=\frac{\mathrm{G}_{\textsc{N}}}{c^2}\,M\;; \qqquad r_\textsc{q}^2\=\frac{\mathrm{G}_{\textsc{N}}}{4\pi\,c^4}\,Q^2\;.
\end{equation}
\endgroup
This solution has two horizons at
\begin{equation}
r_{{}_\pm} \= r_\textsc{m} \,\pm\, \sqrt{r_\textsc{m}^2-r_\textsc{q}^2}
\qqquad\; \text{if}\quad r_\textsc{m}>r_\textsc{q}\;,
\end{equation}
while it is \emph{singular} (curvature singularity not hidden inside an horizon) if \,$r_\textsc{m}<r_\textsc{q}$\,.\par\smallskip
In 1963 R. Kerr generalized Schwarzschild's solution to describe a spinning particle, and this solution was further generalized by E. Newman et al.\ in 1965 \cite{Newman:1965my} to describe a charged spinning particle (\emph{Kerr-Newman solution}). This represents the most general asymptotically flat, axisymmetric  solution to Einstein's theory of gravity coupled to an electromagnetic field, or Einstein-Maxwell theory.

\paragraph{Black hole thermodynamics.}
It is possible to formulate a formal analogy between the principles of thermodynamics and black hole properties (related to surface gravity, absorption of a particle, horizon area) calculated from pure, classical general relativity analysis \cite{Hawking:1971vc,Bardeen:1973gs}. In particular, the following general properties were found:
\begin{enumerate}[label=\roman*),itemsep=0.55em,after=\vspace{0.6ex}]
\item  {the surface gravity $\kappa$ is uniform over the horizon;}
\item  {the rest energy variation of a black hole due to the absorption of
        a spinning, charged particle can be written%
        \footnote{%
        in the presence of scalar fields coupled to the solution, which is typical of supergravity black holes, a further term should be added, which depends on the \emph{scalar charges}, defined in terms of the radial derivatives of the scalar fields at spatial infinity \cite{Gibbons:1996af}
        }:
        \begin{equation}
        \delta M \= \frac{\kappa}{8\pi\GN}\,\delta A_\textsc{h}
            +\frac{1}{c^2}\,\Omega_\textsc{h}\;\delta\mathcal{J}_\textsc{h}
            +\Phi\,\delta Q\;,
        \end{equation}
        in terms of the horizon area $A_\textsc{h}$, angular momentum at the horizon $\mathcal{J}_\textsc{h}$, charge $Q$, angular velocity at the horizon $\Omega_\textsc{h}$, electric potential $\Phi$;}
\item  {the total area of the black hole horizons can not decrease:
        \;\;$\delta A_\textsc{h}\ge0$\;;}
\item  {the extremal solution ($\kappa=0$) can not be reached through a
        finite process.}
\end{enumerate}
If we identify $\kappa$ with the temperature and $A_\textsc{h}$ with the entropy of the solution, we can recognize an analogy between these properties and the zeroth, first, second and third laws of thermodynamics. The fact that the analogy is not just formal -- and that these are the actual laws of thermodynamics applied to a black hole -- was proven in 1974 by Hawking \cite{Hawking:1974sw}, whose quantum analysis showed that black holes can emit black-body radiation at a temperature
\begin{equation}
T\=\frac{\kappa\,\hbar}{2\pi\,k_\textsc{b}\,c}\;,
\label{temperature}
\end{equation}
where $k_\textsc{b}$ is the Boltzmann constant. Since general relativity tells us that the total area $A_\textsc{h}$ of a black hole horizon cannot decrease in our Universe, we can see the correspondence with the classical second law of thermodynamics provided we identify the entropy of the solution with \cite{Hawking:1971vc,Bekenstein:1973ur}:
\begin{equation}
\mathcal{S}\=\frac{k_\textsc{b}}{4\,\ell_\textsc{p}^2}\,A_\textsc{h}\;,
\label{entropy}
\end{equation}
where $\ell_\textsc{p}$ is the Planck length. This is the famous \'area law" or Bekenstein-Hawking formula for the entropy. One of the main  successes of superstring theory has been indeed the derivation of the Bekenstein-Hawking formula \eqref{entropy} from a microstate counting \cite{Strominger:1996sh}, as expected for a microscopic quantum description of gravity.

\subsubsection{Black holes in Supergravity}
We have seen that a static, asymptotically flat, charged black hole configuration is described by the Reissner-Nordström solution.
If we want the solution not to be singular, its spatial singularity must be hidden inside an event horizon, so that it does not pose problems of predictability outside the black hole. To satisfy this requirement, its mass $M$, electric and magnetic charges $q$ and $p$, should obey a \emph{regularity bound} that, in natural units, reads \cite{Andrianopoli:2006ub}:
\begin{equation}
M^2\,\ge\,\frac{p^2+q^2}{2}\;.
\label{regbound0}
\end{equation}
In general relativity there is a \emph{cosmic censor conjecture} \cite{Wald:1984rg}, according to which the above condition is satisfied by all black hole solutions in nature, that is our Universe is clear of naked singularities which would make it unpredictable. However, there is no definite proof of this conjecture in classical theory.

\paragraph{Supersymmetry and black holes.}
Some things change in a supergravity theory, due to the presence of supersymmetry \cite{Kallosh:1992ii}. As solutions to a supersymmetric theory, supergravity black holes must belong to massive representations of the super-Poincaré algebra.
In general, supersymmetric field theories can have multiple $Q^A$ generators, where $A=1,\,\dotsc\,,\;\N$\, and where $\N$ is the number of supersymmetries in the theory. In this case, one can write the SUSY algebra as:
\begin{equation}
\{Q_\alpha^\ms{A}\,,\;\bar{Q}_\beta^\ms{B}\} \= 2\,\delta^\ms{AB}\,P_\mu\,\Gamma_{\alpha\beta}^\mu + Z^\ms{AB}\,\delta_{\alpha\beta}\;,
\label{algZ}
\end{equation}
where we have considered also the action of the \emph{central charge operator} $Z^\ms{AB}$.
%
If computed on a black hole background, the central charges of the algebra \eqref{algZ} have non-vanishing values which depend on the electric and magnetic charges of the theory; they can in fact be considered as topological quantities associated with the solution \cite{Witten:1978mh}. \par
In an $\N$-extended theory, the central charges are entries of an \;$\N\times\N$ antisymmetric matrix $Z^\ms{AB}$
\begin{equation}
Z^\ms{AB}~=\,-Z^\ms{BA}\;, \qqquad  A,B=1,\dotsc,\,\N\;.
\end{equation}
It can be easily shown that supersymmetry implies that the mass $M$ of the solution must be greater than the modulus of all the skew-eigenvalues $z_\ell$ of $Z^\ms{AB}$:
\begin{equation}
M~\ge~|z_\ell|\;, \qqquad \ell=1,\dotsc,\frac{\N}{2}\;,
\label{regcond2}
\end{equation}
and these can be thought as the supergravity analogue of the so-called Bogomol'nyi-Prasad-Sommerfield condition (BPS bound) for solitonic solutions to gauge theories \cite{Gibbons:1982fy}.
On the Reissner-Nordstr\"om solution the above condition implies the regularity bound \eqref{regbound0}. We have obtained that, at least for static solutions, supersymmetry acts as a cosmic censor: it naturally provides a general principle which rules out the existence of naked singularities. \par
If the inequalities \eqref{regcond2} are not saturated for any $\ell$, the solution is \emph{non-extremal} and has a non-vanishing Hawking temperature. By quantum mechanical effects it radiates (Hawking-evaporation) until its mass equals the largest $|z|_\text{max}$ of the $|z_\ell|$ eigenvalues and the temperature drops to zero. The resulting solution is called \emph{extremal} and preserves a fraction of the $\N$ supersymmetries (at least $1/\N$).\par
Supersymmetric black holes are called BPS (i.e.\ saturating the BPS bound) and are solutions to a set of first-order differential equations, the \emph{Killing spinor equations}, which imply the second-order field equations. BPS-solutions have played an important role in the study of superstring non-perturbative dualities, since $|z_\ell|$ are duality-invariant quantities and are protected, to a certain extent, from quantum corrections by supersymmetry.\par
Supergravity has more general solutions than the above Reissner-Nordstr\"om one, featuring a non-trivial interplay between scalar and vector fields of the theory. These new solutions belong to different topological sectors of the theory and, after evaporating, the described black holes reach a lowest mass, zero-temperature (extremal) state in which $M$ equals a new characteristic quantity $M'>|z_\ell|$. A remarkable feature of these extremal solutions is that, although they do not preserve any supersymmetry and thus are non-BPS, they are still described by a set of first-order differential equations which imply the second order field equations \cite{Andrianopoli:2006ub}.\par

\paragraph{Attractor mechanism.}
In the case of extremal black hole configurations, i.e.\ solutions with vanishing Bekenstein-Hawking temperature, either static or under-rotating (rotating with no ergosphere), the entropy only depends on the quantized charges of the theory and not on the values of scalar fields at infinity. This reflects a general property of these solutions known as \emph{attractor mechanism} \cite{Ferrara:1995ih,Ferrara:1996dd,Ferrara:1996um}, according to which the scalar \'hair" of the black hole runs into a fixed value on the horizon, independently of the boundary conditions at spatial infinity. For static, spherically symmetric black holes, the fixed values of the scalars at the horizon are determined in terms of the quantized electric and magnetic charges characterizing the solution, as extrema of some suitable effective potential $\VBH(\phi^s,\,e,m)$, function of the scalars and of the electric and magnetic charges of the theory \cite{Ferrara:1997tw,Ceresole:2007wx,Andrianopoli:2007gt}.


\newpage

\section{Ungauged extended supergravities} \label{sec:ungsugra}

\subsection{Overview}
Stationary black holes are time-independent solutions of Einstein theory of gravity which exhibit a space-time singularity hidden by an event horizon. The fact that classical black hole solutions satisfy the laws of thermodynamics (with a well-defined expression for the entropy, given by the Bekenstein--Hawking formula) suggests that we can think of them as macroscopic ensembles of microscopic
states, pertaining to some fundamental quantum field theory of gravity. Supergravity, as a theory of gravity, admits black hole solutions. We shall now restrict to ungauged supergravity models and study their stationary black hole solutions.\par
An \emph{ungauged supergravity} is a supergravity model in which the vector fields are not minimally coupled to any other field in the theory. The vectors of the theory transform under an abelian group and there are no charged fields. Moreover, the only admitted vacuum of the theory is a supersymmetric Minkowski vacuum: black hole solutions are therefore (locally) asymptotically flat.\par
We will focus on the study of the bosonic sector of the theory, the total structure being to a large extent determined by supersymmetry. The bosonic sector consists of the graviton field, $\nv$ vector fields and $\ns$ scalar fields. The possible couplings are constrained by the request of supersymmetry and diffeomorphism invariance. The scalar fields are described by a non-linear $\sigma$-model, that is they are coordinates of a non-compact target space, a Riemannian differentiable manifold. The $\sigma$-model action turns out to be invariant under the action of global isometries of the scalar manifold, i.e.\ the isometry group of the manifold is a global symmetry.\par
In extended ($\N>1$) supergravities, multiplets start becoming large enough as to accommodate both scalar and vector fields. This feature has important implications on the mathematical structure of the models, since it poses strong constraints on the (non-minimal) scalar-vector couplings in the Lagrangian. Given the scalar manifold, supersymmetry fixes the couplings, up to a choice of the frame related to the geometric structure of the scalar manifold. Moreover, with each point on the manifold a symmetric symplectic matrix is associated, and with each isometry transformation on the same manifold is associated a corresponding constant symplectic matrix. Finally, global isometry transformations on the scalar fields
induce, by supersymmetry, global transformations on the vector fields that act as electric-magnetic transformations on the vector field strengths and their magnetic duals, defining the on-shell global symmetries of the theory.\par

\paragraph{Bosonic Lagrangian}
Let us consider stationary solutions in an extended ungauged $D=4$ supergravity theory. The bosonic sector consists in $\ns$ scalar fields $\phi^s(x)$, $\nv$ vector fields $A^\Lambda_\mu(x)$ ({\small$\Lambda$}\;$=1,\,\dotsc,\,\nv$), and the graviton $g_{\mu\nu}(x)$. The physical configuration is described by the four-dimensional Lagrangian%
\footnote{%
we will use the ``mostly minus'' convention, with \;$8\pi\GN=c=\hbar=1$\; and \;$\veps_{0123}=-\veps^{0123}=1$\,.
}%
:
\begin{equation}
\frac{1}{\eD}\;\Lagr_{(4)}~=\,
    -\frac{R}{2}\+\frac{1}{2}\,\Gm[s][u](\phi)\,\dm\phi^s\,\dmup\phi^u
    \+\frac{1}{4}\;\II_{\Lambda\Sigma}(\phi)\,F^\Lambda_{\mu\nu}\,F^{\Sigma\,\mu\nu}
    \+\frac{1}{8\,\eD}\,\RR_{\Lambda\Sigma}(\phi)\,\veps^{\mu\nu\rho\sigma}\,F^\Lambda_{\mu\nu}
    \,F^{\Sigma}_{\rho\sigma}
\label{boslagr}
\end{equation}
where
\begin{equation}
F^\Lambda_{\mu\nu}=\dm A^\Lambda_\nu -\dm[\nu]A^\Lambda_\mu\;,\qqquad \eD=\sqrt{|\Det(g_{\mu\nu})|}\,.
\end{equation}
The scalar fields $\phi^s$ are described by a non-linear $\sigma$-model, that is they are coordinates of a non-compact, Riemannian $\ns$-dimensional differentiable manifold $\Mscal$ (target space). The positive-definite metric on the manifold is $\Gm[s][u](\phi)$, and the $\sigma$-model action is invariant under the action of global (i.e.\ space-time independent) isometries of the scalar manifold.
We will see that the group $G$ can be promoted to a global symmetry group of the field equations and Bianchi identities, if its action on the scalar fields is combined with a suitable electric-magnetic duality transformation on the vector field strengths and their magnetic duals.

\subsection{Scalar manifolds of extended supergravities} \label{sec:scalmanif}
In all $\N>2$ models, supersymmetry constrains the scalar manifold
to be \emph{homogeneous symmetric}, while the $\N=2$ models also allow for a class of manifolds that are only homogeneous or even non-homogeneous. The scalar manifolds for $\N=2$ supergravities have been studied in \cite{deWit:1983xhu,deWit:1984rvr,Cremmer:1984hc,Castellani:1990tp,Strominger:1990pd,Andrianopoli:1996cm,Craps:1997gp}.
\par\smallskip
A manifold $\Ms$ is said \emph{homogeneous} if any couple of points is connected by an isometry, that is, the isometry group $G$ has a transitive action on  $\Ms$. This means that any point $p$ can be reached from the origin of the manifold $\OO$ through a (not necessarily unique) element of the isometry group $G$.
%
Let us denote by $H$ the isotropy group of the origin $\OO$, i.e.\ $H\star\OO=\OO$.
If we denote by \,$gH=\{gh\in G,\;\;h\in H\}$\, the \emph{left coset} of $H$ in $G$, there is a one-to-one correspondence between the points of the homogeneous manifold $\Ms$ and left cosets $gH$:
\begin{equation}
p \in \Ms \qlq g_p\,H\subset G \;.
\label{gp}
\end{equation}
The set of all left cosets of $H$ in $G$ is denoted by $G/H$, and there is a bijection (or diffeomorphism) between $\Ms$ and $G/H$ so that one can make the identification:
\begin{equation}
\Ms ~\sim~ G/H \;,
\end{equation}
where $\sim$ stands for diffeomorphic. The set $G/H$ is called a \emph{coset manifold}, thus homogeneous spaces can be described using correspondent coset manifolds.
Since $\Ms$ is a metric manifold and $G$ its isometry group, $\Ms$ and  $G/H$ are isometric: all geometric quantities of $\Ms$ like its connection, curvature or geodesics, can be computed on $G/H$.%
\par
A generic element $g$ of $G$ is defined by a number of continuous parameters given by $\dim(G)$. Through right multiplication by an element of $H$, we may fix a number \,$\dim(H)$\, of these parameters, so that each left-coset depends on a minimum number of parameters given by $\dim(G)-\dim(H)$; this number turns out to be the dimension of manifold $\Ms$:
\begin{equation}
\dim(\Ms)\=\dim(G)-\dim(H)\,.
\end{equation}
Let $\phi^s$ denote the $\dim(\Ms)$ parameters obtained upon fixing the right-action of $H$. The corresponding representative of each coset space is denoted by \,$\LL(\phi^s)\in G$, and each point of $\Ms$ can be described in terms of a \emph{coset representative} $\LL(\phi^s)$:
\begin{equation}
p \in \Ms \qlq \LL(\phi^s)~\in~g_p\,H~\subset~G\,.
\end{equation}
The parametrization is provided once this fixing is performed, namely when a specific representative $\LL(\phi^s)$ of each coset $g_p\,H$ is taken to represent the corresponding point $p$ of $\Ms$. \par\smallskip
Let $g\in G$ be an isometry of the manifold, $p$ a point of coordinates $\phi=\phi^s$\, and \,$p'=g\star p$\, the transformed of \,$p$\, through \,$g$, of coordinates $\phi'=g\star\phi=\phi^{\prime s}(\phi^u)$. Now, since both \,$\LL(g\star\phi)$\, and \,$g\,\LL(\phi)$\, represent the same point $p'$, they must belong to the same left-coset so that we have:
\begin{equation}
g\,\LL(\phi)\=\LL(g\star\phi)\;h(\phi,\,g)\;,
\label{gLh}
\end{equation}
where the element $h(\phi,\,g)$ of $H$ is called \emph{compensator} and in general depends on $g$ and on the point $p$ of coordinates $\phi$.\par \smallskip
In general, $G$ may not be a semisimple Lie group. Homogeneous manifolds occurring in supergravity theories are non-compact, simply-connected spaces. Let $\mathfrak{g}$ and $\mathfrak{H}$ denote the Lie algebras of the groups $G$ and $H$, respectively. The Lie algebra of $G$ can be splitted as
\begin{equation}
\mathfrak{g}\=\mathfrak{H} \oplus \mathfrak{K} \;,
\end{equation}
and, being $\mathfrak{H}$ a Lie algebra, we must have:
\begin{equation}
[\mathfrak{H},\,\mathfrak{H}]~\subseteq~\mathfrak{H} \;.
\end{equation}
Now, one can always define the subset $\mathfrak{K}$ so that:
\begin{equation}
[\mathfrak{H},\,\mathfrak{K}]~\subseteq~\mathfrak{K} \;.
\label{reductive}
\end{equation}
The above adjoint action of $\mathfrak{H}$ on $\mathfrak{K}$ defines a \emph{representation} of $H$.
The previously introduced space $\mathfrak{K}$ can be viewed as the \emph{tangent space} to $G/H$ at the origin. In general, however, one has:
\begin{equation}
[\mathfrak{K},\, \mathfrak{K}]~\subseteq~\mathfrak{K}\oplus\mathfrak{H}\;,
\end{equation}
and it can be proven that, if it is possible to define a $\mathfrak{K}$ so that
\begin{equation}
[\mathfrak{K},\,\mathfrak{K}]~\subseteq~\mathfrak{H} \,;
\label{symmetric}
\end{equation}
the homogeneous space is also \emph{symmetric} \cite{Gallerati:2016oyo}. A symmetric space is defined in general as a space invariant under parallel translations (curvature covariantly constant). Symmetric, simply-connected spaces are also homogeneous.

\paragraph{Cartan decomposition.}
For non-compact, simply-connected symmetric spaces with negative curvature there exists a transitive semisimple, non-compact isometry group $G$, with $H$ as \emph{maximal compact subgroup}. In any given matrix representation of $G$, one can choose a basis in which $\mathfrak{H}$ is represented by anti-hermitian matrices and $\mathfrak{K}$ by hermitian ones:
\begin{equation}
H\in\mathfrak{H} \;\;\Rightarrow\;\; H^\dagger=-H\;,
\qqquad
K\in\mathfrak{K} \;\;\Rightarrow\;\; K^\dagger=K\;.
\end{equation}
This basis is called the \emph{Cartan basis}. Properties \eqref{reductive} and \eqref{symmetric} directly follow from commutation rules%
. In the corresponding basis \,$T_\mathcal{A}=\{H_q,\,K_{\und{s}}\}$\, of generators of $\mathfrak{g}$, condition \eqref{symmetric} reads:
\begin{equation}
[K_{\und{s}},\,K_{\und{u}}] \= \mathcal{C}_{{\und{s}}\,{\und{u}}}{}^q\,H_q \;.
\end{equation}
If $\{K_s\}$ denote a basis of $\mathfrak{K}$ of hermitian matrices we can write the coset representative $\LL(\phi)$ as:
\begin{equation}
\LL(\phi^s)=\exp(\phi^s\,K_s)\;,
\label{cartpar}
\end{equation}
and the parametrization is called \emph{Cartan parametrization}. It is defined in terms of the coordinates  $\phi^s$, that transform linearly under $H$, namely in the representation $\Rs_\mathfrak{K}$ defined by the adjoint action of $H$ on the space $\mathfrak{K}$.

\paragraph{Solvable decomposition.}
We already said that, in general, \,$\N=2$ supergravity admits non-homogeneous, homogeneous and homogeneous-symmetric scalar manifolds, while the scalar manifolds of $\N>2$ supergravities are only of homogeneous-symmetric type.
All homogeneous scalar manifolds (symmetric or not) are of \emph{normal type}, that is they admit a transitive \emph{solvable Lie group} of isometries whose action on $\Ms$ is free.
A solvable Lie group $G_\text{solv}$ can be locally described as the Lie group generated by the \emph{solvable Lie algebra} $\mathcalboondox{s}$:
\begin{equation}
G_\text{solv}=\exp(\mathcalboondox{s}) \;.
\end{equation}
A Lie algebra $\mathcalboondox{s}$ is solvable if
\;$\textbf{D}^k\mathcalboondox{s}=0$, for some $k>0$. The derivative ${\bf D}$ of a Lie algebra $\mathfrak{g}$ is defined as
\begin{equation}
\textbf{D}\mathfrak{g} ~\equiv~ [\mathfrak{g},\mathfrak{g}] \;,\qqquad
\textbf{D}^n\mathfrak{g} ~\equiv~
    [\textbf{D}^{n-1}\mathfrak{g},\,{\bf D}^{n-1}\mathfrak{g}]\;.
\end{equation}
and, in a suitable basis of a given representation, all the elements of the solvable Lie group or algebra are described by upper (or lower) triangular matrices.\par
Since there is a transitive solvable group \,$G_\text{solv}$\, of isometries with a free action on $\Ms$, we can choose a coset representative $\LL_s(\phi_p)$ fixing a suitable right-action of $H$, so that
\begin{equation}
\left\{\LL_s(\phi_p)\right\} \= G_\text{solv}\;,  \qquad  p\in\Ms \;,
\end{equation}
and this means that the manifold $\Ms$ is isometric to a solvable Lie group,
\begin{equation}
\Ms~\sim~G_\text{solv} \;,
\end{equation}
once fixed, on the tangent space at the origin of $G_\text{solv}$, the metric of the tangent space at the corresponding point of $\Ms$. This procedure defines a parametrization $\phi=\phi^s$ called \emph{solvable parametrization} of $\Ms$. In all parameterizations, the origin $\OO$ is defined as the point in which the coset representative equals the identity element of $G$ and thus the $H$-invariance of $\OO$ is manifest, $\LL(\OO)=\Id$.\par
Both the solvable and the Cartan parameterizations (for symmetric cosets) are global parameterizations of the scalar manifold. For symmetric manifolds, the solvable Lie group $G_\text{solv}$ is defined by the \emph{Iwasawa decomposition} of the non-compact semisimple group $G_\text{semi}$ with respect to $H$, according to which there is a unique decomposition of a generic element $g$ of $G_\text{semi}$ as the product of an element $s$ of $G_\text{solv}$ and an element $h$ of $H$:
\begin{equation}
\forall g\in G_\text{semi}  \;\;\Rightarrow\;\;
g=s\,h \qqquad\quad
\text{with}\;\; s\in G_\text{solv}\,,\,\; h\in H \;,
\end{equation}
and this defines a unique coset representative $\LL_s$ for each point of the manifold $\Ms$.\par
The solvable parametrization is very useful when the $D=4$ dimensional supergravity comes from a Kaluza-Klein reduction (on some internal compact manifold) of an higher dimensional theory: the solvable coordinates can in fact be directly used to describe the dimensionally reduced fields, the parametrization making manifest the shift symmetries of the metric%
\footnote{the drawback of this description is that $\mathcalboondox{s}$ does not define the carrier of a representation of $H$ as $\mathfrak{K}$ does: now the above eq.\ \eqref{reductive} does not hold for $\mathcalboondox{s}$,\; i.e. $[\mathfrak{H},\,\mathcalboondox{s}]\nsubseteq\mathcalboondox{s}$}%
.
In the following sections we will restrict ourselves to \emph{symmetric cosets}, of which we can give a description either in terms of Cartan or solvable coordinates.

\paragraph{Vielbein and connection.}
Let $\LL(\phi)$ be a coset representative corresponding to a generic parametrization. We can construct the \emph{left-invariant one form} on $G/H$:
\begin{equation}
\Omega\=\LL^{-1}\,d\LL \;,
\label{omegapro}
\end{equation}
with value in the Lie algebra $\mathfrak{g}$. The above one-form can be expanded in the Cartan basis $\{T_A\}=\{H_q,\,K_{\und{s}}\}$:
\begin{equation}
\Omega(\phi)\=\sigma^A(\phi)\,T_A\=\LL(\phi)^{-1}\,d\LL(\phi)\=
V^{\und{s}}(\phi)\,K_{\und{s}}+\ww^r(\phi)\,H_r\=\pp(\phi)+\ww(\phi)\;,
\label{Om}
\end{equation}
where the above quantities can be written
\begin{equation*}
\Omega(\phi)=\Omega_s(\phi)\,d\phi^s\,,\;\quad\;
V^{\und{u}}(\phi)=V_s{}^{\und{u}}(\phi)\,d\phi^s\,,\;\quad\;
\pp(\phi)=V^{\und{s}}(\phi)\,K_{\und{s}}\,,\;\quad\; \ww(\phi)=\ww^q(\phi)\,H_q\;.
\end{equation*}
We use the underlined indices ($\und{s},\,\und{u},\,\dotsc$) as \emph{rigid indices} that label the basis components $\{K_{\und{s}}\}$ of the tangent space to the group manifold defining a representation $\Rs_\ms{K}$ of $H$, while we denote the remaining non-underlined indices ($s,\,u,\,\dotsc$) as \emph{curved indices} labeling the coordinates $\phi^s$, that is the scalar fields. We emphasize that the scalar fields carry rigid indices only in the Cartan parametrization.\par\smallskip
The exterior derivative of the left-invariant one form $\Omega$ gives
\begin{equation}
d\Omega\=d\LL^{-1}\wedge d\LL=d\LL^{-1}\,\LL\,\LL^{-1}\wedge d\LL
~=\,-\LL^{-1}d\LL\wedge \LL^{-1}d\LL~=\,-\Omega\wedge\Omega\;,
\end{equation}
the above relation being the Maurer-Cartan equations for the group $G$:
\begin{equation}
d\Omega+\Omega\wedge\Omega\=0\;.
\label{MCeq}
\end{equation}
%
Now we want to evaluate how the previously defined quantities transform under the action of $G$. For any $g\in G$, using eq.\ \eqref{gLh}, we can write $\LL(g\star\phi)=g\,\LL(\phi)\,h^{-1}$, so that:
\begin{equation}
\Omega(g\star\phi)\=h\,\LL(\phi)^{-1}\,g^{-1}\;d\left(g\,\LL(\phi)\,h^{-1}\right)
\=h\,\LL(\phi)^{-1}\big(d\LL(\phi)\big)\,h^{-1}+h\,dh^{-1} \;.
\end{equation}
From \eqref{Om} we find:
\begin{equation}
\begin{split}
\Omega(g\star\phi)&\=\pp(g\star\phi)\;+\;\ww(g\star\phi)\=
  V^{\und{s}}(g\star\phi)\,K_{\und{s}}\;+\;\ww^r(g\star \phi)\,H_r\=\\
  &\=h\,(V^{\und{s}}(\phi)\,K_{\und{s}})\,h^{-1}+h\,(\ww^u(\phi)\,H_u)\,h^{-1}+h\,dh^{-1}\=\\
  &\=h\,\pp(\phi)\,h^{-1}+h\,\ww(\phi)\,h^{-1}+h\,dh^{-1}\,.
\end{split}
\end{equation}
Since $h\,dh^{-1}$ is the left-invariant 1-form on $\mathfrak{H}$, it has value in this algebra. Projecting the above equation over $\mathfrak{K}$ and $\mathfrak{H}$, we find:
\begin{subeqs} 
\begin{align}
\pp(g\star\phi)&\=h\,\pp(\phi)\,h^{-1}\;,\label{Ptra}\\
\ww(g\star\phi)&\=h\,\ww(\phi)\,h^{-1}+h\,dh^{-1}\;.\label{omtra}
\end{align}
\end{subeqs}
%
In analogy with the standard description of curved space-time, we see that here $V^{\und{s}}$ plays the role of the vielbein 1-form, and $\ww$ is identified with the $H$-connection.\par
For symmetric spaces, from \eqref{MCeq} it follows that $\ww$ and $\pp$ satisfy the conditions
\begin{subeqs} 
\begin{align}
\mathscr{D}\pp& ~\equiv~ d\pp+\ww\wedge\pp+\pp\wedge\ww\=0\;,\label{DP}
\\
R(\ww)& ~\equiv~ d\ww+\ww\wedge\ww~=\,\pp\wedge\pp\;,\label{RW}
\end{align}
\end{subeqs}
where we have defined the $H$-covariant derivative $\mathscr{D}\pp$ of $\pp$ and the $\mathfrak{H}$-valued curvature $R(\ww)$ of the manifold, that can be written in components as:
\begin{equation}
R(\ww)\=\frac{1}{2}\,R_{su}\,d\phi^s\wedge d\phi^u \qRq
R_{su}~=\,-[\pp_s,\,\pp_u]\;\in\; \mathfrak{H}\;.
\label{Rcompo}
\end{equation}

\paragraph{Metric on \,$\Ms$.}
Now we want to construct a $G$-invariant metric on the scalar manifold $\Ms$ in terms of $V^{\und{s}}$. In analogy with the definition of a (local) Lorentz invariant metric $\eta_{ab}$ on the tangent space of a curved space-time, here we want to define on the tangent space to $\Ms$ an $H$-invariant (positive definite) metric $\kappa_{\und{s}\und{u}}$. With reference to a matrix representation of $G$, we define $\kappa_{\und{s}\und{u}}$ as the restriction of the Cartan-Killing metric of $\mathfrak{g}$ to $\mathfrak{K}$:
\begin{equation}
\kappa_{\und{s}\und{u}}\equiv k\,\Tr\left(K_{\und{s}}\,K_{\und{u}}\right)\;,
\end{equation}
where $k$ is a representation-dependent normalization constant. The metric on $\Ms$ is defined as:
\begin{equation}
\Gm[s][u](\phi)=V_s{}^{\und{s}}(\phi)\,V_u{}^{\und{u}}(\phi)\,\kappa_{\und{s}\und{u}}
\quad\;\Leftrightarrow\quad\;
ds^2(\phi)=\Gm[s][u](\phi)\,d\phi^s\,d\phi^u=k\,\Tr\left(\pp(\phi)^2\right)\;,
\end{equation}
where $\pp=\pp_s\,d\phi^s$.
The $G$-invariance of this metric immediately follows from equation \eqref{Ptra} 
%
and the $\sigma$-model Lagrangian density can be written in the form
\begin{equation}
\Lagr_{\text{scal}}
    \=\frac{\eD}{2}\,\Gm[s][u](\phi)\,\dm\phi^s\,\dmup\phi^u
    \=\frac{\eD}{2}\,k\;\Tr\big[\pp_s(\phi)\,\pp_u(\phi)\big]\;
                    \dm\phi^s\,\dmup\phi^u\;,
\label{Lscal}
\end{equation}
and, just as the metric $ds^2$, is manifestly invariant under global $G$ and local $H$-transformations acting on $\LL$ as in \eqref{gLh}.

\paragraph{Killing vectors.}
Let us denote by $t_\alpha$ the infinitesimal generators of $G$, defining a basis of its Lie algebra $ \mathfrak{g}$ and satisfying the corresponding commutation relations
\begin{equation}
[t_\alpha,\,t_\beta]\=f_{\alpha\beta}{}^\gamma\,t_\gamma\;,
\label{talg}
\end{equation}
$f_{\alpha\beta}{}^\gamma$ being the structure constants of $\mathfrak{g}$. Under an infinitesimal $G$-transformation generated by $\epsilon^\alpha\,t_\alpha$ (with $\epsilon^\alpha \ll 1$)
\begin{equation}
g ~\approx~ \Id+\epsilon^\alpha\,t_\alpha\;,
\end{equation}
the scalars transform as
\begin{equation}
\phi^s\rightarrow \phi^s+\epsilon^\alpha\,k^s_\alpha(\phi)\;,
\end{equation}
$k^s_\alpha(\phi)$ being the Killing vector associated with $t_\alpha$ satisfying the algebraic relations (note the minus sign):
\begin{equation}
[k_\alpha,\,k_\beta]~=\,-f_{\alpha\beta}{}^\gamma\,k_\gamma\;.
\end{equation}

\subsubsection{Equations of motion}
Consider an extended ungauged supergravity theory with homogeneous symmetric scalar manifold described in terms of the bosonic Lagrangian \eqref{boslagr}.
Let us define the \emph{dual field strengths}
\begin{equation}
\Gdual_{\Lambda\mu\nu} ~\equiv\,
 -\eD\;\LCTd\;\frac{\partial\Lagr_{(4)}}{\partial F^\Lambda_{\rho\sigma}}
 \=\RR_{\Lambda\Sigma}\,F^\Sigma_{\mu\nu}-\II_{\Lambda\Sigma}\;\*F^\Sigma_{\mu\nu}\;,
\label{GF}
\end{equation}
where the $\*$ operation means:
\begin{equation}
\*F^\Lambda_{\mu\nu} ~\equiv~
\frac{\eD}{2}\,\LCTd\;F^{\Lambda\,\rho\sigma} \;.
\end{equation}
%
The bosonic part of the equations of motion for the scalar fields can be derived from the Lagrangian \eqref{boslagr} and reads%
\footnote{%
here and in the following we ignore fermion-terms
} \cite{Gallerati:2016oyo}%
\begin{equation}
\tilde{\mathcal{D}}_\mu\left(\partial^\mu\phi^s\right)\=
    \frac{1}{4}\,\Gmu[s][u]\,\left(F^\Lambda_{\mu\nu}\;
    \partial_u\,\II_{\Lambda\Sigma}\;F^{\Sigma\,\mu\nu} +F^\Lambda_{\mu\nu}\;\partial_u\,
    \RR_{\Lambda\Sigma}\;\*F^{\Sigma\,\mu\nu}\right)\;,
\label{scaleqs}
\end{equation}
while the vector Maxwell equations have the form
\begin{equation}
\nabla_\mu\left(\*F^{\Lambda\mu\nu}\right)=0\,, \qquad\;
\nabla_\mu\left(\*\Gdual^{\Lambda\mu\nu}\right)=0\;,
\label{Maxweqs}
\end{equation}
where $\nabla_\mu$ is the covariant derivative containing the Levi-Civita connection on space-time, while $\tilde{\mathcal{D}}$ can be defined through
\begingroup
\abovedisplayshortskip=0pt
\belowdisplayshortskip=12pt
\begin{equation}
\tilde{\mathcal{D}}_\mu(\partial_\nu\phi^s) ~\equiv~ \nabla_\mu(\partial_\nu\phi^s)+\tilde{\Gamma}^s_{vu}\,\partial_\mu\phi^{v}\,\partial_\nu\phi^{u}\,.
\end{equation}
\endgroup
and also contains the Levi-Civita connection $\tilde{\Gamma}$\, on \,$\Ms$.
Using the definition \eqref{GF} for the dual field strengths and the property \,${\*\*F^\Lambda=-F^\Lambda}$, %
we obtain for $\*F^\Lambda$ and $\*\Gdual_\Lambda$ the expressions
\begin{equation} \label{GF2}
\begin{split}
\*F^\Lambda&\=
\left(\II^{-1}\right)^{\Lambda\Sigma}\,\left(\RR_{\Sigma\Pi}\,F^\Pi-\Gdual_\Sigma\right)\;,\\[\jot]
\*\Gdual_\Lambda&\=
\left(\RR\;\II^{-1}\,\RR+\II\right)_{\Lambda\Sigma}\;F^\Sigma\,-\,\left(\RR\,\II^{-1}\right)_\Lambda{}^\Sigma\;\Gdual_\Sigma\;,
\end{split}
\end{equation}
linear functions of $F^\Lambda$ and $\Gdual_\Lambda$ (we have omitted space-time indices).\par
The field strengths can be arranged in a single $2\,\nv$-dimensional vector $\FF\equiv\FF^M$
of two-forms:
\begin{equation}
\FF^M ~\equiv~
\left(\begin{matrix}
F^\Lambda_{\mu\nu}\cr
\Gdual_{\Lambda\mu\nu}
\end{matrix}\right) \;,
\label{bbF}
\end{equation}
in terms of which eq.s \eqref{GF2} are easily rewritten in the compact form
\begin{equation}
\*\FF~=-\,\Cc\;\M(\phi^s)\;\FF \;,
\label{FCMF}
\end{equation}
where the matrix $\Cc$ is defined as
\begin{equation}
\Cc~\equiv~\Cc^{MN}~\equiv~
\left(\begin{matrix}
\Zero & {\Id\,} \cr
-\Id & {\Zero\,}
\end{matrix}\right)~\equiv~
\left(\begin{matrix}
\Zero_{\nv} & \Id_{\nv} \cr
-\Id_{\nv} & \Zero_{\nv}
\end{matrix}\right) \;,
\label{Cc}
\end{equation}
and where $\M(\phi)$ reads:
\begin{equation}
\M(\phi)~\equiv~\M(\phi)_{MN}~\equiv~
\left(\begin{matrix}
(\RR\;\II^{-1}\,\RR\,+\,\II)_{\Lambda\Sigma} & -(\RR\,\II^{-1})_\Lambda{}^\Gamma \cr
-(\II^{-1}\,\RR)^\Xi{}_\Sigma & (\II^{-1})^{\Xi \Gamma}
\end{matrix}\right)\;,
\label{M}
\end{equation}
resulting in a symmetric, negative-definite matrix, function of the scalar fields.\par
Now, in matrix notation, the Maxwell equations can then be recast in the following equivalent forms:
\begin{equation}
\nabla_\mu(\*\FF^{\mu\nu})\=0
\quad\Leftrightarrow\quad
\nabla_\mu(\Cc\,\M(\phi)\,\FF^{\mu\nu})\=0
\quad\Leftrightarrow\quad
d\FF\=0 \;,
\label{Maxweqs2}
\end{equation}
where the symplectic matrix indices {\footnotesize{$M$}}, {\footnotesize{$N$}}, $\dotsc$ have been suppressed.\par\smallskip
The field equations depending on the vector field strengths can be rewritten in terms of the matrix $\M(\phi)$ and of its derivatives. The scalar field equations \eqref{scaleqs} can be rewritten as:
\begin{equation}
\tilde{\mathcal{D}}_\mu\left(\partial^\mu\phi^s\right)\=
\frac{1}{8}\,\Gmu[s][u]\;(\FF_{\mu\nu})^T\;\partial_u\M(\phi)\;\FF^{\mu\nu}\;.
\label{scaleqs2}
\end{equation}
\par

\paragraph{Gravity.}
The Einstein equations have the form:
\begin{equation}
R_{\mu\nu}-\frac{1}{2}\,g_{\mu\nu}\,R
\=T_{\mu\nu}{}^\text{(S)}+T_{\mu\nu}{}^\text{(V)}\,,
\label{Einsteqs}
\end{equation}
in terms of the energy-momentum tensors $T_{\mu\nu}{}^\text{(S)}$ for the scalar fields and $T_{\mu\nu}{}^\text{(V)}$ for the vector fields. The latter can be rewritten in the general form
\begin{equation}
\begin{split}
T_{\mu\nu}{}^\text{(S)}&\=\Gm[s][u](\phi)\,\partial_\mu\phi^s\partial_\nu\phi^u
    -\frac{1}{2}\,g_{\mu\nu}\,\Gm[s][u](\phi)\,\partial_\rho\phi^s\partial^\rho\phi^u\;,\\
T_{\mu\nu}{}^\text{(V)}&\=(F_{\mu\rho})^T\;\II\,F_{\nu}{}^\rho
    -\frac{1}{4}\,g_{\mu\nu}\,(F_{\rho\sigma})^T\;\II\,F^{\rho\sigma}\;,
\end{split}
\end{equation}
and the vector fields energy-momentum tensors can be expressed in terms of $\M(\phi)$ and $\FF$ as
\begin{equation}
T_{\mu\nu}{}^\text{(V)}\=\frac{1}{2}\,\left(\FF_{\mu\rho}\right)^T\M(\phi)\;\FF_{\nu}{}^\rho\;.
\end{equation}
Now, since in \eqref{Einsteqs} we have
\begin{equation}
R\=\Gm[s][u](\phi)\,\partial_\rho\phi^s\,\partial^\rho\phi^u \;,
\end{equation}
the Einstein equation can be finally recast in the form:
\begin{equation}
R_{\mu\nu}\,=~\Gm[s][u](\phi)\,\partial_\mu\phi^s\,\partial_\nu\phi^u
+\frac{1}{2}\,(\FF_{\mu\rho})^T\;\M(\phi)\;\FF_{\nu}{}^\rho\;.
\label{Einsteqs2}
\end{equation}
\par\smallskip
\noindent
Summarizing, the bosonic equations derived in the above discussion can be written, omitting fermion terms, as:
\begin{subeqs} \label{eom} 
\begin{align}
\text{Scalar eqs}\;:&\quad\;
  \tilde{\mathcal{D}}_\mu\left(\partial^\mu\phi^s\right)\,=~\frac{1}{8}\,\Gmu[s][u]\;(\FF_{\mu\nu})^T\;\partial_u\M(\phi)\;\FF^{\mu\nu}\;,\qquad\label{scaleqs3}
  \\[1.25ex]
\text{Einstein eqs}\;:&\quad\;
  R_{\mu\nu}\,=~\Gm[s][u](\phi)\,\partial_\mu\phi^s\,\partial_\nu\phi^u+\frac{1}{2}\,(\FF_{\mu\rho})^T\;\M(\phi)\;\FF_{\nu}{}^\rho\;,\qquad\label{Einsteqs3}
  \\[1.55ex]
\text{Maxwell eqs}\;:&\quad\;
  d\FF\=0 \;\qLq\;
  \nabla_\mu(\Cc\,\M(\phi)\,\FF^{\mu\nu})\=0\;.\label{Maxweqs3}
\end{align}
\end{subeqs}
The isometry group $G$ is a global symmetry only of the scalar kinetic term, since, in general, it alters the action for the vector fields as a consequence of the scalar field-dependence (encoded in the matrices $\II(\phi)$ and $\RR(\phi)$). On the other hand, the Maxwell equations $\nabla_\mu(\*\FF^{M\,\mu\nu})=0$ in \eqref{Maxweqs} are invariant with respect to a generic linear transformation on $\FF$, while the definition of $\Gdual_\Lambda$ and the equations \eqref{FCMF}, \eqref{GF2} are not.\par

\subsection{On-shell duality} \label{subsec:onshellduality}
In extended supergravity models, the global invariance of the scalar kinetic term (expressed through $G$) can be extended to a global symmetry of the full set of equations of motion and Bianchi identities, though not in general of the whole action \cite{Gaillard:1981rj}. This is possible because, in extended supergravities, supersymmetry connects scalar and vector fields and, as a consequence of this, transformations on the scalars imply transformations on the vector field strengths $F^\Lambda$ and their duals $\Gdual_\Lambda$

\paragraph{Symplectic structure.}
On the scalar manifold $\Ms$ it is possible to define a \emph{symplectic} geometric structure%
\footnote{%
at least on the manifold spanned by the scalar fields sitting in the same supermultiplet as the vector ones%
}%
, associating with each point $\phi$ on the manifold the symmetric symplectic $2\,\nv\times2\,\nv$ matrix $\M(\phi)\in\Sp(2\,\nv,\mathbb{R})$ satisfying therefore
\begin{equation}
\M\in\Sp(2\,\nv,\mathbb{R})\;:\quad\;
\M^T\,\Cc\,\M\=\M\,\Cc\,\M^T\=\Cc\;,\qquad
\label{MCM}
\end{equation}
where the symplectic invariant matrix $\Cc$ is defined in \eqref{Cc}. This also tells us that the symmetric matrix $\M(\phi)_{MN}$ satisfies the property
\begin{equation}
\M(\phi)_{MP}\;\Cc^{PL}\,\M(\phi)_{LN}=\Cc_{MN}
\;\qLq\;
\M(\phi)^{-1}=-\,\Cc\,\M(\phi)\,\Cc\;.
\label{CMC}
\end{equation}
Once given the symplectic structure of the manifold through the matrix $\M$, we associate each isometry $g\in G$ on the manifold with a constant symplectic $2\,\nv\times2\,\nv$ matrix \,$\Rsv[g]\equiv\Rsv[g]^M{}_N$\, such that:
\begin{equation}
\M(g\star\phi)\=\Rsv[g]^{-T}\,\M(\phi)\;\Rsv[g]^{-1}\;.
\label{RMR}
\end{equation}
The correspondence between $g\in G$ and $\Rsv[g]$ defines a \emph{symplectic representation} of the group $G$, i.e.\ an embedding $\Rsv$ of the group $G$ inside $\Sp(2\,\nv,\,\mathbb{R})$:
\begin{equation}
g\in G\;\leftrightarrow\;\Rsv[g]\in \Sp(2\,\nv,\mathbb{R})
\;\qLq\;
G \stackrel{\Rsv}{\hookrightarrow}\,\Sp(2\,\nv,\mathbb{R})\;,
\end{equation}
together with the general properties defining a representation and a symplectic matrix, that is
\begin{subeqs} 
\begin{align}
\Rsv[g_1\cdot g_2]&\=\Rsv[g_1]\;\Rsv[g_2]\;,
\\[1.5ex]
\Rsv[g]\;\Cc\;\Rsv[g]^T&\=\Rsv[g]^T\;\Cc\;\Rsv[g]\=\Cc\;.\label{RCR}
\end{align}
\end{subeqs}
\sloppy
The field strengths and their magnetic duals transform under the duality action \eqref{dual} of $G$ in a $2\,\nv$-dimensional symplectic representation.
We denote by ${\Rsvst=\Rsv^{-T}}$ the representation dual to $\Rsv$, acting on covariant symplectic vectors, so that, for any $g\in G$ one has:
\begin{equation}
\begin{split}
&\Rsvst[g]\=\left(\Rsvst[g]_M{}^N\right)\=\Rsv[g]^{-T}~=\,-\Cc\,\Rsv[g]\,\Cc
\\[1ex]
&\Longrightarrow\quad\Rsvst[g]_M{}^N\=\Cc_{MP}\,\Rsv[g]^P{}_Q\,\Cc^{NQ}\;,
\end{split}
\end{equation}
having $\Rsv$ the properties of a symplectic representation.
The above conditions \eqref{RMR} and \eqref{RCR} are verified in extended supergravity models as a consequence of supersymmetry: in these theories, SUSY is large enough as to connect certain scalars to vector fields, so that symmetry transformations on the former imply transformations on the latter (more precisely transformations on the vector field strengths and their duals).\par
The existence of a symplectic representation $\Rsv$ of $G$, together with the properties of the matrix $\M(\phi)$, suggest that the definition of $\M(\phi)$ itself is built-in in the mathematical structure of the scalar manifold. The matrices \,$\II(\phi)$ and $\RR(\phi)$ entering the action can be then defined in terms of \,$\M(\phi)$ by \eqref{M}, and the only freedom left lies in the choice of the basis of the symplectic representation (\emph{symplectic frame}), which amounts to a change in the definition of $\M(\phi)$ by a constant symplectic transformation $E_\textsc{s}$:
\begin{equation}
\M(\phi)\;\stackrel{E_\textsc{s}}{\longrightarrow}\;\;\M'(\phi)=(E_\textsc{s})^T\,\M(\phi)\,E_\textsc{s}\,.
\end{equation}
The action is affected by the above transformation, and in particular we find a change in the coupling of the scalar fields to the vectors. At the \emph{ungauged level}, this only amounts to a (non-perturbative) redefinition of the vector field strengths and their duals, with no physical implication \cite{Gallerati:2016oyo}. If we are dealing with a \emph{gauged theory}, where vectors are minimally coupled to the other fields, the symplectic frame becomes physically relevant and may lead to different vacuum-structures defined by the scalar potential, as we are going to discuss in Sect.\ \ref{sec:gaugsugra}. \par\smallskip
The existence of a symplectic structure on the scalar manifold is a general feature of all extended supergravites, including those $\N=2$ models in which the scalar manifold is not homogeneous (i.e.\ the isometry group does not act transitively on the manifold itself)%
\footnote{%
in the $\mathcal{N}=2$ case, only the scalar fields belonging to the vector multiplets are non-minimally coupled to the vector fields, namely enter the matrices $\II(\phi),\,\RR(\phi)$, and they span a \emph{special K\"ahler}
manifold; on this manifold, a flat symplectic bundle is defined: it fixes the scalar dependence of the matrices $\II(\phi),\,\RR(\phi)$, and the matrix
$ \M(\phi)$ defined in \eqref{M} satisfies the properties
\eqref{CMC} and \eqref{RMR}
}%
. If the scalar manifold is homogeneous, one can study at any point the coset representative $\LL(\phi)\in G$ in the symplectic, $2\,\nv$-dimensional representation $\Rsv$:
\begin{equation}
\LL(\phi) \quad\stackrel{\Rsv}{\longrightarrow}\quad
    \Rsv[\LL(\phi)] \in \Sp(2\,\nv,\,\mathbb{R})\;.
\end{equation}
In general, the representation $\Rsv[H]$ of the isotropy group $H$ may not be orthogonal, $\Rsv[H]\nsubseteq\SO(2\,\nv)$. In this case, one can always change the basis of the representation by means of a matrix $\mathcal{A}$
\begin{equation}
\mathcal{A}~\equiv~\mathcal{A}^N{}_{\underline{M}}
    \;\in\;\Sp(2\,\nv,\,\mathbb{R})/\U(n)\;,
\end{equation}
where underlined indices label the new basis. Now, in the transformed representation, we have that \;$\Rsvu[H]\equiv\mathcal{A}^{-1}\,\Rsv[H]\,\mathcal{A}\,\subset\,\SO(2\,\nv)$\; and therefore $\Rsvu[h]$ is orthogonal.\par\smallskip
For any point $\phi$ on the scalar manifold, let us define the \emph{hybrid coset-representative matrix} \;$\LLb(\phi)\equiv\LLb(\phi)^M{}_{\underline{N}}$\, as:
\begin{equation}
\LLb(\phi)\,\equiv\,\Rsv[\LL(\phi)]\,\mathcal{A}
\qLq
\LLb(\phi)^M{}_{\underline{N}}~\equiv~
    \Rsv[\LL(\phi)]^M{}_N\;\,\mathcal{A}^N{}_{\underline{N}}\;,
\label{hybrid}
\end{equation}
and introduce also the matrix
\begin{equation}
\LLb(\phi)_M{}^{\underline{N}}~\equiv~
    \Cc_{MP}\;\;\Cc^{\,\underline{NQ}}\;\;\,\LLb(\phi)^P{}_{\underline{Q}}\;.
\end{equation}
Note that, since the indices of $\LLb$ refer to two different symplectic bases, $\LLb$ itself is not a matrix representation of the coset representative $\LL$.
Using now \eqref{gLh}, the property of $\Rsv$ of being a representation and the definition \eqref{hybrid} we find:
\begin{equation}
\forall g\in G \;:\qquad
    \Rsv[g]\,\LLb(\phi)\=\LLb(g\star\phi)\;\Rsvu[h]\;,
\label{gLh2}
\end{equation}
where $h\equiv h(\phi,g)$ is the compensating transformation. The above equation \eqref{gLh2} clarifies the hybrid index structure of $\LLb$, being the coset representative acted on to the left by group $G$ and to the right by group $H$, respectively (in our notations, underlined symplectic indices {\footnotesize $\underline{M},\,\underline{N},\dots$} are acted on by $H$ while non-underlined ones by $G$).
The matrix $\M(\phi)$ is then expressed in terms of the coset representative as:
\begin{equation}
\M(\phi)_{MN}\=\Cc_{MP}\;\LLb(\phi)^P{}_{\underline{L}}\;\LLb(\phi)^R{}_{\underline{L}}\;\Cc_{RN}
\qLq
\M(\phi)\=\Cc\;\LLb(\phi)\;\LLb(\phi)^T\;\Cc\;,
\label{Mcos}
\end{equation}
where summation over the index {\footnotesize $\underline{L}$} is understood.\par
The above definition of the matrix $\M(\phi)$ is $H$-invariant and thus only depends on the point $\phi$, transforming according to \eqref{RMR}:
\begin{equation}
\forall g\in G \;:\quad \M(g\star\phi)\=\Cc\;\LL(g\star\phi)\;\LL(g\star\phi)^T\;\Cc
    \=\Rsv[g]^{-T}\;\M(\phi)\;\Rsv[g]^{-1}\;,
\end{equation}
where we have used eq.\ \eqref{gLh2}, the symplectic property of $\Rsv[g]$ and orthogonality property of $\Rsvu[h]$.\par

\paragraph{On shell invariance.}
We can now study the simultaneous action of $G$ on the scalar fields
and on the field strength vector $\FF^M_{\mu\nu}$:
\begin{align}
g\in G &:\quad
\begin{cases}
\quad \phi^s\,&\stackrel{g}\longrightarrow\;\;\;g\star\phi^s
\cr
\,\FF^M_{\mu\nu}\,&\stackrel{g}\longrightarrow\quad
\FF^{\prime M}_{\mu\nu}=\Rsv[g]^M{}_N\,\FF^N_{\mu\nu}
\end{cases}
\label{dualityaction}
\end{align}
and we can easily verify that it is a symmetry of the field equations (we ignore fermion terms). The Maxwell equations are in fact clearly invariant under \eqref{dualityaction} if $F^\Lambda$ and $\Gdual_\Lambda$ were independent, since the latter are invariant with respect to any linear transformation on $\FF$. However, one must show that the definition of $\Gdual_\Lambda$ in \eqref{GF} is invariant under the above transformation or, equivalently, that the form \,$\*\FF=-\,\Cc\,\M\,\FF$\, of eq.\ \eqref{FCMF} in the transformed fields holds as well; this can be proven using the inverted \eqref{dualityaction}, \eqref{RMR} and symplectic properties of $\Rsv[g]$.
At the same time, the invariance of the scalar and Einstein equations is manifest if we look at their expressions \eqref{scaleqs2} and \eqref{Einsteqs2}, and follows from the invariance of the quantity \,$(\FF_{\mu\nu})^T\,\M(\phi)\,\FF_{\rho\sigma}$ \cite{Gallerati:2016oyo}.
Moreover, the duality invariance of the space-time metric and of the scalar action under \eqref{dualityaction} implies the same property for the Einstein tensor and for the scalar energy-momentum tensor $T_{\mu\nu}{}^\text{(S)}$.\par\medskip
Summarizing, we found that the bosonic equations derived in the previous subsection are written in a manifestly $G$-invariant formulation: in extended supergravity models the global symmetry group $G$ of the scalar action can be promoted to a global invariance of, at least, the field equations and the Bianchi identities \cite{Gaillard:1981rj}, provided its (non-linear) action on the scalar fields is associated with a linear transformation on the vector field strengths $F^\Lambda_{\mu\nu}$ and their magnetic duals $\Gdual_{\Lambda\mu\nu}$:
\begin{equation}
g\in G\;:\quad
\left\{
    \begin{aligned}
    \phi^s &~\rightarrow~~g\star\phi^s &\text{(non-linear)}\;,
\\[\jot]
    \left(\begin{matrix}
    F^\Lambda \cr \Gdual_\Lambda
    \end{matrix}\right)
        &~\rightarrow~~\Rsv[g]\cdot
    \left(\begin{matrix}
        F^\Lambda\cr \Gdual_\Lambda
    \end{matrix}\right)
    =\left(\begin{matrix}
            {A_g}^\Lambda{}_\Sigma & {B_g}^{\Lambda\Sigma}\cr
            C_{g\,\Lambda\Sigma} & D_{g\,\Lambda}{}^\Sigma
      \end{matrix}\right)\,
      \left(\begin{matrix}
        F^\Sigma\cr \Gdual_\Sigma
      \end{matrix}\right)
    &\text{(linear)}\;.
\end{aligned}
\right.
\label{dual}
\end{equation}
The action of $G$ on the field strengths and magnetic duals is defined by the symplectic embedding $\Rsv$, and can be seen as a \emph{generalized electric-magnetic duality transformation} promoting the isometry group of the scalar manifold to a global symmetry of the field equations and Bianchi identities. It is a generalization of the $\U(1)$-duality invariance of the standard Maxwell theory, that schematically reads:
\begin{equation}
\left(\begin{matrix}
F_{\mu\nu} \cr \*F_{\mu\nu}
\end{matrix}\right)
\quad\xrightarrow{\U(1)}\quad
\left(\begin{matrix}
F'_{\mu\nu}\cr
\*F'_{\mu\nu}
\end{matrix}\right)
\=\left(\begin{matrix}
\cos(\theta)& \sin(\theta) \cr
-\sin(\theta) & \cos(\theta)
\end{matrix}\right)
\left(\begin{matrix}
F_{\mu\nu} \cr \*F_{\mu\nu}
\end{matrix}\right)\;.
\end{equation}
For this reason $G$ is denoted as the \emph{duality group} of the classical theory. If electric and magnetic sources are present, the symplectic action of $G$ is extended to the charges themselves (as in the Maxwell theory). We emphasize that, as shown in \eqref{dual}, $G$ determines general non-perturbative $g$-transformations under which
\begin{equation}
\left(\begin{matrix}
F^{\Lambda} \cr \Gdual_{\Lambda}
\end{matrix}\right)
\quad\xrightarrow{{\Rsv[g]}}\quad
    \left(\begin{matrix}
    F^{\prime\,\Lambda} \cr \Gdual'_{\Lambda}
    \end{matrix}\right)
\=\left(\begin{matrix}
{{A_g}^\Lambda}_\Sigma\;F^\Sigma &\+ {B_g}^{\Lambda\Sigma}\;\Gdual_{\Sigma} \cr
C_{g\,\Lambda\Sigma}\;F^{\Sigma} &\+ D_{g\,\Lambda}{}^\Sigma\;\Gdual_{\Sigma}
\end{matrix}\right)\;,
\end{equation}
and these are not a symmetry of the action but only of the field
equations and Bianchi identities (\emph{on-shell symmetry}). The
duality group is important because it is believed to encode the
known string/M-theory dualities \cite{Hull:1994ys}.

\paragraph{Lagrangian.}
From the definition \eqref{Mcos} of $\M$ in terms of the coset representative, it follows that, for symmetric scalar manifolds, the scalar Lagrangian \eqref{Lscal} can be written in the equivalent form:
\begin{equation}
\Lagr_{\text{scal}}\,=~\frac{\eD}{2}\,\Gm[s][u](\phi)\partial_\mu\phi^s\,\partial^\mu\phi^u
    \=\frac{\eD}{8}\,k\,\Tr\big(\M^{-1}\,\partial_\mu\M\;\M^{-1}\,\partial^\mu\M\big)\;,
\label{lagrscalM}
\end{equation}
where $k$ depends on the representation $\Rsv$ of $G$.\par
\sloppy
The transformation properties under $G$ of the matrices \,$\II_{\Lambda\Sigma}(\phi)$\, and \,$\RR_{\Lambda\Sigma}(\phi)$, encoding {non-minimal} couplings in $\Lagr_{\text{vect}}, $ can be inferred from \eqref{RMR} and are conveniently described by defining the complex symmetric matrix
\begin{equation}
\NN_{\Lambda\Sigma} ~\equiv~ \RR_{\Lambda\Sigma}+i\,\II_{\Lambda\Sigma}\;.
\label{NIR}
\end{equation}
Under the action of a generic element $g\in G$, the matrix $\NN$ transforms as:
\begin{equation}
\NN(g\star\phi)\=\Big(C_g+D_g\,\NN(\phi)\Big)\,\Big(A_g+B_g\,\NN(\phi)\Big)^{-1}\;,
\label{Ntra}
\end{equation}
where $A_g,\,B_g,\,C_g,\,D_g$ are the $\nv\times \nv$ blocks of the matrix $\Rsv[g]$ defined in \eqref{dual}.\par

\paragraph{Electric and magnetic charges.}
Ungauged supergravities only contain neutral fields w.r.t.\ the $\U(1)^{\nv}$ gauge-symmetry of the vector fields. These models, however, feature \emph{solitonic solutions}, namely configurations of neutral fields which carry $\U(1)^{\nv}$ electric-magnetic charges. These solutions are typically black holes in four dimensions or black branes in higher.\par
On a charged dyonic solution, we can define the electric and magnetic charges as%
\footnote{%
we are using the rationalized Heaviside-Lorentz (HL) units in which \;$\varepsilon_0=1$\,, that fixes the choice of electric/magnetic charge units%
}:
\begin{equation}
\begin{split}
e_\Lambda&~\equiv~\frac{1}{4\pi}\,\int_{\mathbb{S}^2}\Gdual_{\Lambda}\=
    \frac{1}{8\pi}\,\int_{\mathbb{S}^2}\Gdual_{\Lambda\mu\nu}\,dx^\mu\wedge dx^\nu\;,
\\[\jot]
m^\Lambda&~\equiv~\frac{1}{4\pi}\,\int_{\mathbb{S}^2}F^{\Lambda}\=
    \frac{1}{8\pi}\,\int_{\mathbb{S}^2} F^\Lambda_{\mu\nu}\,dx^\mu\wedge dx^\nu\;,
\end{split}
\label{emcharges}
\end{equation}
where $\mathbb{S}^2$ is a spatial two-sphere. They define a symplectic vector $\Gamma^M$:
\begin{equation}
\Gamma^M\=
\left(\begin{matrix}
m^\Lambda \cr e_\Lambda
\end{matrix}\right)
\=\frac{1}{4\pi}\,\int_{\mathbb{S}^2} \FF^M \;.
\end{equation}
These are the \emph{quantized charges}, namely they satisfy the Dirac-Schwinger-Zwanziger quantization condition for dyonic particles \cite{Dirac:1931kp,Schwinger:1966,Zwanziger:1969by}:
\begin{equation}
({\Gamma_2}^M)^T\;\Cc\;{\Gamma_1}^M\=
     m_2{}^\Lambda\;e_{1\,\Lambda}-m_1{}^\Lambda\;e_{2\,\Lambda}\= \frac{1}{2\pi}\,\hbar\,c\,n\;;\qquad (n\in\mathbb{Z})\;.
\label{DZS}
\end{equation}
It must be noticed that, going at the quantum level, the dyonic charges belong to a symplectic lattice: this breaks the duality group $G$ to a suitable discrete subgroup $G_\text{d}(\mathbb{Z})$ which leaves this lattice  invariant (see also next Subsect.s \ref{subsubsec:symplframes}, \ref{subsubsec:dualizdual}).\par
Finally, let us note that, due to the non-minimal couplings of the scalar fields to the vectors in the Lagrangian \eqref{boslagr}, the electric and magnetic fields that one would actually measure at spatial infinity on a solution are not given directly by the field strengths $F^\Lambda$ and $\Gdual_{\Lambda}$, and thus the measured electric and magnetic charges are not the quantized charges ($e$, $m$). In fact, their values also depend on the scalar fields at infinity and are expressed in terms of composite fields, depending on the scalar fields as well as on the field-strengths \cite{Gallerati:2016oyo}.

\subsubsection{Symplectic frames} \label{subsubsec:symplframes}
The duality action $\Rsv[g]$ of $G$ depends on which elements, in the basis of the $2\,\nv$ representation, are chosen to be the $\nv$ electric vector fields appearing in the Lagrangian and which their magnetic duals: this is equivalent to choosing the \emph{symplectic frame} which determines the embedding of the group $G$ inside $\Sp(2\,\nv,\mathbb{R})$. Different choices of the symplectic frame may yield inequivalent Lagrangians (i.e.\ not related by local field redefinitions), with different global symmetries. Indeed, the global symmetry group of the Lagrangian%
\footnote{%
here we only consider \emph{local} transformations on the fields
}
is defined as the subgroup $\Gel\subset G$, whose duality action is linear on the electric field strengths
\begin{flalign}
\qqqquad
g\in \Gel\;:\qquad
\Rsv[g]\=
\left(\begin{matrix}
A^\Lambda{}_\Sigma & \Zero \cr
C_{\Lambda\Sigma} & D_{\Lambda}{}^\Sigma
\end{matrix}\right)\;,&&
\label{ge}
\end{flalign}
where the symplectic condition fixes $D=A^{-T}$, so that one has
\begin{flalign}
\qqqquad
\begin{split}
g\in \Gel\;:\qquad
F^\Lambda&\;\rightarrow\;\,F^{\prime\Lambda}=A^\Lambda{}_\Sigma\,F^\Sigma\;,
\\
\Gdual_\Lambda&\;\rightarrow\;\,\Gdual'_{\Lambda}
        =C_{\Lambda\Sigma}\,F^\Sigma+D_{\Lambda}{}^\Sigma\,\Gdual_\Sigma\;.
\label{Gel}
\end{split}
&
\end{flalign}
As the reader can verify using eq.\ \eqref{Ntra}, under the above transformation the matrices $\II,\,\RR$ transform as follows:
\begin{equation}
\II_{\Lambda\Sigma} \rightarrow D_{\Lambda}{}^\Pi\,D_{\Sigma}{}^\Delta\,\II_{\Pi\Delta}\;;\qquad
\RR_{\Lambda\Sigma} \rightarrow D_{\Lambda}{}^\Pi\,D_{\Sigma}{}^\Delta\,\RR_{\Pi\Delta}+C_{\Lambda\Pi}\,D_{\Sigma}{}^\Pi\;,
\end{equation}
and the consequent variation of the Lagrangian reads
\begin{equation}
\delta\Lagr_\textsc{bos}\,=~
\frac{1}{8}\,C_{\Lambda\Pi}\,D_{\Sigma}{}^\Pi\veps^{\mu\nu\rho\sigma}\,F^\Lambda_{\mu\nu}F^\Sigma_{\rho\sigma}\;,
\label{deltaLC}
\end{equation}
which is a total derivative since $C_{\Lambda\Pi}\,D_{\Sigma}{}^\Pi$ is constant. These transformations are called \emph{Peccei-Quinn transformations} and follow from shifts in certain axionic scalar fields. These transformations are symmetries of the classical action, while invariance of the perturbative path-integral requires the variation \eqref{deltaLC}, integrated over space-time, to be proportional through an integer to $2\pi\hbar$. This constrains the symmetries to close to a discrete subgroup $G(\mathbb{Z})$ of $G$ whose duality action is implemented by integer-valued matrices $\Rsv[g]$. Such restriction of $G$ to $G(\mathbb{Z})$ in the quantum theory was discussed earlier as a consequence of the Dirac-Schwinger-Zwanziger quantization condition for dyonic particles \eqref{DZS}.\par
From \eqref{Gel} we see that, while the vector field strengths $F^\Lambda_{\mu\nu}$ and their duals $\Gdual_{\Lambda\mu\nu}$ transform together under $G$ in the $2\,\nv$-dimensional symplectic representation $\Rsv[g]$, the vector field strengths alone transform linearly under the action of $\Gel$ in a smaller representation $\nv$, defined by the $A$-block in \eqref{ge}.\par\smallskip
Different symplectic frames of the same ungauged model may originate from different compactifications. In $\N\geq 3$ theories, scalar fields always enter the same multiplets of the vector fields. Supersymmetry then implies their non-minimal coupling to the vectors and also that the scalar manifold is endowed with a symplectic structure, associating with each isometry a constant symplectic matrix. In $\N=2$ theories, scalar fields may sit in vector multiplets or hypermultiplets. The former span a \emph{special K\"ahler manifold}, the latter a \emph{quaternionic K\"ahler} one, so that the scalar manifold is always factorized in the product
\begin{flalign}
\qqqquad\qqquad
\N=2\;:\qquad
\Mscal\=\Ms_\textsc{sk}\times\Ms_\textsc{qk}\;.&&
\label{SKQK}
\end{flalign}
The scalar fields in the hypermultiplets are not connected to the vector fields through supersymmetry and consequently they do not enter the matrices $\II(\phi)$ and $\RR(\phi)$. As a consequence of this the isometries of the quaternionic-K\"ahler manifolds spanned by these scalars are associated with trivial duality transformations
\begin{equation}
g\,\in\,\text{isom. of}\;\Ms_\textsc{qk}
\;\quad\Longrightarrow\quad\;
\Rsv[g]\=\Id\;,
\label{qisom}
\end{equation}
while only $\Ms_\textsc{sk}$ features a flat symplectic structure which defines the embedding of its isometry group inside $\Sp(2\,\nv,\mathbb{R})$ and the couplings of the vector multiplet-scalars to the vector fields through the matrix $\M(\phi)$%
\footnote{%
we remark that such structure on a special K\"ahler manifold exists even if the manifold is not homogeneous; this means that one can still define the symplectic matrix $\LLb(\phi)$ and, in terms of the components $\II_{\Lambda\Sigma}(\phi)$ and $\RR_{\Lambda\Sigma}(\phi)$, also the matrix $\M(\phi)$ as in \eqref{Mcos}, although $\LLb(\phi)$ has no longer the interpretation of a coset representative for non-homogeneous manifolds%
}.\par
The transformation properties of the bosonic fields under group $G$ can be rewritten in the infinitesimal form:
\begin{equation}
G\;:\quad
\begin{cases}
\;\delta\,\LLb \= \Lambda^\alpha\,t_\alpha\,\LLb\;,
\\
\;\delta \FF^{M}_{\mu\nu}~=\,
    -\Lambda^{\alpha}\;(t_{\alpha})_{N}{}^{M}\;\FF_{\mu\nu}^{N}\;,
\end{cases}
\end{equation}
in terms of the infinitesimal generators $t_\alpha$ of $G$, defining a basis of its Lie algebra $ \mathfrak{g}$ and satisfying the corresponding commutation relations
\begin{equation}
[t_\alpha,\,t_\beta]\=f_{\alpha\beta}{}^\gamma\,t_\gamma\;,
\label{talg}
\end{equation}
$f_{\alpha\beta}{}^\gamma$ being the structure constants of $\mathfrak{g}$.
The matrices $(t_{\alpha})_{M}{}^{N}$ define the infinitesimal duality action of $G$ and are symplectic generators
\begin{equation}
(t_{\alpha})_M{}^N\,\Cc_{NP} \= (t_{\alpha})_P{}^N\,\Cc_{NM}
\qqquad
\mbox{{\footnotesize$M$},\,{\footnotesize$N$}},\dotsc=1,\dotsc,2\,\nv\;,
\end{equation}
that is equivalently stated as the property of the tensor $t_{\alpha\,MN}\equiv (t_{\alpha})_{M}{}^{P}\,\Cc_{PN}$ of being symmetric in {\footnotesize $M\,N$}:
\begin{equation}
(t_\alpha)_{MN}\=(t_\alpha)_{NM}\;.
\end{equation}

\subsubsection{Fermion fields}
We have seen that the vector fields and the scalar fields transform under the action of the group $G$, isometry group of the scalar manifold. This group has a global (symplectic) action on the vector of electric and magnetic field strengths, while it acts on the scalar fields as an isometry group.\par
We know that fermion fields transform covariantly with respect to the group of local Lorentz transformations (isotropy group of space-time). In the same way, they have a well defined transformation property only with respect to the isotropy group $H$ of the scalar manifold. In all extended supergravity models, this group has the form \cite{Andrianopoli:1996ve}:
\begin{equation}
H \= H_\tts{R} \times H_\text{matt}\;,
\label{HHH}
\end{equation}
where $H_\tts{R}$ is the the R--symmetry group (automorphism of the supersymmetry algebra), while $H_\text{matt}$ is a compact Lie group acting on the matter multiplets. Aside from the gravitino, the other fermion fields consist in \emph{dilatinos} $\chi_{ijk}$ which are spin-$1/2$ fields  belonging to the gravitational supermultiplet for $\N\ge 3$, and spin-$1/2$ fields ${\lambda_i}^A$ (where $A$ is a vector field label) called \emph{gauginos}, belonging to the vector multiplets, i.e.\ supermultiplets in which the highest spin field has spin 1. In the $\N=2$ we also have spin-$1/2$ fields $\varkappa^a$ in the hypermultiplets called \emph{hyperinos}.\par
The coupling of the bosons to the fermionic fields is also fixed by the geometry of the scalar manifold $\Mscal$. In particular, in the models with an homogeneous scalar manifold, this coupling is fixed by the coset representative $\LLb(\phi)$.\par
Let us recall that \eqref{gLh} states that the matrix $\LLb(\phi)$ is acted to the left by $G$ and to the right by the compensator element in $H$:
\begin{equation}
G\;\;\rightsquigarrow\;\;\LLb(\phi)\;\;
\reflectbox{$\rightsquigarrow$}\;\;H \;.
\label{intertwine}
\end{equation}
The matrix $\LLb(\phi)$ therefore can \'intermediate'' between objects transforming directly under $G$ and other objects transforming only under $H$, namely between bosons and fermions. This means that it is possible to construct $G$-invariant quantities, coupling in a suitable way bosonic fields {\bf b} (and their derivatives) to the fermionic fields ${\bf f}$ through $\LLb(\phi)$, considering the contraction
\begin{equation}
(\partial {\bf b})\cdot\LLb(\phi)\cdot{\bf f}
\={\bf d}(\phi,\,\partial{\bf b})\cdot{\bf f}\;.
\end{equation}
This scalar-dependent matrix determines the coupling of bosons and fermions in the Lagrangian and in the equations of motion. The fermions, in other words, couple to composite objects -- that we denoted
${\bf d}(\phi,\,\partial{\bf b})$ -- obtained by \'dressing'' the derivatives of bosonic fields by scalar fields through the matrix $\LLb(\phi)$. Then, these objects transform only through the corresponding compensating transformations $h(\phi,\,g)\in H$, as the scalars and vectors transform under $G$, see \eqref{gLh}. This tell us that the trasformations of all fermion fields is obtained by means of $h(\phi,\,g)$, namely we can define the action of $G$ over all the fields of the theory as:
\begin{align}
g\in G &:\quad
\begin{cases}
\;\phi^s &\stackrel{g}\longrightarrow\quad g\star\phi^s \cr
\;\FF^M_{\mu\nu}&\stackrel{g}\longrightarrow\quad\FF^{\prime M}_{\mu\nu}\=
    \Rsv[g]^M{}_N\;\FF^N_{\mu\nu}\cr
\;{\bf f}&\stackrel{g}\longrightarrow\quad{\bf f}'\=h(\phi,\,g)\star{\bf f}
\end{cases}
\label{duality2}
\end{align}
Now one can construct a manifestly $H$-invariant Lagrangian using the fermion fields and the composite fields ${\bf d}(\phi,\,\partial{\bf b})$. Moreover, $H$-covariance of the standard supersymmetry transformations
\begin{equation}
\delta_\epsilon{\bf b}\=\bar{\epsilon}\,{\bf f}\;,
\qqquad
\delta_\epsilon{\bf f}\=\epsilon\;\partial{\bf b}\;,
\label{fb}
\end{equation}
implies that the supersymmetry variations for the fermion fields can be written as:
\begin{equation}
\delta_\epsilon{\bf f}\={\bf d}(\phi,\,\partial{\bf b})\;\epsilon\;.
\end{equation}
The fields transforming in  representations of $H_\tts{R}$ are therefore either the fermions or the composite fields ${\bf d}(\phi,\,\partial{\bf b})$, but not the scalar fields $\phi^s$ and the vector fields $A^\Lambda_\mu$ directly, since the latter are always real fields. The composite objects ${\bf d}(\phi,\,\partial{\bf b})$ can be imagined as the actual bosonic fields that can be measured, at spatial infinity, on a solution.

\subsubsection{Dualization of dualities}\label{subsubsec:dualizdual}
We briefly mentioned in previous paragraphs the concept of string-duality, namely the idea that different superstring theories on various backgrounds can be thought of as different realizations of a unique fundamental quantum theory, the correspondences among them being called \emph{dualities}. These dualities are conjectured to be encoded into the global symmetries of the resulting (ungauged) supergravity \cite{Hull:1994ys}.\par
A wide class of ungauged extended supergravities feature, at the
classical level, a continuous group of global symmetries acting as a generalized electric-magnetic duality. At the quantum level, Dirac-Zwanziger quantization condition \eqref{DZS} on the charges causes the breaking of this global symmetry group to some suitable discrete subgroup. The latter discrete subgroup is conjectured to describe the above string-dualities.\par
In four-dimensional ungauged supergravity, an important feature of the theory is that antisymmetric tensor fields and scalar fields are related by Poincaré duality. The amount of global symmetry of the theory depends on the number of antisymmetric tensor fields which have been dualized into scalars. It is maximal when all antisymmetric tensors are dualized into scalar fields. The latter phenomenon is called \emph{dualization of dualities} \cite{Cremmer:1997ct}.


\newpage

\section{Black hole configurations} \label{sec:bhconfig}

We shall now restrict our discussion to static black hole solutions, with spherical symmetry and asymptotically flat.

\subsection{General properties of the solution} \label{subsec:bhsol}
The general ansatz for the black hole metric has the form:
\begin{equation}
ds^2\=f(r)^2\,dt^2-f(r)^{-2}\,dr^2-h(r)^2\,(d\theta^2+\sin^2(\theta)\,d\varphi^2)\;,
\label{metrans}
\end{equation}
where $f(r)$, $h(r)$ are functions of the radial variable to be determined by the equations of motion. Moreover we also set, for fermion and scalar fields,
\begin{equation}
\text{fermions}=0\;, \qqquad \phi^s=\phi^s(r)\;.\quad
\end{equation}

\paragraph{Equations of motion.}
\sloppy
If we consider dyonic solution, with quantized electric and magnetic charges ${\Gamma^M\equiv(m^\Lambda,\,e_\Lambda)}$, one can verify that the following expression for $\FF^M$
\begin{equation}
\FF^M\=
\left(
\begin{matrix}
F^\Lambda_{\mu\nu} \cr \Gdual_{\Lambda\mu\nu}
\end{matrix}
\right)\,
\frac{dx^\mu\wedge dx^\nu}{2}\=
\frac{1}{h^2}\,\Cc\,\M(\phi)\,\Gamma^M\;dt\wedge dr + \Gamma^M\,\sin(\theta)\;d\theta\wedge d\varphi\;,
\label{FFans}
\end{equation}
satisfies the Maxwell equations \eqref{Maxweqs3}.\par\smallskip
The scalar field equations \eqref{scaleqs3} can be recast using \eqref{FFans}. The right hand side is rewritten as:
\begin{equation}
\begin{split}
(\FF_{\mu\nu})^T\,\partial_s\M\;\FF^{\mu\nu}
&\=2\,(\FF_{tr})^T\;\partial_s\M\;\FF_{tr}\;g^{tt}g^{rr}
+2\,(\FF_{\theta\varphi})^T\;\partial_s\M\;\FF_{\theta\varphi}\;g^{\theta\theta}g^{\varphi\varphi}=
\\
&~=-\frac{2}{h^4}\,\Gamma^T\M\;\Cc^T\;\partial_s\M\;\Cc\;\M\;\Gamma
+\frac{2}{h^4}\,\Gamma^T\;\partial_s\M\;\Gamma=
\\
&\=\frac{4}{h^4}\,\Gamma^T\;\partial_s\M\;\Gamma
~=-\frac{8}{h^4}\,\partial_s \VBH\;,
\end{split}
\end{equation}
where we have introduced the \emph{black hole effective potential} $\VBH$
\begin{equation}
\VBH(\phi,\,\Gamma) ~\equiv\, -\frac{1}{2}\,\Gamma^T\M(\phi)\Gamma\,>0\;.
\label{VBH}
\end{equation}
The scalar field equation now reads:
\begin{equation}
\left(f^2\,h^2\,\phi^{\prime s}\right)'
+\tilde{\Gamma}^s{}_{uv}\,\phi^{\prime u}\,\phi^{\prime v}\=
    \frac{1}{h^2}\,\Gmu[s][u]\,\partial_u\VBH
\;,\label{scaleqs4}
\end{equation}
where the prime stands for the derivative with respect to the radial variable $f'(r)\equiv\frac{d}{dr}$, while $\partial_{s,\,u,\,v,\,\dotsc}$ indicates $\frac{\partial}{\partial\phi^{s,\,u,\,v,\,\dotsc}}$\;.\par\smallskip
It is useful to introduce a new radial variable $\tau=\tau(r)$ defined by the condition:
\begin{equation}
\frac{d\tau}{dr}\=\frac{1}{f^2\,h^2} \;.
\label{tau}
\end{equation}
Using the notation $\dot{f}(\tau)\equiv\dfrac{df(\tau)}{d\tau}$, equation \eqref{scaleqs4} becomes:
\begin{equation}
\ddot{\phi}^s+\tilde{\Gamma}^s{}_{uv}\,\dot{\phi}^{u}\dot{\phi}^{v}
\=f^2\,\Gmu[s][u]\,\partial_u\VBH \;.
\label{scaleqs5}
\end{equation}
The above equation -- where the radial variable $\tau$ has the role
of time -- describes the motion of a particle, subject to a
potential $\VBH$, in the manifold $\Mscal$%
\footnote{%
a geodesic motion
corresponds to $\VBH=\mathrm{const.}$
}%
.\par\smallskip
Let us consider now the Einstein equations \eqref{Einsteqs2}. Using \eqref{metrans} and \eqref{FFans}, the equations can be rewritten:
\begin{equation}
\begin{split}
&R_{rr}\=\Gm[s][u]\,\phi^{\prime s}\,\phi^{\prime u}
    -\frac{1}{f^2\,h^4}\,\VBH\;,
\\[\jot]
&R_{tt}\=\frac{f^2}{h^4}\,\VBH\;, \qquad\;
    R_{\theta\theta}\=\frac{1}{h^2}\,\VBH\;, \qquad\;
    R_{\varphi\varphi}\=\frac{\sin^2(\theta)}{h^2}\,\VBH\;,
     \;,
\label{Einsteq4}
\end{split}
\end{equation}
from which we get
\begin{equation}
R^t{}_t\=\frac{1}{f^2}\,R_{tt}\=\frac{1}{h^4}\,\VBH\=
    \frac{1}{h^2}\,R_{\theta\theta}~=\,-R^\theta{}_\theta\;.
\label{Ricci}
\end{equation}
Comparing the above results with the expressions for the Ricci tensor that one gets doing an explicit calculation from the metric form \eqref{metrans}, we find the relations:
\begin{equation}
\begin{split}
R^t{}_t\,=\,-R^\theta{}_\theta
\quad&\Rightarrow\quad
\frac{(f\,f'\,h^2)'}{h^2}
    =\frac{1}{h^2}\left(1-(f^2\,h\,h')'\right)
\\
&\Rightarrow\quad (f^2\,h^2)^{\prime\prime}=2 \;.
\end{split}
\end{equation}
This condition, which is implied on the ansatz by the Einstein equation, is solved in general by setting:
\begin{equation}
\begin{split}
f^2\,h^2&\=(r-r_0)^2-\cex^2\=(r-r_+)(r-r_-)\,;
\\[\jot]
r_{{}_\pm}& ~\equiv~ r_0 \,\pm\, \cex\;,
\label{f1f2cond}
\end{split}
\end{equation}
where we have introduced the integration constant $\cex$, called \emph{extremality parameter}, which is assumed to have a positive square, $\cex^2\ge 0$. If this is not the case, i.e.\ $\cex^2<0$, the two roots $r_{{}_\pm}$ are imaginary. As we shall see, $r_{{}_\pm}$ can be identified with an inner and outer horizon of the black hole, and thus, if $\cex^2<0$, the solution has no horizon to hide its singularity and turns out to be not regular.\par\smallskip
The above equation \eqref{tau}, defines the affine parameter $\tau$:
\begin{equation}
\frac{d\tau}{dr}\=\frac{1}{f^2\,h^2}\=\frac{1}{(r-r_0)^2-\cex^2}
\;\qLq\;
r-r_0~=-\cex\,\coth(\cex\tau)\;,
\label{tau1}
\end{equation}
from which we get the explicit expression
\begin{equation}
\tau\=\frac{1}{2\,\cex}\,\log\left(\frac{r-r_+}{r-r_-}\right)\,.
\label{tau2}
\end{equation}
The coordinate $\tau$ turns out to be non-positive and  runs from $-\infty$ when $r=r_+$ (outer horizon of the black hole) to $\tau=0$ at radial infinity $r=+\infty$.\par
The above eq.\ \eqref{tau2} can be also rewritten
\begin{equation}
\frac{d\tau}{dr}\=\frac{1}{(r-r_0)^2-\cex^2}\=
    \frac{\sinh^2(\cex\tau)}{\cex^2}\;.
\label{tau3}
\end{equation}
Making use of \eqref{f1f2cond}, we can simplify the notation and write the functions $f(r)$, $h(r)$ in terms of a single function $U(r)$ as:
\begin{equation}
\begin{split}
f(r)^2&\=e^{2U(r)}\;,
\\
h(r)^2&\=e^{-2U(r)}\,(r-r_+)(r-r_-)\=e^{-2U(r)}\,\frac{\cex^2}{\sinh^2(\cex\tau)}\;.
\end{split}
\end{equation}
The metric \eqref{metrans} now reads:
\begin{equation}
ds^2\=e^{2U}\,dt^2-e^{-2U}\,\left(dr^2\,+\,
    (r-r_+)(r-r_-)\,d\Omega^2\right)\;,
\label{dsrr}
\end{equation}
where \,$d\Omega^2\equiv d\theta^2+\sin^2(\theta)\,d\varphi^2$\,.
We can also express it in terms of the new radial variable $\tau$ as \cite{Ferrara:1997tw}:
\begin{equation}
ds^2\=e^{2U}\,dt^2-e^{-2U}\,\left(\frac{\cex^4}{\sinh^4(\cex\tau)}\,d\tau^2+\frac{\cex^2}{\sinh^2(\cex\tau)}\,d\Omega^2\right)\,,
\label{dstau}
\end{equation}
where $U=U(\tau)$. Using then the property 
\begin{equation}
f\,f'\,h^2\=\frac{\dot{f}}{f}\=\dot{U}\;,
\end{equation}
from \eqref{Ricci} we also find
\begin{equation}
\ddot{U}\=e^{2U}\,\VBH \;.
\label{Udot}
\end{equation}
Finally, the Ricci tensor in the new radial coordinate has non-vanishing entries
\begin{equation}\label{Ricci2}
R_{tt}\=\frac{1}{h^4}\,\ddot{U}\;,\qquad\;
R_{\tau\tau}\=2\,\cex^2-2\,\dot{U}^2+\ddot{U}\;,\qquad\;
R_{\theta\theta}\=\frac{R_{\varphi\varphi}}{\sin^2(\theta)}\=\frac{1}{f^2\,h^2}\,\ddot{U}\;.
\end{equation}
From the first of eq.s \eqref{Einsteq4} and using the above \eqref{Ricci2}, we find also \cite{Ferrara:1997tw}
\begin{equation}
\dot{U}^2+\frac{1}{2}
\,\Gm[s][u]\,\dot{\phi}^s\dot{\phi}^u-e^{2U}\,\VBH\=\cex^2\;,
\label{Udotdot}
\end{equation}
where we have used the previous result \eqref{Udot}.\par\medskip
Summarizing, we have found that the most general ansatz for the static solution depends on $\ns+1$ independent functions of the radial variable $\tau$, that we denoted as $U(\tau)$ and $\phi^s(\tau)$. The latter are subject to the equations:
\begin{subeqs} 
\begin{align}
&\ddot{U}\=e^{2U}\,\VBH\;,\label{Udot2}
\\
&\ddot{\phi}^s+\tilde{\Gamma}^s{}_{uv}\,\dot{\phi}^{u}\dot{\phi}^{v}\=
    e^{2U}\,\Gmu[s][u]\,\partial_u\VBH\;,\label{scaleqs6}
\\
&\dot{U}^2+\frac{1}{2}\,\Gm[s][u]\,\dot{\phi}^s\dot{\phi}^u
    -e^{2U}\,\VBH\=\cex^2\;.\label{Udotdot2}
\end{align}
\end{subeqs}
\par

\paragraph{Effective action.}
The presence of the scalar fields in non-minimal coupling to the vectors (typical of supergravity black holes) determines their participation in the solution, together with the form of the effective potential $\VBH(\phi,e,m)$. On the other hand, the scalar fields which do not couple to any electric-magnetic charges, do not affect the effective potential and thus do not exhibit a radial evolution.\par
The above equations \eqref{Udot2}, \eqref{scaleqs6} can be derived from a suitable \emph{effective action} of the form:
\begin{equation}
\mathscr{S}_\text{eff}\=
    \int\Lagr_\mathrm{eff}\;d\tau\=
    \int\left(\dot{U}^2+\frac{1}{2}\,\Gm[s][u](\phi)\,\dot{\phi}^s\,\dot{\phi}^u
    +e^{2U}\,\VBH(\phi,\,\Gamma^M)\right)\,d\tau\;.
\label{effact}
\end{equation}
This action describes a Lagrangian system in which the radial coordinate $\tau$ plays the role of the standard time variable. The corresponding Hamiltonian $\Ham$ exhibits the property:
\begin{equation}
\frac{d\Ham}{d\tau}\=0 \qquad\Rightarrow\qquad
\Ham=\mathrm{const}\;,
\end{equation}
that is, in analogy with the standard Hamiltonian formalism, it is \'conserved'' with respect to the dependence on the radial variable $\tau$ (and not on time $t$). The Hamiltonian constraint, expressed in terms of the functions $U(\tau)$, $\phi^s(\tau)$, is nothing but eq.\ \eqref{Udotdot2}:
\begin{equation}
\Ham\=\dot{U}^2+\frac{1}{2}\,\Gm[s][u](\phi)\,\dot{\phi}^s\,\dot{\phi}^u-e^{2U}\,\VBH(\phi,\,\Gamma)
\=\cex^{\,2} \;,
\label{Hamconstr}
\end{equation}
where, in this case, the integration constant $\cex^{\,2}$ plays the
role of the energy.
\par

\paragraph{Physical properties of the solution.}
Our solution has a globally-defined time-like Killing vector of the
form $\xi^\mu\partial_\mu=\partial_t$\,. The ADM-mass is given by the
Komar integral \cite{Wald:1984rg} over the sphere $\mathbb{S}^2_{\infty}$ at
radial infinity (i.e.\ $\tau=0$):
\begin{equation}
\MADM\=\frac{c^2}{8\pi\,\GN}\,\int_{\mathbb{S}^2_{\infty}}
\eD\;\veps_{\theta\varphi\mu\nu}\,\nabla^\mu\xi^\nu\,d\theta\,d\varphi\;,
\end{equation}
and, on our general solution, it can be proven the relation
\begin{equation}
\MADM\=\frac{c^2}{\GN}\,\lim_{\tau\rightarrow 0^-}\dot{U}\;,
\end{equation}
using the explicit expression of the covariant derivative built from the previous metric expressions.\par \smallskip
The boundary conditions on the scalar fields at radial infinity ($\tau=0$) defining our solution are written as:
\begingroup
\abovedisplayskip=8pt
\begin{equation}
U(0)=0\;; \qquad
\dot{U}(0)=\frac{\GN}{c^2}\,\MADM\;; \qquad
\phi^s(0)=\phi^s_0\;; \qquad
\dot{\phi}^s(0)=\dot{\phi}^s_0\;,\quad
\end{equation}
\endgroup
while the boundary conditions on the vector fields have already been fixed by the values of the electric and magnetic charges ($e$, $m$). Moreover, one can note that the first condition \,$U(0)=0$\, is nothing but the requirement of asymptotic flatness of the metric.\par
We can write the Hamilton constraint \eqref{Hamconstr} at radial infinity (restoring the constants%
\footnote{%
all terms in the constraint equation \eqref{Hamconstr} have the dimension of a squared length; since the scalar potential has dimension of a squared charge in HL-units, when restoring the constants we need the replacement $\VBH\rightarrow\frac{8\pi\GN}{c^4}\,\VBH$
}%
) in terms of the above boundary data:
\begin{equation}
\tau\rightarrow 0\;:\qquad\;
\frac{\GN^2}{c^4}\,\MADM^2+\frac{1}{2}\,\Gm[s][u](\phi_0)\,\dot{\phi}_0^s\,\dot{\phi}_0^u
-\frac{8\pi\,\GN}{c^4}\,\VBH(\phi_0,\,\Gamma)\=\cex^{\,2}\;.
\label{Hamconstrinfty}
\end{equation}
Regularity of the solution requires the existence of the two horizons, corresponding to $r_{{}_\pm}$, that in turn requires \,$\cex^{\,2}\ge0$\, and a related condition on the boundary data, according to \eqref{Hamconstrinfty}. The two horizons may coincide ($r_+=r_-$) when the extremality parameter goes to zero ($\cex=0$).

\paragraph{No scalar hair.}
The radial derivatives \,$\dot{\phi}^s_0$\, of the scalar fields of a solution are called \emph{scalar charges}. In the black hole solutions present in the known literature, these quantities evaluated at infinity are not independent (boundary) data, but can be written in terms of the other quantities at infinity, namely the ADM-mass, the electric and magnetic charges and the values $\phi_0^s$\,.
The dependence occurs since, in this class of solutions, the radial
evolution \,$\dot{\phi}^s_0$\, of the scalar fields is only due to their
non-minimal coupling to the electric-magnetic charges. This means
that their values are forced by the vector fields and do not exhibit
independent dynamics.\par
Even if there is no general proof of this characteristic, this reported behavior seems to indicate that the most general static black hole solution can be completely determined by its ADM-mass, electric-magnetic charges, and, for non-static stationary solutions, angular momentum%
\footnote{%
here we are just considering the physical quantities related to the radial
derivatives of the fields at infinity; the boundary values of the scalar fields do not have a physical meaning in an ungauged supergravity
}%
. So, there seems to be a generalization to supergravity black holes of the general relativity \emph{``no-hair'' theorem} for ordinary black holes \cite{ruffini:1971}. The theorem states that the most general axisymmetric, asymptotically flat, black hole solution in the  Einstein-Maxwell theory is the Kerr-Newman solution \cite{israel:1967,Robinson:1975bv,Carter:1971zc,Mazur:1982db}: the latter is totally defined by its mass, electric-magnetic charges and angular momentum. The power of the statement lies in the fact that any system, containing charged matter, that collapses into a black hole, loses any other physical property (hair): for example, multipole moments, baryon or lepton numbers, are physical features that disappear with the collapse.\par
We said above that a proof of an analogous theorem for scalars coupled
to supergravity black holes is still missing. However, if one considers extended supergravity models with homogeneous-symmetric scalar manifold, the use of an effective three dimensional description of the solution -- in which a larger global symmetry group connecting $D=4$ stationary solutions is manifest -- gives some argument in support of the hypothesis of an analogous behavior \cite{Breitenlohner:1987dg}.
\par
Finally, the fact that on a black hole solution the radial evolution of the scalar fields is completely determined by their boundary values $\phi_0^s$ (for fixed ADM-mass and electric-magnetic charges) suggests that, for the scalar fields, an effective description can be given in terms of a system of first-order differential equations.

\subsection{Near-horizon behavior} \label{subsec:nhbehav}
The two zeros of the metric \eqref{dsrr} are located at $r_{{}_\pm}=r_0\,\pm\,\cex$. These are coordinate singularities representing an inner and outer horizons, as in Reissner-Nordstr\"om solution \eqref{RN}.\par
Consider the 2-sphere $\mathbb{S}^2$, and require that it has a finite, positive area  $A_\textsc{h}=4\pi\,r_\textsc{h}^2$\, when $r\rightarrow r_\textsc{h}=r_+$. The area $A_\textsc{h}$ can be evaluated as:
\begin{equation}
A_\textsc{h}\,=\,\lim_{\tau\rightarrow -\infty} \int_{S_2}
\sqrt{g_{\theta\theta}\,g_{\varphi\varphi}}\,d\theta\,d\varphi
~=\,\lim_{\tau\rightarrow-\infty}
  4\pi\,e^{-2U}\frac{\cex^{\,2}}{\sinh^2(\cex\tau)}\;.
\end{equation}
The above request of a finite and positive area implies, for the warp factor $e^U$, the near-horizon behaviour
\begin{flalign}
\qquad
r\;\rightarrow\;r_\textsc{h}=r_+\;\;:\qquad
e^{-2U}\,\sim\,\frac{A_\textsc{h}}{4\pi}\;\frac{\sinh^2(\cex\tau)}{\cex^{\,2}}
    \=\frac{r_\textsc{h}^2}{(r-r_+)(r-r_-)}\;.&&
\label{nhbehav}
\end{flalign}
Now, for $r\rightarrow r_+$\,, the metric \eqref{dstau} becomes:
\begin{equation}
ds^2\=\frac{(r-r_+)(r-r_-)}{r_\textsc{h}^2}\;dt^2
    -\frac{r_\textsc{h}^2}{(r-r_+)(r-r_-)}\;dr^2
    -r_\textsc{h}^2\;d\Omega^2 \;,
\label{dsrrH}
\end{equation}
that is the near-horizon geometry of a non-extremal Reissner-Nordstr\"om \eqref{RN} solution. This justifies the identification of $r_{{}_\pm}$ with the outer and inner horizons of the black hole solution, and the condition $\cex^{\,2}\ge 0$ as the regularity condition for the existence of these horizons.\par \smallskip
From the behavior of the solution in the near-horizon limit, we can deduce the thermodynamic quantities like the temperature and the entropy. To this end, we use the general formula for the surface gravity \cite{Wald:1984rg}:
\begin{equation}
\kappa^2~=\,-\,\frac{c^4}{2}\,\nabla^\mu\xi^\nu\,\nabla_\mu\xi_\nu\;.
\end{equation}
Then, making use of the explicit expression for the covariant derivatives and killing vectors and taking into account \eqref{nhbehav}, we rewrite it in the following form (restoring the constants):
\begin{equation}
\kappa\=\frac{c^2\,\cex}{r_\textsc{h}^2}\;.
\label{kappagen}
\end{equation}
Now, the temperature is given by the equation \eqref{temperature} in terms of the surface gravity as:
\begin{equation}
T\=\frac{\hbar\,c}{2\pi\,k_\textsc{b}}\,\frac{\cex}{r_\textsc{h}^2}\;,
\end{equation}
while the entropy reads
\begin{equation}
\Sentr\=\frac{k_\textsc{b}\,c^3\,A_\textsc{h}}{4\,\GN\,\hbar}\;.
\end{equation}
This tells us that we can identify the extremality parameter with the quantity:
\begin{equation}
\cex\=2\,\frac{\GN}{c^4}\,\Sentr\,T \;,
\end{equation}
and it is zero if and only if the temperature is zero, namely when the solution is \emph{extremal}. This is the case of the Reissner-Nordstr\"om extremal solution in which the two horizons coincide ($r_+=r_-$).\par

\paragraph{Extremal solutions and the attractor mechanism.}
In addition to the regularity condition $\cex^{\,2}\ge 0$, we also require the scalar fields to have a regular behavior at the horizon. For this purpose, we define the \emph{proper distance} $\rho$ from the horizon by the equation
\begin{equation}
d\rho^2\=e^{-2U}\,dr^2\;,
\label{rhort}
\end{equation}
and require that the scalar fields, rewritten as functions of $\rho$, run to finite values in the near horizon region, located at $\rho=\rho_\textsc{h}$\,:
\begin{equation}
\lim_{\rho\rightarrow\rho_\textsc{h}}\phi^s(\rho)\=
    \phi_*^s\;;\qqquad
    |\phi_*^s|<\infty \;.
\label{regscal}
\end{equation}
Consider now extremal solutions, defined by the property $\cex=0$. If we send $\cex\rightarrow 0$, from eqs.\ \eqref{tau1}--\eqref{tau2} we get:
\begin{flalign}
\qqquad\qquad
\cex\,\rightarrow\,0\;:\;\qquad
\tau=-\frac{1}{r}\;,&&
\end{flalign}
where we have redefined \,$(r-r_0)\rightsquigarrow r$\,.\par
With the above redefinition, the horizon is located at $r=r_\textsc{h}=0$, or, correspondingly, at $\tau\rightarrow -\infty$, and the near-horizon behavior of the warp function $U(\tau)$ of an extremal solution is given by \eqref{nhbehav}:
\begin{flalign}
\qqquad\qquad
\tau\rightarrow-\infty\;:\;\qquad
e^{-2U}\sim\;\lim_{\cex\rightarrow0}\,r_\textsc{h}^2\,\frac{\sinh^2(\cex\tau)}{\cex^{\,2}}
    \=r_\textsc{h}^2\tau^2\;,&&
\end{flalign}
that also implies, in the extremal near-horizon limit $\tau\rightarrow-\infty$
\begin{flalign}
\qqquad\qquad
\tau\rightarrow-\infty\;:\;\qquad
e^{-U}\sim -\,\tau\,r_\textsc{h}\,,\qquad
\dot{U}\sim -\frac{1}{\tau}\,,\qquad
\ddot{U}\sim \frac{1}{\tau^2} \;.&&
\label{nhbehavU}
\end{flalign}
The proper distance $\rho$ is then defined by the condition \eqref{rhort}:
\begin{flalign}
\qqquad\qquad
\begin{split}
\tau\rightarrow-\infty\;:\;\qquad
d\rho&\=e^{-U}\,dr\=
    \lim_{\cex\rightarrow0}\,e^{-U}\,\cex^{\,2}\frac{d\tau}{\sinh^2(\cex\tau)}\=
\\
&\=e^{-U}\,\frac{d\tau}{\tau^2}~\sim~ -r_\textsc{h}\,\frac{d\tau}{\tau}\;,
\end{split}
&
\end{flalign}
from which we get
\begin{flalign}
\quad
\cex\rightarrow0\,,\;\;\tau\rightarrow-\infty\;:\qquad
\rho~=\,-r_\textsc{h}\,\log(-\tau)\;,&&
\end{flalign}
and, with respect to the proper distance, the horizon is located at $\rho_\textsc{h}=-\infty$\,.\par\smallskip
The regularity request for the scalars \eqref{regscal} is now rewritten at the horizon
\begin{equation}
\lim_{\rho\rightarrow-\infty}\phi^s(\rho)\=\phi_*^s\;,\qquad\quad
|\phi_*^s|<\infty\;,
\label{regscal2}
\end{equation}
and this implies the vanishing of the derivatives of the scalar fields with respect to $\rho$ in this limit:
\begin{equation}
\lim_{\rho\rightarrow-\infty}\,\frac{d^\ell}{d\rho^\ell}\,\phi(\rho)\=0\;.
\label{dkrho}
\end{equation}
Explicitly, for \,$\ell=1$\, and \,$\ell=2$, one has:
\begin{equation}
\lim_{\tau\rightarrow-\infty}\tau\,\dot{\phi}^s\=
\lim_{\tau\rightarrow-\infty}\tau^2\,\ddot{\phi}^s\=0\;.
\label{tauphi0}
\end{equation}
The scalar field equations \eqref{scaleqs6} near the horizon have
the form:
\begin{flalign}
\qqquad\qquad
\tau\rightarrow-\infty\;:\;\qquad
\tau^2\ddot{\phi}^s+\tilde{\Gamma}^s{}_{uv}\,(\tau\dot{\phi}^{u})(\tau\dot{\phi}^{v})
    \=\frac{1}{r_\textsc{h}^2}\,\Gmu[s][u]\;\partial_u\VBH\;.&&
\end{flalign}
where we have used \eqref{nhbehavU}. Now, taking the horizon limit of the previous expression and using \eqref{tauphi0}, the left hand side vanishes and we get
\begin{equation}
\lim_{\phi\rightarrow\phi_*}\,\partial_u\VBH
\=\partial_s\VBH(\phi^s_*,e,m)\=0\;.
\label{dVfix}
\end{equation}
This means that, going from radial infinity to the horizon of an extremal static black hole, the scalar fields of the solution flow toward fixed values $\phi_*^s$\,, which define an extremum of the potential.\par
In general $\VBH$ may not depend on all the scalars, that is it can have the so-called \emph{flat directions}. These correspond to scalar fields which are not effectively coupled to the black hole solution. So, the above \eqref{dVfix} will only  fix scalars along the non-flat directions as functions of the electric and magnetic charges only
\begin{equation}
\phi_*^s\=\phi_*^s(e,m)\;,
\label{scalem}
\end{equation}
and, as a consequence, the potential $\VBH$ at the extremum $\phi_*^s$ will only depend on the electric and magnetic charges:
\begin{equation}
{\VBH}^\text{(ex)}\=\VBH(\phi_*,e,m)\={\VBH}^\text{(ex)}(e,m)\,.
\end{equation}
Using \eqref{nhbehavU}, the equation \eqref{Udot2} evaluated in the near horizon region gives
\begin{flalign}
\qqquad
\begin{split}
\tau\rightarrow-\infty\;:\;\qquad
&\frac{1}{\tau^2}\=\ddot{U}\=e^{2U}\,{\VBH}^\text{(ex)}\=\frac{1}{r_\textsc{h}^2\,\tau^2}\,{\VBH}^\text{(ex)}
\\[\jot]
&\Rightarrow\quad {\VBH}^\text{(ex)}\=r_\textsc{h}^2\,.
\end{split}
&
\end{flalign}
This means that the area of the horizon can be expressed through
the value of ${\VBH}^\text{(ex)}(e,m)$ as:
\begin{equation}
A_\textsc{h}\=4\pi\,{\VBH}^\text{(ex)}(e,m)\=A_\textsc{h}(e,m)\;,
\label{AV}
\end{equation}
in terms of the electric and magnetic charges only%
\footnote{%
restoring the constants we would write:
$A_\textsc{h}=4\pi\,\frac{8\pi \GN}{c^4}\,{\VBH}^\text{(ex)}(e,m)$
}%
.\par\smallskip
The near horizon metric can be easily computed from the previous form \eqref{dsrrH} and reads:
\begin{equation}
ds^2\={\left(\frac{r}{r_\textsc{h}}\right)}^2\,dt^2
    \,-\,{\left(\frac{r}{r_\textsc{h}}\right)}^{-2}\,dr^2
    \,-\,r_\textsc{h}^2\,d\Omega^2\;.
\label{dsrr3}
\end{equation}
This metric describes a \emph{Bertotti-Robinson} solution, that is an $\AdS_2\times \mathbb{S}^2$ space, whose geometry only depends on the area $A_\textsc{h}$ of the horizon which, in turn, only depends on the quantized charges of the solution (as relation \eqref{AV} states) and not on the boundary values $\phi_0$ of the scalar fields. This condition goes under the name of \emph{attractor mechanism} \cite{Ferrara:1995ih}: the scalars non-trivially coupled to the black hole (non-flat directions of the potential) flow from their values at radial infinity $\phi_0$ towards fixed values at the horizon $\phi_*$. The latter are solution to \eqref{dVfix} and only depend on the quantized charges as stated in \eqref{scalem}.\par\smallskip
We can notice that the extremal black holes interpolate between two vacua of the ungauged $\N$-extended supergravity, the $\mathbb{M}_4$ Minkowski space-time and  $\AdS_2\times \mathbb{S}^2$:
\begin{equation}
\text{$\mathbb{M}_4$ at radial infinity}
\quad\longleftrightarrow\quad
\AdS_2\times \mathbb{S}^2\;\;\text{at the horizon}
\end{equation}
similarly to solitonic solutions in ordinary field theory, interpolating between different vacua. In this sense, extremal black hole are viewed as \emph{solitons} of the ungauged supergravity theories. \par
If we consider extremal dyonic black holes, for a given set of charges $\Gamma^M=(e,m)$ it is always possible to find boundary conditions on the scalar fields for which the scalars themselves are constant in the whole space. In fact, it suffices to take:
\begin{equation}
\phi^s(\tau=0)\=\phi_*^s\;,
\end{equation}
and, being also
\begin{equation}
\partial_s\VBH(\phi_*^s,e,m)\=0\;,
\end{equation}
the scalar field equations are solved by $\phi^s(\tau)\equiv \phi_*^s$\,. These solutions, i.e.\ extremal solutions with constant scalar fields, are called \emph{double extremal}.

\paragraph{Non-extremal case.}
If we repeat the above analysis for the non-extremal case, we find for the proper distance $\rho$ in the near-horizon region:
\begin{flalign}
\qquad
\begin{split}
\tau\,\rightarrow\,-\infty\;:\;\qquad
d\rho\=e^{-U}\,dr
    &~\sim~-r_\textsc{h}\,\frac{\sinh(\cex\tau)}{\cex}\,\frac{dr}{d\tau}\,d\tau~\sim~
\\
&~\sim~-\,\frac{\cex}{\sinh(\cex\tau)}\,d\tau~\sim~2\,\cex\,e^{\cex\tau}\,d\tau\;,
\end{split}
&
\end{flalign}
from which we get this time
\begin{flalign}
\qquad
\cex\neq0\,,\quad\tau\rightarrow-\infty\;\;:\qqquad
\rho(\tau)\=2\,e^{\cex\tau} \;.&&
\label{rhotau}
\end{flalign}
Now the horizon is located at $\rho_\textsc{h}=0$, thus the regularity condition on the scalar fields no longer implies the vanishing \eqref{dkrho} of the derivatives of the scalar fields $\phi^s$ with respect to $\rho$\,:
\begingroup
\belowdisplayskip=15pt
\belowdisplayshortskip=15pt
\begin{equation}
\lim_{\rho\rightarrow0}\phi^s(\rho)\=\phi_*^s\;,\qquad
|\phi_*^s|<\infty \qquad
\xcancel{
\Rightarrow\quad
\lim_{\rho\rightarrow-\infty}\,\frac{d^\ell}{d\rho^\ell}\,\phi(\rho)\=0
}\;.
\end{equation}
\endgroup
Moreover, equation \eqref{scaleqs6} no longer implies that $\phi_*^s$\, is an extremum for the black hole potential.\par\medskip

\subsection{Black holes and duality}
We have seen in Subsect.\ \ref{subsec:onshellduality} that the on-shell global symmetries of an extended supergravity theory are encoded -- at the classical level -- in the isometry group $G$ of the scalar manifold. The non-linear action of this duality group on the scalar fields $\phi^s$ is combined with a simultaneous linear symplectic action on the field strengths $F^\Lambda$ and their duals $\Gdual_\Lambda$. This duality action of $G$ is defined by a symplectic representation $\Rsv$ of $G$.\par
We have also studied how fermion fields transform under the compensating transformation $h(g,\,\phi)\in H$ in \eqref{duality2}. Under this action, static black hole solutions, defined by \eqref{metrans}, are mapped into solutions of the same kind.\par
Let us see more in detail this duality transformation. A transformation given by $g\in G$ maps a black hole solution into a new solution as:
\begin{align}
g\in G ~:\qquad
\begin{cases}
\;U(\tau) \cr
\;\phi^s(\tau)\cr
\;\Gamma^M\cr
\;\MADM
\end{cases}
\,\stackrel{g}\longrightarrow\quad
\begin{cases}
\;U^\prime(\tau)\=U(\tau)\cr
\;\phi^{\prime\,s}(\tau)\=g\star\phi^s(\tau)\cr
\;\Gamma^{\prime\,M}\=\Rsv[g]\,\Gamma\cr
\;\MADM
\end{cases}
\label{duality3}
\end{align}
the ADM-mass remaining the same being a property of the  metric of the solution, and hence not affected by duality transformations which leave the metric unaltered.\par
The above properties tell us that, for given charges $\Gamma$ and ADM-mass, the solution \,$\Xi=\{U(\tau),\,\phi^s(\tau)\}$\, is uniquely defined by the boundary condition $\phi^s_0$ for the scalar fields, while \,$\Xi'=\{U^\prime(\tau)=U(\tau),\,\phi^{\prime\,s}(\tau)\}$\, is the unique solution with charges $\Gamma^\prime$ defined by the boundary condition $\phi_0^\prime=g\star\phi_0$.\par
Using eq.s \eqref{RMR} and \eqref{duality3}, we see that the effective potential
\begin{equation}
\VBH(\phi,\;\Gamma)~\equiv\,-\frac12\,\Gamma^T\M(\phi)\,\Gamma\;,
\end{equation}
function of the scalar fields and quantized charges, turns out to be invariant under the simultaneous action \eqref{duality3}:
\begin{equation}
\begin{split}
\VBH(\phi,\;\Gamma)
\quad\stackrel{g}\longrightarrow\quad
\VBH'(g\star\phi,\;\Rsv[g]\,\Gamma)&\,=\,-\frac12\Gamma^T\,\Rsv^{T}\,\Rsv^{-T}\,\M(\phi)\,\Rsv^{-1}\,\Rsv\,\Gamma\,=
\\
&\,=\VBH(\phi,\;\Gamma)\;.
\label{Vhinv}
\end{split}
\end{equation}
This implies that $\VBH$ is $G$-invariant. From this property of invariance, it follows that the effective action \eqref{effact} and the extremality constraint \eqref{Hamconstr} are both manifestly duality-invariant expressions. A remarkable consequence of this, is that black holes in extended supergravities can be classified in \emph{orbits} with respect to the duality action \eqref{duality3} of the global symmetry group $G$.\par
If now we denote by $\phi_*^s(\Gamma)=\phi_*^s(e,m)$ the extremum of the potential $\VBH(\phi,\;\Gamma)$:
\begingroup
\belowdisplayskip=2pt
\belowdisplayshortskip=10pt
\begin{equation}
\partial_s\VBH\left(\phi_*(\Gamma),\;\Gamma\right)\=0\;,
\end{equation}
\endgroup
from \eqref{Vhinv} we find
\begin{equation}
\partial_s\VBH(\phi_*\,,\;\Gamma)\=0
\quad\Leftrightarrow\quad
\partial_s\VBH(g\star\phi_*\,,\;\Rsv[g]\,\Gamma)\=0\;,
\end{equation}
that is, the point $g\star\phi_*$ extremizes the potential $V(\phi'\,,\;\Rsv[g]\,\Gamma)$. However, such extremum was denoted by $\phi_*(\Rsv[g]\,\Gamma)$, so we can write:
\begin{equation}
g\star\phi^s_*(\Gamma)\=\phi^s_*(\Rsv[g]\,\Gamma)\;.
\label{gphistar}
\end{equation}
If we consider extremal solutions, the above property \eqref{gphistar}, together with \eqref{Vhinv}, implies:
\begin{equation}
\begin{split}
{\VBH}^\text{(ex)}(\Gamma)&\=\VBH\left(\phi_*(\Gamma)\,,\;\Gamma\right)
\=\VBH\left(g\star\phi_*(\Gamma)\,,\;\Rsv[g]\,\Gamma\right)
\=\VBH\left(\phi_*(\Rsv[g]\,\Gamma)\,,\;\Rsv[g]\,\Gamma\right)\=
\\[\jot]
&\={\VBH}^\text{(ex)}(\Rsv[g]\,\Gamma)\;.
\end{split}
\end{equation}
In other words, in the extremal case, the scalar potential at the
extremum -- which defines the horizon area $A_\textsc{h}$ and thus the entropy of the solution -- is a $G$-invariant function of the quantized
charges only. This implies that the entropy of the extremal solution
is a $G$-invariant function of the charges $\Gamma^M$.\par

\paragraph{Quartic invariant.}
In all the extended supergravity models with homogeneous-symmetric scalar manifold%
\footnote{%
except the $\N=2$ ones with $G=\U(1,n)$ and the $\N=3$ supergravity
}%
, the representation $\Rsv$ of $G$ (under which the electric and magnetic charges transform) has a single invariant quantity
\begin{equation}
I_4(\Gamma)\=I_4(e,m)\;,
\end{equation}
function of the electric-magnetic charge vector $\Gamma$. This is called the
\emph{quartic invariant} and has degree four in the charges \cite{Kallosh:1996uy,Fre:2011uy}.\par
If we denote by $\Rsv[T_\A]\equiv(T_\A)_M{}^N$ the matrices representing the
generators $T_\A$ of $G$ in the chosen symplectic duality representation $\Rsv$, the quartic invariant of these models has the general form:
\begin{equation}
I_4(\Gamma)~=\,-\frac{\nv(2\,\nv+1)}{6\;\mathrm{dim}(G)}\,
    (T_\A)_{MN}\,(T^\A)_{PQ}\;\Gamma^M\,\Gamma^N\,\Gamma^P\,\Gamma^Q\;,
\end{equation}
where the symplectic indices can raised and lowered using $\Cc^{MN}$ and $\Cc_{MN}$, while the index $A$ is raised by the inverse of \,$\eta_{{}_{\A\B}}\equiv(T_\A)_M{}^N(T_\B)_N{}^M$\,.\par\smallskip
The potential at the extremum can be written in terms of $I_4(e,m)$ as
\begin{equation}
{\VBH}^\text{(ex)}=~\sqrt{\big|I_4\big|}\;,
\end{equation}
while the horizon area $A_\textsc{h}^\text{(ex)}$ reads
\begin{equation}
A_\textsc{h}^\text{(ex)}=~4\pi\,\left(\frac{8\pi\GN}{c^4}\,\sqrt{\big|I_4\big|}\right)\;,
\end{equation}
and, therefore, the entropy of the extremal solution has the form
\begin{equation}
\Sentr^\text{(ex)}=~\frac{k_\textsc{b}}{\ell_\textsc{p}^2}\;\pi\,\left(\frac{8\pi\GN}{c^4}\,\sqrt{\big|I_4\big|}\right)\;.
\end{equation}
In most theories, the orbits of the magnetic charges $\Gamma^M$ of a black hole solution can be classified, with respect to the action of $G$, according to:
\begin{equation}
\begin{split}
\text{Orbit I (BPS)}\;&:\quad I_4>0\;,\qquad\qquad
\\
\text{Orbit II (non-BPS)}\;&:\quad I_4>0\;,\qquad\qquad
\\
\text{Orbit III (non-BPS)}\;&:\quad I_4<0\;,\qquad\qquad
\end{split}
\end{equation}
while orbits of $\Gamma^M$ charges with vanishing quartic invariant ($I_4=0$) define the so-called \emph{small black holes}.
It was shown that orbits I, II and III define all possible orbits of regular black hole configurations in extended supergravity models \cite{Bellucci:2006xz}.


\newpage

\section{Constructing black hole solutions} \label{sec:constrbhsol}
The study of stationary black holes solutions in supergravity is a field of research of great interest because of its theoretical and phenomenological implications. The latter, in particular, can have a profound impact on our comprehension of particle physics, cosmology and mathematical formulation of fundamental field theories (like superstring or M-theory).\par
Extremal black hole solutions \cite{Ferrara:1995ih,Andrianopoli:2006ub,Ceresole:2007wx} feature an universal near-horizon behavior due to the attractor phenomenon \cite{Ferrara:1996dd,Ferrara:1996um,Kunduri:2007vf} that we introduced in Subsect.\ \ref{subsec:nhbehav}. Non-extremal, stationary solutions exhibit a less constrained form of the metric \cite{Bergshoeff:1996gg,Galli:2011fq,Chow:2013tia}, the known examples typically obtained through the so-called \emph{solution-generating techniques} \cite{Geroch:1970nt,Cvetic:1995kv}. The idea underlying this approach is that stationary solutions to $D=4$ supergravity are also solutions to an Euclidean theory in three dimensions, formally obtained by compactifying the $D=4$ correspondent model along time-direction \cite{Breitenlohner:1987dg} and dualizing the vectors of the theory into scalars. The resulting $D=3$ theory is a sigma-model coupled to gravity and features a global $\G$ symmetry group larger than the original $\G[4]$ group of the $D=4$ model. The obtained extra symmetries can be used to generate new (hidden) four-dimensional solutions from known ones.
These symmetries include, for instance, \emph{Harrison transformations}, which can generate electric and magnetic charges acting on a neutral solution (like the Schwarzschild or Kerr black hole).
The physical properties of stationary black holes in four dimensions can be then classified in orbits w.r.t.\ the action of the three-dimensional global symmetry group $\G$. \par
Extremal solutions of supergravity black holes can be obtained as limits of the previous non-extremal ones, where the extremality parameter, related to the Hawking temperature of the black hole solution, is sent to zero \cite{Cvetic:1996kv,Rasheed:1995zv,Bena:2009ev}.
Another non-trivial example of extremal limit (which we shall refer to as the \emph{Rasheed-Larsen limit}) was defined in \cite{Rasheed:1995zv}, and allowed to find the first instance of extremal under-rotating (no ergosphere) solutions from a given non-extremal one in the $D=4$ theory obtained, through Kaluza--Klein reduction, from pure gravity in five dimensions \cite{Bena:2009ev}. The Rasheed-Larsen limit was generalized in \cite{Andrianopoli:2013kya,Andrianopoli:2013jra} to a non extremal stationary black hole in the so-called $T^3$ model, obtaining the non-BPS under-rotating solution through a singular Harrison transformation applied on a non-extremal Kerr black hole. \par
In this Section, we will focus on generic symmetric, extended supergravity models, in order to obtain the form of the most general, single center extremal solution, modulo the action of the global symmetry group $\G$.
Non-extremal rotating, asymptotically-flat black hole solutions can be obtained by acting with a suitable Harrison transformations on the non-extremal neutral Kerr solution.\par
Representatives of the $\G$-orbits of regular, extremal solutions in supergravity theories, can be obtained as limits of a single non-extremal rotating solution of the so-called STU-model (see App.\ \ref{app:coset}).
A broad class of symmetric, extended supergravities share the STU-model as a common universal truncation, and comprise all the extended $D=4$ models whose scalar manifold is symmetric of the form $\G[4]/\HH[4]$, where the four dimensional isometry group $\G[4]$, is a non-degenerate group of type-$\text{E}_7$ \cite{brown1969groups}%
\footnote{%
in the $\N=2$ case, the condition is referred to the special K\"ahler
manifold spanned by the scalar fields in the vector multiplets
}%
. These models include the maximal $\N=8$ and half-maximal $\N=4$ supergravity, as well as \,$\N=2$  models with rank-3 symmetric special K\"ahler manifold. At least as far as the single-center solutions are concerned, the $\G$-orbits of regular black holes in all these models have a representative in the STU-truncation.\par
%

\subsection{Solution-generating technique}
Let us consider stationary solutions in an extended, ungauged  $D=4$ supergravity, whose bosonic sector consists in $\ns$ scalar fields $\phi^s(x)$, $\nv$ vector fields $A^\Lambda_\mu(x)$ ({\small$\Lambda$}\;$=1,\,\dotsc,\,\nv$), and the graviton $g_{\mu\nu}(x)$. The solution is described by the four-dimensional Lagrangian \eqref{boslagr} introduced in Sect.\ \ref{sec:ungsugra}, that reads%
\footnote{%
in the ``mostly minus'' convention and \,$8\pi \GN=c=\hbar=1$
} %
:
\begin{equation}
\frac{1}{\eD}\Lagr_{(4)}~=\,-\frac{R}{2}
\+\frac{1}{2}\,\Gm[s][u](\phi)\,\partial_\mu\phi^s\,\partial^\mu\phi^u
\+\frac{1}{4}\,\II_{\Lambda\Sigma}(\phi)\,F^\Lambda_{\mu\nu}\,F^{\Sigma\,\mu\nu}
\+\frac{1}{8\,\eD}\,\RR_{\Lambda\Sigma}(\phi)\,\veps^{\mu\nu\rho\sigma}\,F^\Lambda_{\mu\nu}\,F^\Sigma_{\rho\sigma}\;,
\end{equation}
The four-dimensional scalar fields $\phi^s$ parameterize an homogeneous, symmetric scalar manifold of the form:
\begin{equation}
\Ms^{(4)}_\text{scal}\=\frac{\G[4]}{\HH[4]}\;,
\label{M4}
\end{equation}
where $\G[4]$ is the semisimple isometry group and $\HH[4]$ its maximal compact subgroup. As we have seen in Sect.\ \ref{subsec:onshellduality}, the group $\G[4]$ also defines the global on-shell symmetry of the theory, through its combined action on the scalars and on vector field strengths and their magnetic duals, as an electric-magnetic duality group.\par\smallskip
The $D=4$ stationary, axisymmetric metric can be cast in the general form
\begin{equation}
ds^2\=e^{2U}\,(dt+\omega_\varphi\,d\varphi)^2\-e^{-2U}\,g_{ij}^{(3)}\,dx^i\,dx^j\,;
\label{ds2}
\end{equation}
where $i,\,j=1,\,2,\,3$\, label the spatial coordinates $x^i=(r,\,\theta,\,\varphi)$\, and \,$U$, $\omega_\varphi$, $g_{ij}^{(3)}$ are functions of the coordinates ($r,\,\theta$). The metric \eqref{ds2} has two Killing vectors, \,$\xi=\partial_t$\, and \,$\psi=\partial_\varphi$\,.

\paragraph{Dimensional reduction.}
The previous stationary metric solution can be formally reduced to three dimensions, compactifying along the time direction and dualizing the vectors of the theory to scalar fields \cite{Breitenlohner:1987dg}. The result gives an effective description of the theory in an euclidean $D=3$ model, where gravity is coupled to \,$n=2+ \ns+2\,\nv$\, scalar fields $\Phi^I(r,\,\theta)$, see App.\ \ref{app:dimred}.\par
After the 3D Hodge-dualization, the propagating degrees of freedom are reduced to the following scalar fields:
\begin{enumerate}[$\circ$,topsep=0.75ex,labelsep=1.15em]
  \item {$\ns$ four-dimensional scalars $\phi^s$;}
  \item {the warp function $U$;}
  \item {$2\,\nv$ scalars \,$\Z^M=\{\Z^\Lambda,\,\Z_\Lambda\}$\, from the dimensional reduction of the four-dimensional vectors fields;}
  \item {the scalar $a$ from the dualization of the Kaluza-Klein vector $\omega_\varphi$\,.}
\end{enumerate}
The relations between the scalars $a$, $\Z^M$ and the four dimensional fields can be written
\begin{equation}
\begin{split}
A_{(4)}^\Lambda&\,=\,A_0^\Lambda(dt+\omega)\+A_{(3)}^\Lambda\;,
    \qquad\; A_{(3)}^\Lambda\,\equiv\,A_i^\Lambda dx^i\;,
\\[\jot]
\FF^M&\,=\,
\left(\begin{matrix}
F_{(4)}^\Lambda{}\cr
\Gdual_{(4)\Lambda}
\end{matrix}\right)
    \,=\,d\Z^M \wedge(dt+\omega)\+e^{-2U}\,\Cc^{MN}\,\M_{(4)\,NP}\;\,
    {}^{*_3}d\Z^P\;,
\\[\jot]
da&\,=\,-e^{4 U}\;{}^{*_3}d\omega\-\Z^T\,\Cc\,d\Z\;,
\label{dualizform}
\end{split}
\end{equation}
with
\begin{equation}
F_{(4)}^\Lambda\=dA_{(4)}^\Lambda\;,\qquad\quad
\Gdual_{(4)\Lambda}=\,-\frac{1}{2}\;
    \*\!\left(\frac{\partial\Lagr_{(4)}}{\partial F_{(4)}^\Lambda}\right)\;,
\end{equation}
and where $\*$ is the Hodge operation in four dimensions, ${}^{*_3}$ stands for the Hodge operation in the $D=3$ Euclidean space and $\M_{(4)}$ is the symmetric, symplectic matrix characterizing the symplectic structure over the manifold $\Ms^{(4)}_\text{scal}$\,.
The symplectic vector $\FF^M$ transforms, under the duality action of $\G[4]$, in a symplectic representation $\Rs_\text{s}$\,.\par
The final resulting effective $D=3$ Lagrangian $\Lagr_{(3)}$ describes a sigma-model coupled to gravity and reads \cite{Andrianopoli:2013kya}:
\begingroup
\belowdisplayskip=-5pt
\belowdisplayshortskip=-5pt
\begin{equation}
\begin{split}
\frac{1}{\eD^{(3)}}\;\Lagr_{(3)}&~=\,
    -\frac{R^{(3)}}{2}\+\frac{1}{2}\,\hat{\mathpzc{G}}_{ab}(z)\,\partial_i{z}^a\,\partial^i{z}^b\=
\\
    &~=\,-\frac{R^{(3)}}{2}\+\big(\partial_i U\,\partial^i U
    \+\frac{1}{2}\,\Gm[s][u]\,\partial_i{\phi}^s\,\partial^i{\phi}^u\+
    \frac{1}{2}\,e^{-2U}\,\partial_i{\Z}^T\,\M_{(4)}\,\partial^i{\Z}\+
    \\
    &\phantom{\=\frac{R^{(3)}}{2}\-\big(}
    \+\frac{1}{4}\,e^{-4U}\,(\partial_i{a}+\Z^T\,\Cc\,\partial_i{\Z})\,
    (\partial^i{a}+\Z^T\,\Cc\,\partial^i{\Z})\big)\;,
\label{geodaction}
\end{split}
\end{equation}
\endgroup
where $\eD^{(3)}\equiv \sqrt{\text{Det}(g^{(3)}_{ij})}$ and $\Cc$ is the symplectic-invariant, antisymmetric matrix defined in \eqref{Cc}.

\subsubsection{Three-dimensional description}
The $D=3$ scalar fields obtained from the dimensional reduction span an homogeneous, symmetric, pseudo-Riemannian scalar manifold $\Ms^{(3)}_\text{scal}$ of the form
\begin{equation}
\Ms^{(3)}_\text{scal}\=\frac{\G}{\HHstar}\;,
\end{equation}
containing $\Ms^{(4)}_\text{scal}$ as a submanifold
\begin{equation}
\Ms^{(4)}_\text{scal}~\subset~\Ms^{(3)}_\text{scal}\;,
\end{equation}
and where the isometry group $\G$ is a semisimple, non-compact Lie group defining the global symmetry of the model, while $\HHstar$ is a non-compact real form of $\HH[3]$, the semisimple maximal compact subgroup of $\G$.\par\smallskip
The three-dimensional scalar fields $\Phi^I$ define a \emph{local solvable parametrization} of the coset, where the coset representative is chosen to be:
\begin{equation}
\LL\left(\Phi^I\right)\=\exp(-a\,T_\ms{\bullet})\;\,\exp(\sqrt{2}\,\Z^M\,T_M)\;\,
             \exp(\phi^r\,T_r)\;\,\exp(2U\,H_0)\;,
\label{cosetr3}
\end{equation}
where $T_{\mathcal{A}}=\{H_0,\,T_\ms{\bullet},\,T_s,\,T_M\}$ are the solvable generators. The generators $T_M$ transform under the adjoint action of $\G[4]\subset \G$ in the symplectic duality representation $\Rs_\text{s}$ of the electric-magnetic charges, so we can use the notation \,$T_M=\left(T_{q_\Lambda},\,T_{p^\Lambda}\right)$.\par\smallskip
The Lie algebra of $\HHstar$ is denoted by $\halgstar$ and is a subalgebra of $\mathfrak{g}_3$, the Lie algebra of the isometry group $\G$. In the above procedure, we have considered a matrix representation in which $\halgstar$ and its orthogonal complement $\kalgstar$ are defined by a \emph{pseudo-Cartan involution} $\hat{\zeta}$\,. This involutive automorphism, acting on the algebra $\mathfrak{g}_3$ of $\G$, leaves invariant algebra $\halgstar$ generating $\HHstar$\,. The action of $\hat{\zeta}$ on a general matrix $A$ is
\begin{equation}
\hat{\zeta}(A)~=\-\eta\;A^\dagger\;\eta\;,
\label{invol}
\end{equation}
where $\eta$ is a suitable $\HHstar$-invariant metric.\par

\paragraph{Physical quantities.}
Stationary axisymmetric black hole solutions can be described by $n$ functions $\Phi^I(r,\,\theta)$ that come from the solutions of the sigma model equations. They are characterized by an \'initial point'' \;$\Phi_0\equiv\Phi_0^I$\; at radial infinity \cite{Andrianopoli:2013ksa}
\begin{equation}
\Phi_0~=\lim_{r\rightarrow \infty} \Phi^I(r,\,\theta)\;,
\end{equation}
and an \'initial velocity'' $\Q$, at radial infinity, in the tangent space $T_{\Phi_0}[\Ms^{(3)}]$. This matrix $\Q$ is the \emph{Noether charge matrix}, belonging to the Lie algebra $\mathfrak{g}_3$ of $\G$.\par
Since the action of \,$\G/\HHstar$\, on \,$\Phi_0$\, is transitive, we can always fix $\Phi_0$ to coincide with the origin of the manifold $\OO$ (defined by the vanishing values of all the scalars) and then classify the orbits of the solutions under the action of $\G$ (maximal sets of solutions connected through the action of $\G$) in terms of the orbits of the velocity vector $\Q\in T_{\OO}[\Ms^{(3)}]$ under the action of $\HHstar$:
\begin{equation}
\frac{\G}{\HH[3]}\;\;
\text{trans.\ on}\;\Phi_0\;:\quad\;
\text{orbits of }\G\;
\xrightarrow{\Phi_0\equiv\OO}\;
\text{orbits of }\HH[3]\;.
\end{equation}
We can now introduce the hermitian, $\HHstar$-invariant matrix $\M_{(3)}$ which, in a chosen matrix representation, reads:
\begin{equation}
\M_{(3)}~\equiv~\M_{(3)}\left(\Phi^I\right)~\equiv~\LL\,\eta\,\LL^\dagger\=\M_{(3)}^\dagger\;.
\label{cm}
\end{equation}
The three-dimensional \emph{Noether currents} associated with a stationary solution $\Phi^I(x^i)$ can be written in terms of $\M_{(3)}$ as:
\begin{equation}
\hat{J}_i~\equiv~\frac{1}{2}\,\partial_i\,\Phi^I\,\M_{(3)}^{-1}\;\partial_I\M_{(3)}\;.
\label{curr}
\end{equation}
In terms of the above currents, the $\mathfrak{g}_3$-valued Noether-charge matrix $\Q$ reads:
\begin{equation}
\Q\=\frac{1}{4\pi}\int_{S_2} {}^{*_{3}}\hat{J}\=\frac{1}{4\pi}\,
    \int\sqrt{\eD^{(3)}}\;\hat{J}^r\,d\theta\,d\varphi\;,
\label{Q}
\end{equation}
the index of $\hat{J}_i$ being raised using $g^{(3)\,ij}$. \par\smallskip
If we restrict to axisymmetric solutions, we find an angular Killing vector \,$\psi=\partial_\varphi$\,, and all the fields will only depend on the spatial variables $(r,\,\theta)$. The global rotation of the solution can be described by means of the $\mathfrak{g}_3$-valued matrix $\Qpsi$ \cite{Andrianopoli:2012ee,Andrianopoli:2013kya,Andrianopoli:2013jra}, derived from the standard Komar-integral definition of $\J$ in $D=4$, having the form \cite{Wald:1984rg}:
\begin{equation}
\Qpsi~=\,-\frac{3}{4\pi}\,\int_{S_2}^\infty
        \psi_{[i}\,\hat{J}_{j]}\;dx^i\wedge dx^j
    \=\frac{3}{8\pi}\,\int_{S_2}^\infty
        g^{(3)}_{\varphi\varphi}\;\hat{J}_\theta\,d\theta\,d\varphi\;.
\label{Qpsi}
\end{equation}
\sloppy
The ADM-mass, NUT-charge, scalar charges $\Sigma_s$, electric and magnetic charges ${\Gamma^M=(p^\Lambda,\,q_\Lambda)}$ and angular momentum $\J$ of the solution can be obtained as components of $\Q$ and $\Qpsi$:
\begin{equation}
\begin{split}
\MADM&=k\,\Tr\left(H_0^\dagger\,\Q\right)\,,\qquad
\Nnut=-k\,\Tr\left(T^\dagger_\ms{\bullet}\,\Q\right)\,,\qquad
\Sigma_s=k\,\Tr\left(T_s^\dagger\,\Q\right)\,,
\\[1.4ex]
\Gamma^M&=\sqrt{2}\;k\;\Cc^{MN}\,\Tr\left(T_N^\dagger\,\Q\right)\,,\qquad\;
\J=k\,\Tr\left(T^\dagger_\ms{\bullet}\,\Qpsi\right)\,.\quad
\label{quantsol}
\end{split}
\end{equation}
Since $\G$ is the global symmetry group of the effective three-dimensional model, a generic element $g\in G$ maps a solution $\Phi^I(r,\,\theta)$ into an other solution $\Phi^{\prime\,I}(r,\,\theta)$ according to the matrix equation:
\begin{flalign}
\quad\;\;
\forall g\in\G\;:\qquad
\M_{(3)}\left(\Phi^{I}\right)\;\xlongrightarrow{g}\;
\M_{(3)}\left(\Phi^{\prime I}\right)
    =g\;\M_{(3)}\left(\Phi^{I}\right)\;g^\dagger\;.&&
\label{gMg}
\end{flalign}
From the definitions \eqref{Q}, \eqref{Qpsi} and from \eqref{gMg}, one finds that $\Q$ and $\Qpsi$ transform under the adjoint action of $\G$ as:
\begin{flalign}
\quad\;\;
\forall g\in\G\;:\qquad
\Q\;\xrightarrow{g}\;\Q'=(g^{-1})^\dagger\,\Q\,g^\dagger\;,\qquad\;
\Qpsi\;\xrightarrow{g}\;\Q'_\psi=(g^{-1})^\dagger\,\Qpsi\,g^\dagger\;.&&
\label{gQg}
\end{flalign}
The angular momentum of the transformed solution can be easily obtained from eqs.\ \eqref{quantsol}, without computing the corresponding transformed Komar integral from \eqref{gMg}. The presence of a non-vanishing $\Qpsi$ is a characteristic of the $\G$-orbits of rotating solutions, and this tells us that is not possible to generate rotation on a static  $D=4$ solution using $\G$-transformations.\par\smallskip
Since it is always possible to map point at radial infinity $(\Phi_0^I)$ into the origin $\OO$ of the manifold by means of a $\G/\HHstar$-transformation, the group $\G$ is broken to the isotropy group $\HHstar$ and, as a consequence of this, the two matrices $\Q$, $\Qpsi$ always lie in the coset space $\kalgstar$.

\paragraph{Harrison transformations.}
The so-called Harrison transformations are $\HHstar$ transformations generated by the non-compact generators $\Jb_M$ of $\halgstar$:
\begin{equation}
\Jb_M ~\equiv~ \frac{1}{2}\,(T_M+T_M^\dagger)\;.
\label{JM}
\end{equation}
The space $\Jb^{(R)}=\Span\left(\Jb_M\right)$ is the carrier of a representation%
\footnote{%
the symplectic duality representation $\Rs_\text{s}$ of $\G[4]$ and the corresponding representation of $\Hc$ are both related to the electric and magnetic charges
}
$\Rs$ with respect to the adjoint action of the maximal compact subgroup $\Hc$ of $\HHstar$. This group has the general form
\begingroup
\begin{equation}
\Hc\=\U(1)_\textsc{e}\times\HH[4]\;,
\label{Hc}
\end{equation}
\endgroup
where $\U(1)_\textsc{e}$ belongs to the Ehlers group $\SL(2,\mathbb{R})_\textsc{e}$\,.\par
We can also define the subspace $\mathbb{K}^{(R)}$ of the coset space $\kalgstar$ spanning the negative-signature directions of the metric; this space defines the support of a representation $\Rs$ of $\Hc$, just as we did with $\Jb^{(R)}$.
The compact generators $\mathbb{K}_M$ of $\kalgstar$ can be written, in the chosen matrix representation, as
\begingroup
\abovedisplayskip=5pt
\abovedisplayshortskip=5pt
\belowdisplayskip=15pt
\belowdisplayshortskip=15pt
\begin{equation}
\mathbb{K}_M ~\equiv~ \frac{1}{2}\,(T_M-T_M^\dagger)\;.
\label{KM}
\end{equation}
\endgroup

\subsection{The Kerr Family} \label{sec:Kerr}
In the seminal paper \cite{Breitenlohner:1987dg}, it was proven that the most general non-extremal (or extremal over-rotating) stationary,  axisymmetric single center black hole solution to the model can be obtained from the non-extremal (or extremal) Kerr solution through the action of $\G$, more precisely through an Harrison transformation. This can be considered as an equivalent version of the ``no-hair theorem'' for this class of theories. In fact, the scalar charges of a generic stationary solution are functions of the Harrison parameters, mass and angular momentum of the original Kerr solution: since the Harrison parameters are in one-to-one correspondence with electric/magnetic charges, the most general solution is uniquely defined by {$\MADM,\,\J,\,\Gamma$}, the scalar charges being dependent on these.\par\smallskip
The matrices $\Q$ and $\Qpsi$ for the Kerr solution are characterized by two parameters, a mass $m$ and an angular-momentum parameter $\alpha$. Since they are diagonalizable, their $\G$-orbits are uniquely characterized by their eigenvalues. In the pure Kerr solution, $\Q$ and $\Qpsi$ belong to the same $\G$-orbit, modulo multiplication by $\alpha$. We find:
\begin{equation}
\Qpsi\,=\,\alpha\;h^{-1}\Q\,h\;;\qqqquad
h\in\U(1)_\textsc{e}\;,
\label{QQpsi}
\end{equation}
where $\U(1)_\textsc{e}$ stands for the compact Ehlers transformation group. This will no longer be the case in the extremal limit.\par
The matrix  $\Q$ belongs to the \emph{Schwarzschild orbit} \cite{Bossard:2009at,Andrianopoli:2013kya,Andrianopoli:2013jra}, characterized by the matrix equation%
\footnote{%
the constant $\bar{q}^2$, in the case of the Kerr-Newmann-NUT black hole with e/m charges $\Gamma=(q,\,p)$ and NUT-charge $\Nnut$, reads: $\bar{q}^2=\frac{k}{2}\Tr(\Q^2)=m^2+\Nnut^2-\frac{p^2+q^2}{2}$
}
\begin{equation}
\Q^3\=\bar{q}^2\,\Q\;,\qqquad\;
\bar{q}^2=\frac{k}{2}\Tr(\Q^2)=m^2\;,
\label{eqa1}
\end{equation}
where $\Q$  is in the fundamental representation%
\footnote{%
this is true if \,$\G\neq\text{E}_{8(8)},\text{E}_{8(-24)}$\,;\; if $\G$ is a real form of $\text{E}_{8}^{\Cc}$ the fundamental and the adjoint representation coincide, and the matrix equation becomes quintic in $\Q$ \cite{Bossard:2009at}
}
of $\G$. From the above equation \eqref{QQpsi}, we find that
\begin{equation}
\Qpsi^3\=\alpha^2\,\bar{q}^2\,\Qpsi\;,\qqquad\;
\alpha^2=\frac{\Tr(\Qpsi^2)}{\Tr(\Q^2)}\;,
\label{eqa2}
\end{equation}
and the following equations holds:
\begin{equation}
\Qpsi^2\,\Q\=\alpha^2\,\bar{q}^2\,\Q\;;\qqquad\;
\Q^2\,\Qpsi\=\bar{q}^2\,\Qpsi\;.
\label{eqa3}
\end{equation}
Equations \eqref{eqa1}, \eqref{eqa2} and \eqref{eqa3}, together with the trace expression for the parameters $m$ and $\alpha$, are $\G$-invariant and thus hold for any representative of the Kerr $\G$-orbit.\par\smallskip
We can define the extremality parameter $\cex$ in terms of the following $\G$-invariant quantity \cite{Andrianopoli:2013kya,Andrianopoli:2013jra}:
\begin{equation}
\cex^2\=m^2-\alpha^2\=\frac{k}{2}\,\Tr(\Q^2)-\frac{\Tr(\Qpsi^2)}{\Tr(\Q^2)}\;.
\label{eqa4}
\end{equation}
The Hawking temperature of the black hole can be now written in terms of the extremality parameter as:
\begin{equation}
T\=\frac{\cex}{2\pi\,\alpha\,\left|\omega_\textsc{h}\right|}
    \=\frac{\cex}{2\,\Sentr}\;,
\label{temperatureH}
\end{equation}
where $\Sentr$ stands for the Bekenstein-Hawking entropy of the solution, that, in turn, can be expressed in terms of the horizon area $A_\textsc{h}$ as
\begin{equation}
\Sentr\=\frac{k_\textsc{b}\,c^3}{\GN\,\hbar}\,\frac{A_\textsc{h}}{4}\=\frac{A_\textsc{h}}{4}
    \=\pi\,\alpha\,\left|\omega_\textsc{h}\right|\;,
\label{entropy2}
\end{equation}
while $\omega_\textsc{h}$ is defined as
\begin{equation}
\omega_\textsc{h}~=\lim_{r\rightarrow r_+\;}\omega_\varphi\;;
\qqquad   r_+=m+\cex\;.
\end{equation}
Using the above expression, one can rewrite the \emph{regularity bound} $\cex^2>0$ for the Kerr solution in a $G$-invariant form:
\begin{equation}
m^2\ge\alpha^2
\qLq
\frac{k}{2}\,\Tr(\Q^2)~\ge~\frac{\Tr(\Qpsi^2)}{\Tr(\Q^2)}\;,
\label{regbound}
\end{equation}
which thus holds for any representative of the Kerr-orbit.

\subsubsection{Angular momentum and duality} \label{subsubsec:angmomdual}
Now we want to study the properties of the angular momentum $\J$ with respect to the four-dimensional duality symmetry group $\G[4]$. To this purpose, we relax the previous assumption to fix the transitive action of $\G/\HHstar$ on the solution choosing the scalars at infinity to correspond to the origin $\OO$ .\par
In general, for a rotating black hole, the angular momentum depends on the boundary values $\phi_0^s$ of the scalars and on the electric-magnetic charges $\Gamma^M$; equation \eqref{quantsol} shows how to express the angular momentum in terms of the matrix $\Qpsi$.\par
Suppose now we transform the solution by means of an element $g\in\G[4]$ into another one with boundary values $\phi_0^{\prime\,s}$ and charges $\Gamma^{\prime M}$:
\begin{equation}
\forall g\in\G[4]\;:\qquad
\begin{cases}
\;\phi_0^s\,\stackrel{g}\longrightarrow\;\phi_0^{\prime\,s}\cr
\;\Gamma^M\stackrel{g}\longrightarrow\;\Gamma^{\prime M}
\end{cases}
\quad.
\end{equation}
Using definitions \eqref{quantsol}, it is possible to show that $\J$ is not affected by the action of $g\in\G[4]$. In fact, the matrix $\mathcal{Q^{\,\prime}_\psi}$ associated with the new solution is related to $\Qpsi$ by eq.\ \eqref{gQg}, so that for the corresponding angular momentum one has:
\begin{equation}
\forall g\in\G[4]\;:\qquad
\J\left(\phi_0^s,\;\Gamma\right)
\;\;\stackrel{g}\longrightarrow\;\;
\J\left(\phi_0^{\prime\,s},\;\Gamma'\right)
\quad.
\end{equation}
with
\begin{equation}
\begin{split}
\J\left(\phi_0^{\prime\,s},\;\Gamma'\right)&\=
    k\,\Tr\left(T^\dagger_\ms{\bullet}\,\mathcal{Q^{\,\prime}_\psi}\right)\=
    k\,\Tr\left(T^\dagger_\ms{\bullet}\,(g^{-1})^\dagger\,\Qpsi\,g^\dagger\right)\=
\\
    &\=k\,\Tr\left(T^\dagger_\ms{\bullet}\,\Qpsi\right)\=\J\left(\phi_0^s\,,\;\Gamma\right)\;,
\label{Jinvar}
\end{split}
\end{equation}
where we have used the property that $\G[4]$ commutes with the Ehlers group  $\SL(2,\mathbb{R})_E$ inside $\G$, so that its elements commute with the $\mathfrak{sl}(2,\mathbb{R})_E$ generators $\{H_0,\,T_\ms{\bullet},\,T^\dagger_\ms{\bullet}\}$. We conclude that $\J$ is a $\G[4]$-invariant function of the scalars at radial infinity and electric-magnetic charges. This is what one would expect for the angular momentum of a solution: being a quantity related to spatial rotation, $\J$ should not be affected by a $D=4$ duality transformation. \par
The above derivation does not hold for a generic global symmetry transformation in $\G$. In fact, in the under-rotating limit $\J$ is independent of \,$\phi_0^s$\, and thus is expressed in terms of the $\G[4]$-invariant of the electric-magnetic charges alone, namely in terms of the quartic invariant function $I_4(e,m)$ \cite{Andrianopoli:2013jra}. We find a similar behaviour for the horizon area (i.e.\ the entropy) by virtue of the attractor mechanism: from this, we conclude that there seems to be some kind of ``attractor phenomenon'' at work also for the angular momentum.\par
Finally, let us notice that the simple proof \eqref{Jinvar} of invariance under $\G[4]$ also applies to the ADM-mass and the NUT-charge, both given in \eqref{quantsol}.

\subsubsection{Extremal Limits} \label{subsec:extrlim}
The regularity bound $\cex^2\ge 0$ is saturated for the extremal solutions, which are thus characterized by a vanishing Hawking temperature \eqref{temperatureH}. This bound can be saturated in essentially two ways:
\begin{enumerate}[label=$\circ$,topsep=0.6ex,itemsep=0.2ex,after=\vspace{0.4ex}]
 \item {both sides of \eqref{regbound}, stay different from zero, so that
        the extremality condition becomes a constraint on the two non-vanishing $G$-invariants: the resulting solution is called \emph{extremal over-rotating} and retains, in this limit, the presence of an ergosphere; the matrices $\Q$ and $\Qpsi$ are still diagonalizable;}
 \item {both sides of \eqref{regbound} vanish separately and the resulting
        solution can either be \emph{extremal under-rotating} \cite{Rasheed:1995zv,Larsen:1999pp,Goldstein:2008fq,Bena:2009ev,Bena:2009en} or \emph{extremal-static} and has no ergosphere in both cases; both $\Q$ and $\Qpsi$ become nilpotent, belonging to different $G$-orbits \big(in particualr $H^{^\mathlarger{*}}$ orbits on $T_\OO(\Mscal)\sim \kalgstar$\big) \cite{Andrianopoli:2013kya,Andrianopoli:2013jra}.}
\end{enumerate}
The second limit has been considered, for example, in Heterotic theory \cite{Cvetic:1995kv,Astefanesei:2006dd} or Kaluza-Klein supergravity \cite{Rasheed:1995zv,Larsen:1999pp}.

\paragraph{Singular Harrison transformations.}
A geometric procedure for connecting the non-extremal Kerr-orbits to extremal static or under-rotating cases can be performed in a frame-independent way making use of \emph{singular} Harrison transformations \cite{Andrianopoli:2013kya,Andrianopoli:2013jra}. The latter effect an In\"on\"u--Wigner contraction on the matrices $\Q$ and $\Qpsi$, resulting in the transformed nilpotent matrices $\Q^{(0)}$ and $\QpsiZ$
associated with extremal static or under-rotating black hole
configurations.\par
Harrison transformations \cite{Breitenlohner:1987dg} are $\HHstar$-transformations  that are not present among the global symmetries of the  $D=4$ theory and have the distinctive property of switching on electric or magnetic charges when acting on neutral solutions (like the Kerr or Schwarzshild ones). Their generators \,$\Jb_M=(\Jb_\Lambda,\,\Jb^\Lambda)\,\in\,\halgstar$\, are in one-to-one correspondence with the electric and magnetic charges $\Gamma^M=(p^\Lambda,\,q_\Lambda)$ and are non-compact generators, that is they are represented, in a suitable basis,  by hermitian matrices.\par
The space \,$\Span\left(\Jb_M\right)$\, is the coset space of the symmetric manifold $\HHstar/\Hc$, where $\Hc$ is the maximal compact subgroup of $\HHstar$. It is the carrier of a representation of $\Hc$, the same in which the charges $\Gamma^M$ transform with respect to the group $\Hc$ itself, that has the structure of eq.\ \eqref{Hc}.

\paragraph{Maximal abelian subalgebra.}
Let us consider the space $\Span\left(\Jb_M\right)$. The \emph{maximal abelian subalgebra} (MASA) of this space is a subspace whose generators \,$\Jb^{(N)}=\{J_\ell\}$\, are defined by the \emph{normal form} of the electric and magnetic charges, i.e.\ the minimal subset of charges into which the charges of the most general solution can be rotated by means of $\Hc$, its dimension $\texttt{p}$ being therefore the rank of the $\HH[3]/\Hc$ coset. In the maximal supergravity, for example, one has
\begin{flalign}
\qqquad
\N=8\;:\qquad\;
\texttt{p}\=\rank\left(\frac{\SO^*(16)}{\U(8)}\right)\=4\;,&&
\end{flalign}
and the same result is found for the half-maximal theory, where
\begin{flalign}
\qqquad
\N=4\;:\qquad\;
    \texttt{p}\=\text{rank}\left(\frac{\SO(6,2)\times\SO(2,6+n)}{\SO(2)^2\times\SO(6)\times\SO(6+n)}\right)
    \=4\;.&&
\end{flalign}
If one considers the $\N=2$ symmetric models with rank-3 scalar special K\"ahler manifold in $D=4$, one gets \;$\texttt{p}=\text{rank}\big(\HH[3]/\Hc\big)=4$\,, since, for this class of theories, one has \,$\texttt{p}=\text{rank}+1$. The simplest representative of the latter class of models is the STU model, which is a consistent truncation of all the others, being a truncation of the maximal and half-maximal theories. This means that its space $\Span\left(\Jb^{(N)}\right)$ is contained in the spaces of Harrison generators of all the above mentioned symmetric models.\par\smallskip
As a consequence of the previous discussion, we can now restrict ourselves to the simplest STU model since the $\G$-orbits of non-extremal and extremal regular solutions to the broad class of the above symmetric models mentioned have a representative in the common STU-truncation%
\footnote{%
if one considers the restricted number of $\N=2$ symmetric models for which the rank of  $\Ms^{(4)}_\text{scal}$ is less than 3, the subsequent discussion has a straightforward generalization
}%
.\par
The higher-dimensional origin of the four-dimensional theory is encoded in the chosen symplectic frame. The latter determines the set of charges constituting the normal form, that can be geometrically characterized.
Let us express the Harrison generators in the form:
\begin{equation}
\Jb_M\=\frac{1}{2}\left(T_M+(T_M)^\dagger\right)
    \=\frac{1}{2}\left(E_{\gamma_M}+(E_{\gamma_M})^\dagger\right)\;,
\end{equation}
where $\gamma_M$ are the $2\,\nv$ roots of $\mathfrak{g}_3$, such that $\gamma_M[H_0]=1/2$. Now, the $\texttt{p}$  generators $J_\ell$ are defined by a maximal set $\{\gamma_\ell\}$ of mutually orthogonal roots
among the $\gamma_M$ roots:
\begingroup
\abovedisplayskip=10pt
\abovedisplayshortskip=10pt
\belowdisplayskip=25pt
\belowdisplayshortskip=25pt
\begin{flalign}
\qqquad
\gamma_{\ell_1}\cdot\gamma_{\ell_2}~\propto~\delta_{\ell_1 \ell_2}\;:
\qquad\;
J_\ell\=\frac{1}{2}\left(E_{\gamma_\ell}+(E_{\gamma_\ell})^\dagger\right)\;.&&
\end{flalign}
\endgroup

\subsubsection{Symplectic frames and normal forms}
For all the symmetric models mentioned above, the normal form of the electric and magnetic charges with respect to the group $\Hc$ is contained in the STU truncation. For this reason, it is useful to study the relevant STU symplectic frames .\par

\paragraph{STU model.} The so-called STU model is a \,$\N=2$ supergravity model coupled to three vector multiplets, whose three complex scalars \{S,T,U\} span a special K\"ahler manifold of the form
\begin{equation}
\Ms^{(4)}_\text{scal}\=\frac{\G[4]}{\HH[4]}
    \=\frac{\SL(2,\mathbb{R})^3}{\SO(2)^3}\;.
\end{equation}
Upon time-like reduction to $D=3$, the scalar manifold is enlarged to
\begin{equation}
\Ms^{(3)}_\text{scal}\=\frac{\G}{\HH[3]}\=\frac{\SO(4,4)}{\SO(2,2)\times\SO(2,2)}\;.
\end{equation}
If the STU model originates from Kaluza--Klein reduction from $D=5$ dimensions, the resulting symplectic frame corresponds to a particular ordering of the roots $\gamma_M$ (${\footnotesize M}\;=1,\dotsc,8$):
\begin{equation}
\Gamma_M\=\Cc_{MN}\,\Gamma^N=~(q_\Lambda,\,-p^\Lambda)
\;\quad\longleftrightarrow\quad\;
\{\gamma_M\}\;\;.
\label{1frame}
\end{equation}
Each root $\gamma_M$ can be represented by its component vector $\vec{\gamma}_M$ in a Cartan subalgebra of $\mathfrak{so}(4,4)$. The first component of this vector is the grading $\gamma_M[H_0]$ with respect to the $\text{O}(1,1)$ generator $H_0$ in the Ehlers group $\SL(2,\mathbb{R})_E$, while the other entries are the components $\gamma_M[H_{\alpha_i}]/2$ with respect to the Cartan generators $H_{\alpha_i}$ of $\G[4]$:
\begin{equation}
\vec{\gamma}_M\=
    \left(\gamma_M[H_0]\,,\;\frac{\gamma_M[H_{\alpha_1}]}{2}\,,\;\frac{\gamma_M[H_{\alpha_2}]}{2}\,,\;\frac{\gamma_M[H_{\alpha_3}]}{2}\right)\;,
\end{equation}
and we find for the STU model
\begin{subeqs}\label{gammaord} 
\begin{align}
\{\vec{\gamma}_a\}&= \left\{
    \left(\frac{1}{2},\,-\frac{1}{2},\,-\frac{1}{2},\,-\frac{1}{2}\right),\,
    \left(\frac{1}{2},\,\frac{1}{2},\,-\frac{1}{2},\,-\frac{1}{2}\right),\,
    \left(\frac{1}{2},\,-\frac{1}{2},\,\frac{1}{2},\,-\frac{1}{2}\right),\,
    \left(\frac{1}{2},-\frac{1}{2},-\frac{1}{2},\frac{1}{2}\right)\right\}\;,
\\[1.2ex]
\{\vec{\gamma}_{a+4}\}&= \left\{
    \left(\frac{1}{2},\frac{1}{2},\frac{1}{2},\frac{1}{2}\right),\,
    \left(\frac{1}{2},\,-\frac{1}{2},\,\frac{1}{2},\,\frac{1}{2}\right),\,
    \left(\frac{1}{2},\,\frac{1}{2},\,-\frac{1}{2},\,\frac{1}{2}\right),\,
    \left(\frac{1}{2},\,\frac{1}{2},\,\frac{1}{2},\,-\frac{1}{2}\right)\right\}\;,
\end{align}
\end{subeqs}
where \,$a=1,\dotsc,4$. We see that there are two maximal sets of $\texttt{p}=4$ mutually orthogonal roots, corresponding to two different normal forms of the charge vector. In particular we have in the first case
\begin{flalign}
\qqquad
\begin{split}
\{\gamma_{\ell}\}&\=\{\gamma_1,\,\gamma_6,\,\gamma_7,\,\gamma_8\}\;,
\\[\jot]
\Gamma_M&\=(0,\,p_1,\,p_2,\,p_3,\,q_0,\,0,\,0,\,0)~\equiv~\{q_0,\,p^i\}\;,
\qqquad (i=1,\,2,\,3)\;,
\label{normform1}
\end{split}
&
\end{flalign}
while in the other case
\begin{flalign}
\qqquad
\begin{split}
\{\gamma_{\ell'}\}&\=\{\gamma_2,\,\gamma_3,\,\gamma_4,\,\gamma_5\}\;,
\\[\jot]
\Gamma_M&\=(p_0,\,0,\,0,\,0,\,0,\,q_1,\,q_2,\,q_3)~\equiv~\{p^0,\,q_i\}\;,
\qqquad (i=1,\,2,\,3)\;.
\label{normform2}
\end{split}
&
\end{flalign}
If we embed the STU model in toroidally compactified Heterotic theory \cite{Cvetic:1995kv}, one of the $\SL(2,\mathbb{R})$ factors in $\G[4]$ has a non-perturbative (i.e.\ not block-diagonal) duality action in the $\Rs_\text{s}={\bf (2,\,2,\,2)}$, while the remaining two factors have a block diagonal symplectic representation.
The corresponding symplectic frame is characterized by the following order of the roots $\gamma_M$:%
\footnote{%
this ordering is related to the property that, in this frame, the Cartan generator of the non-perturbative $\SL(2,\mathbb{R})$ degenerate over the electric (and thus over the magnetic) charges
}
\begin{equation}
\Gamma_M^{\prime}  \qlq
\{\gamma_1,\gamma_6,\gamma_3,\gamma_4,\gamma_5,\gamma_2,\gamma_7,\gamma_8\}\;.
\end{equation}
The two normal forms of the charge vector, being identified by the same sets of roots $\{\gamma_\ell\}$ and $\{\gamma_{\ell'}\}$, now correspond to two electric and two magnetic charges, $\{p^{\prime 2},\,p^{\prime 3},\,q'_0,\,q'_1\}$ and $\{p^{\prime 0},\,p^{\prime 1},\,q'_2,\,q'_3\}$. \par\smallskip
Finally, one can consider the frame in which the generators of $\G[4]$ can be chosen to be represented by symplectic matrices which are either block diagonal or completely block-off-diagonal (i.e. having entries only in the off-diagonal blocks). This is the frame originating from direct truncation of the $\mathcal{N}=8$ theory in which the $\SL(8,\mathbb{R})$ subgroup of $\text{E}_{7(7)}$ has a block-diagonal embedding in $\Sp(56,\mathbb{R})$.  It corresponds to the following order of the roots $\gamma_M$:
\begin{equation}
\Gamma_M^{\prime\prime}
\quad\leftrightarrow\quad \{\gamma_5,\gamma_2,\gamma_3,\gamma_4,\gamma_1,\gamma_6,\gamma_7,\gamma_8\}\;.
\end{equation}
The two normal forms of the charge vector now correspond to either all electric or all magnetic charges: $\{p^{\prime\prime\,\Lambda}\}$ and $\{q''_\Lambda\}$. \par\smallskip
In all these cases, the MASAs of $\Span\left(\Jb_M\right)$ are always defined by the same sets of generators
\begin{equation}
\{J_\ell\}_{{}_{\ell\,=\,1,6,7,8}}\;\;; \qquad
\{J_{\ell'}\}_{{}_{\ell'\,=\,2,3,4,5}}\;\;.
\end{equation}
\par\medskip

\subsubsection{From Kerr to extremal solutions}
Now we summarize the procedure to connect the Kerr orbit to orbits of extremal under-rotating and static solutions. First, we transform the Kerr solution by means of an Harrison transformation generated by the chosen MASA $\Jb^{(N)}$ of $\Span\left(\Jb_M\right)$:
\begin{equation}
\mathcal{H}\in \exp\left(\Jb^{(N)}\right)\;:\qquad
\mathcal{H}\=
    \begin{cases}
    \;\exp\Big(\sum_{\ell}\log(\beta_\ell)\,J_\ell\Big) &
    \{q_0,\,p^i\} \text{ - case}\;; \cr \;\exp\Big(\sum_{\ell'}\log(\beta_{\ell'})\,J_{\ell'}\Big) &
    \{p^0,\,q_i\} \text{ - case}\;,
    \end{cases}
\end{equation}
where $\ell=\{1,\,6,\,7,\,8\}$\, and \,$\ell'=\{2,\,3,\,4,\,5\}$. \par\smallskip
The matrices $\Q$, $\Qpsi$ transform according to eq.\ \eqref{gQg}:
\begin{alignat}{2} 
\Q\;\;\stackrel{\mathcal{H}} \longrightarrow
&\;\; &\Q'&=(\mathcal{H}^{-1})^\dagger\;\Q\;\mathcal{H}^\dagger\;,
\\[\jot]
\Qpsi\;\;\stackrel{\mathcal{H}}\longrightarrow
&\;\;\; &\mathcal{Q^{\,\prime}_\psi}&=(\mathcal{H}^{-1})^\dagger\;\Qpsi\;\mathcal{H}^\dagger\;.
\end{alignat}
Next we perform the rescalings:
\begin{equation}
\beta_\ell\,\rightarrow\,m^{\sigma_\ell}\,\beta_\ell\;,
\qquad\alpha\,\rightarrow\,m\,\Omega\;,
\qqquad (\ell=1,\,6,\,7,\,8)\;,
\end{equation}
or, in the other case,
\begin{equation}
\beta_{\ell'}\,\rightarrow\,m^{\sigma_{\ell'}}\,\beta_{\ell'}\;,
\qquad\alpha\,\rightarrow\,m\,\Omega\;,
\qqquad (\ell'=2,\,3,\,4,\,5)\;,
\end{equation}
where $\sigma_\ell,\,\sigma_{\ell'}=\;\pm1$. Then, we send $m$ to zero.\par
The above limits correspond to an In\"on\"u-Wigner contraction of $\Q'$ and $\mathcal{Q^{\,\prime}_\psi}$, which become nilpotent matrices $\Q^{(0)}$, $\QpsiZ$ with a \emph{different degree of nilpotency}. This means they belong to different $\HHstar$-orbits: $\Q^{(0)}$ with nilpotency degree three, while $\QpsiZ$ either vanishes or has degree two. This explains why, in the $m\rightarrow 0$ limit, the ratio on the right hand side of eq.\ \eqref{regbound} goes to zero, since the numerator $\Tr(\Qpsi^2)$ vanishes faster than the denominator $\Tr(\Q^2)$.

\paragraph{Physical quantities in the extremal limit.}%
The charge vector $\Gamma_M$ of the resulting solution, in the two cases, has 4 non-vanishing charges corresponding to the chosen normal form, i.e.\ $\{q_0,\,p^i\}$ or $\{p^0,\,q_i\}$. Depending on the choice of the gradings ($\sigma_\ell$ or $\sigma_{\ell'}$), the charge vector can belong to any of the $\G[4]$-orbits of regular solutions, characterized in terms of the extremal $\G[4]$-quartic invariant $I_4$ of the representation $\Rs_\text{s}$ as follows%
\footnote{%
see Appendix \ref{stumodel} for the explicit form of $I_4(e,m)$ in the STU model
}
\cite{Bellucci:2006xz}\,:
\begin{equation}
\begin{split}
\text{BPS} & :\quad\; I_4>0 \quad\;\;
\text{$\mathbb{Z}_3$-symmetry on the $p^i$ and the $q_i$}\;,
\\[\jot]
\text{non-BPS}_1 & :\quad\; I_4>0 \quad\;\;
\text{no $\mathbb{Z}_3$-symmetry}\;,
\\[\jot]
\text{non-BPS}_2&:\quad\; I_4<0 \;.
\end{split}
\end{equation}
For those choices of the gradings yielding $I_4>0$, we find both the BPS and a non-BPS solution and the resulting angular momentum is zero ($\QpsiZ=0$) and thus the black hole solution is an extremal-static. Only in the cases for which $I_4<0$ we find a rotating black hole, which is the known extremal under-rotating solution of \cite{Rasheed:1995zv,Larsen:1999pp,Goldstein:2008fq,Bena:2009ev,Bena:2009en} :
\begin{equation}
\begin{split}
I_4>0\;&:\qquad \QpsiZ=0  \;\;\rightarrow\;\;  \Jextr=0
\qquad \text{(BPS and non-BPS)}\;,
\\[1.2ex]
I_4<0\;&:\qquad \QpsiZ\neq0  \;\;\rightarrow\;\;  \Jextr\neq0
\qquad \text{(non-BPS)}\;.
\end{split}
\end{equation}
We find, in general, that the extremal solutions obtained in this way have an angular momentum given by
\begin{equation}
\Jextr\=\frac{\Omega}{4}\;(1-\varepsilon)\;\sqrt{|I_4|}\;.
\label{MI4}
\end{equation}
where $I_4=\varepsilon\,|I_4| (\varepsilon=\pm1)$.
%
In Subsect.\ \ref{subsubsec:angmomdual} we proved the invariance of $\J$ under  $\G[4]$-transformations for a generic solution. Now, the formula \eqref{MI4} makes the invariance manifest, since both \;$I_4(e,m)$\; and \;$\Omega=\J^{\text{Kerr}}/m^2$\; are $\G[4]$-invariants, being the latter related to the original Kerr solution.\par
Actually, one cannot see the dependence of the various quantities on the scalar fields $\Phi_0^I$, and, in particular, on the four-dimensional ones at radial infinity, since these were fixed to zero. However, having proven that $\J$ is a $\G[4]$-invariant function of $\phi_0^s$ and $\Gamma$ and also that it is already an invariant function of the electric-magnetic charges alone, we conclude that for the extremal under-rotating solutions $\Jextr$ only depends on the extremal charges
\begin{equation}
\Gamma_M^{\text{(ex)}}\=\left(p^{(\text{ex})\Lambda},\;q^{(\text{ex})}_\Lambda\right)\;.
\end{equation}
The entropy of the solution, related to the horizon area and expressed in formula \eqref{entropy2}, has the following form in the extremal limit:
\begin{equation}
\begin{split}
\Sextr&\=\pi\,\lim_{m\rightarrow 0}\,\alpha\,|\omega_\textsc{h}|
    \=\pi\,\lim_{m\rightarrow 0}\,m\,\Omega\,|\omega_\textsc{h}|
    \=\pi\,\sqrt{|I_4|-4 \left(\Jextr\right)^2}\=
\\
    &\=\pi\,\sqrt{|I_4|\,\left(1-\frac{\Omega^2}{2}(1-\varepsilon)\right)}\;.
\label{SI4}
\end{split}
\end{equation}
The above expression, obtained by using \eqref{MI4}, makes it manifest that $\Sextr$, as well as the whole near horizon geometry, is $\G[4]$-invariant as $\Jextr$ is. In the rotating extremal case ($\varepsilon=-1$) we further need to impose \,$\Omega<1$\, in order for the solution to be well-behaved.

\paragraph{Attractor mechanism.}
We observe that, \emph{before} the above extremal $m\rightarrow 0$ limit is performed, the expression of  $\Sentr$ is not manifestly $\G[4]$-invariant. This can be explained by the fact that we had generally made a $\G[4]$ ``gauge'' choice, corresponding to fixing at the origin of the moduli space the four dimensional scalar fields at infinity. This has broken the manifest $\G[4]$-invariance to $\HH[4]$.\par
In the extremal under-rotating and static cases, the attractor mechanism is at work \cite{Ferrara:1996dd,Ferrara:1995ih,Gibbons:1996af,Goldstein:2005hq,Tripathy:2005qp,Kallosh:2005ax}. As a consequence of this, the near-horizon geometry becomes independent of the values of the scalar fields at radial infinity (fixed to the origin) and only depends on the extremal quantized charges $\Gamma^{\text{(ex)}}$.\par
In the non-extremal case, $\cex^2>0$, the above discussion do not apply and the near horizon geometry, as well as the entropy, depends on the values of the four dimensional scalar fields at infinity $\phi^s_0$. We can then argue that $\Sentr=\Sentr(e,m,\phi_0^s)$ is still invariant under $\G[4]$, provided we transform both $\Gamma^M$ and $\phi_0^s$ simultaneously, just as it was proven for the angular momentum in \eqref{Jinvar}. In other words, within our choice of scalar boundary conditions, $\Sentr$ is  expressed in terms of $\HH[4]$-invariants and, in the extremal limit, such expression should reduce to the only scalar-independent $\HH[4]$-invariant, namely to the above \eqref{SI4}.

\newpage

\section{Gauged supergravities} \label{sec:gaugsugra}
We mentioned in Sect.\ \ref{sec:intro} how superstring/M-theory can be a promising candidate for a fundamental quantum theory of gravity. Since these theories are defined in dimensions $D>4$ and since we live in a four dimensional universe, a fundamental requirement for any predictable model is the presence of a mechanism of dimensional reduction from ten or eleven dimensions to four. The simplest mechanism of this type is ordinary Kaluza-Klein compactification of string/M-theory on solutions with geometry of the form
\begin{equation}
\mathbb{M}_4^{{}^{(1,3)}}\times\;\Ms\;,
\end{equation}
where $\mathbb{M}_4^{{}^{(1,3)}}$ is the maximally symmetric four dimensional space-time with Lorentzian signature and $\Ms$ is a compact internal manifold. We have also stated that the low-energy dynamics of superstring/M-theory, compactified on a Ricci-flat manifold $\Ms$,  can be well described by a four dimensional (ungauged) supergravity theory, which involves the massless modes on $\mathbb{M}_4^{{}^{(1,3)}}$. \par\smallskip
From a phenomenological point of view, extended supergravity models on four dimensional Minkowski vacua, obtained through ordinary Kaluza-Klein reduction on a Ricci-flat manifold, are not consistent with experimental observations.
These models typically contain a certain number of massless scalar fields -- which are associated with the geometry of the internal manifold $\Ms$ -- whose vacuum expectation values (vevs) define a continuum of degenerate vacua. In fact, there is no scalar potential that encodes any scalars dynamics, so we can not avoid the degeneracy. This turns into an intrinsic lack of predictiveness for the model, in addition to a field-content of the theory which comprises massless scalar fields that are not observed in our universe.
Another feature of these models is the absence of a local internal gauge-symmetry, that is the vector fields are not minimally coupled to any other field in the theory. This means that no matter field is charged under a gauge group, hence the name \emph{ungauged supergravity}.\par
Realistic quantum field theory models in four dimensions need the presence of a non trivial scalar potential, which could solve the problem of the moduli degeneracy and, on the other hand, select a consistent vacuum state for our universe.

\paragraph{Scalar potential.}
We have seen in Sect.\ \ref{sec:ungsugra} the structure of ungauged supergravity theories. In the latter class of models, the presence of a scalar potential in the bosonic Lagrangian \eqref{Lagrscal} is allowed only for the minimal $\N=1$ case and is called \emph{F-term potential}.
%
A realistic and phenomenologically interesting framework requires the presence of a non-trivial scalar potential encoding scalar dynamics, which could lift the moduli-degeneracy and define a suitable vacuum state for our universe featuring desirable physical properties (for instance, mass terms for the scalars and the presence of some effective cosmological constant). Moreover, a scalar potential is an essential ingredient for having spontaneous supersymmetry breaking scenarios in supergravity theories, depending on the choice of the internal gauge symmetry \cite{Nilles:1983ge,Wess:1992cp}.\par
In extended supergravities, the only known mechanism to introduce a non-trivial scalar potential without explicitly breaking supersymmetry is the so-called gauging procedure \cite{deWit:1982ig,deWit:2002vt,deWit:2003hq,deWit:2005hv,deWit:2005ub,Hull:1984vg,Hull:1984qz,Gallerati:2016oyo,Trigiante:2016mnt}. The latter consists in promoting a suitable global symmetry (sub)group to a local symmetry to be gauged by the vector fields of the theory. It could be possible that the gauge group is non-abelian and part of the scalars may be charged under the gauge group; this is achieved introducing proper covariant derivatives in \eqref{Lagrscal} and replacing \eqref{Lagrvect} by the corresponding Yang-Mills terms. Theories in which the scalar potential is generically non-vanishing, are referred to as \emph{gauged supergravities}. In a gauged theory, vectors are minimally coupled to the other fields and the symplectic frame becomes physically relevant, leading to different vacuum-structures defined by the scalar potential.\par
In the gauged theory, the Lagrangian is modified with additional terms: besides the minimal couplings of the gauge fields to the charged ones, some extra contributions come from the requirement of supersymmetry of the action. This determines the presence of additional terms in the supersymmetry transformation rules of the gravitino and fermion fields, together with the introduction of gravitino and fermion mass contributions, as well as the scalar potential in the Lagrangian.

\paragraph{Ungauged vs.\ gauged models.}
We have already mentioned the fact that supergravity can be seen as a consistent and well established low-energy approximation of some fundamental superstring theory, since massless sectors of superstring models can be described by ungauged supergravities.
Global symmetries of the lower dimensional effective supergravity play a relevant role in understanding non–perturbative aspects of superstring: behind the concept of string duality, there is the idea that superstring models (or M-theories) are just different realizations of a fundamental quantum theory, the correspondences among them called \emph{dualities}. After standard dimensional reduction to $D=4$ Minkowski space–time, these dualities are conjectured to be encoded in the global symmetries of the resulting ungauged supergravity \cite{Hull:1994ys}. This means that it could be possible to obtain information about string dualities and non-perturbative string behaviour by studying ungauged SUGRA models.\par
%
Gauged supergravity models satisfy the requirement of gauge and supersymmetry invariance, and can be derived from ungauged models (having same field content and amount of SUSY) through the previously mentioned gauging procedure. The latter can be seen as a \emph{deformation} of an ungauged theory and consists in promoting some suitable subgroup $\Gg$ of the global symmetry group $G$ of the Lagrangian to \emph{local} symmetry. This can be achieved by introducing minimal couplings for the vector fields, mass deformation terms and the scalar potential itself. The coupling of the (formerly abelian) vector fields to the new local gauge group provides matter fields that are charged under the new local gauge symmetry.\par
The gauging procedure, however, will in general break the global symmetry group of the ungauged theory: the latter, acting as a generalized electric-magnetic duality, is broken by the introduced minimal couplings, which only involve the electric vector fields. As a consequence of this, in a gauged supergravity we loose track of the string/M-theory dualities, which were described by global symmetries of the original ungauged theory.
The drawback can be avoided using the \emph{embedding tensor formulation} of the gauging procedure \cite{Cordaro:1998tx,Nicolai:2000sc,deWit:2002vt,deWit:2005ub,deWit:2007mt,Gallerati:2016oyo,Trigiante:2016mnt} in which all deformations involved by the gauging is encoded in a single object (the embedding tensor) which is itself covariant with respect to the global symmetries of the ungauged model. This allows to formally restore symmetries at the level of the gauged field equations and Bianchi identities, provided the embedding tensor is transformed together with the other fields: the global symmetries of the ungauged theory now act as equivalences between gauged models. Since the embedding tensor encodes all background quantities in the compactification describing the fluxes and the structure of the internal manifold, the action of the global symmetry group on it allows to systematically study the effect of dualities on flux compactifications.\par
When originating from superstring/M-theory compactifications, gauged SUGRAs give the possibility to investigate the perturbative low-energy dynamics of the system, since they describe the full non-linear dynamics of the low-lying modes. In general, there is a correspondence between vacua of the microscopic fundamental theory%
\footnote{%
if already formulated, since there are several gauged SUGRAs whose superstring/M-theory origin is unknown}%
 and vacua of the low-energy supergravity.\par
%

\subsection{Gauging of a theory}
Given the Lagrangian symmetry group $\Gel$ (see Subsect.\ \ref{subsubsec:symplframes}),
the gauging procedure consists in promoting a suitable global symmetry subgroup \,$\Gg\subset\Gel$\, to a local symmetry gauged by the vector fields of the theory, implying the preliminary condition
\begin{equation}
\dim(\Gg) ~\le~ \nv\;.
\label{preliminary}
\end{equation}
As already pointed out in Sect.\ \ref{sec:ungsugra}, different symplectic frames correspond to ungauged Lagrangians with different global symmetry groups $\Gel$ and thus to different choices for the possible gauge groups.\par
To become a viable gauge group, the global symmetry subgroup $\Gg$ must admit a subset $\{A^{\hat{\Lambda}}\}$ of the vector fields%
\footnote{%
hatted-indices are those pertaining to the symplectic frame in which the Lagrangian is defined
}
which transform under the co-adjoint representation of the duality action of $\Gg$. These fields will become the \emph{gauge vectors} associated with the generators $X_{\hat\Lambda}$ of the subgroup $\Gg$ itself.
We denote as \emph{electric frame} the symplectic frame defined by our ungauged Lagrangian (labeled by hatted indices).\par
Once the gauge group is chosen within $\Gel$, its action on the various fields is fixed, being defined by the action of $\Gg$ as a global symmetry group of the ungauged theory (i.e.\ duality action on the vector field strengths, non-linear action on the scalars and indirect action through $H$-compensators on the fermionic fields). The fields of the theory are thus automatically associated with representations of $\Gg$.\par

\subsubsection{Gauge algebra. Curvature. Covariant derivatives}
After the initial choice of $\Gg$ in $\Gel$, one has to pursue the construction of the non-abelian gauge theory. First of all, we have to introduce the gauge-connections, gauge-curvatures (i.e.\ non-abelian field strengths) and covariant derivatives. We will also need to introduce an extra topological term needed for the gauging of the Peccei-Quinn transformations \eqref{deltaLC}. This will give us the gauged Lagrangian $\LgaugO[0]$ with manifest local $\Gg$-invariance. Consistency of the construction will imply constraints on the possible choices of $\Gg$ inside $G$. The minimal couplings will however break supersymmetry: the second part of the gauging procedure will consist in further deforming the constructed $\LgaugO[0]$ in order to restore the original supersymmetry of the ungauged theory, preserving, at the same time, local $\Gg$-invariance.
%

\paragraph{Gauge algebra.}
Let us introduce the gauge connection:
\begin{equation}
\Omega_{\bf g}\=\Omega_{{\bf g}\mu}\,dx^\mu\;; \qqquad
\Omega_{{\bf g}\mu}\equiv g\,A^{\hat{\Lambda}}_\mu\,X_{\hat{\Lambda}}\;,
\label{gconnection}
\end{equation}
where $g$ is the coupling constant. The gauge-algebra relations can be written
\begin{equation}
\left[X_{\hat{\Lambda}},\,X_{\hat{\Sigma}}\right]
    \=f_{{\hat{\Lambda}}{\hat{\Sigma}}}{}^{\hat{\Gamma}}\,X_{\hat{\Gamma}}\;,
\label{gaugealg}
\end{equation}
characterized by the structure constants
$f_{{\hat{\Lambda}}{\hat{\Sigma}}}{}^{\hat{\Gamma}}$. The above closure condition results in a constraint on $X_{\hat{\Lambda}}$; the structure constants are fixed in terms of the action of the gauge generators on the vector fields, as global symmetry generators of the original ungauged theory.\par
Since $\Gg\in\Gel$, its electric-magnetic duality-action as a global symmetry group has the form \eqref{ge}. This action of the infinitesimal generators $X_{\hat{\Lambda}}$ on the vector field strengths and their duals is then represented by a symplectic matrix of the form
\begin{equation}
\left(X_{\hat{\Lambda}}\right)^{\hat{M}}{\!}_{\hat{N}}\=
    \left(\begin{matrix}
        X_{\hat{\Lambda}}{}^{{\hat{\Lambda}}}{}_{{\hat{\Sigma}}} & \Zero
        \cr
        X_{{\hat{\Lambda}}\,{\hat{\Gamma}}{\hat{\Sigma}}} & X_{{\hat{\Lambda}}\,{\hat{\Gamma}}}{}^{\hat{\Delta}}
    \end{matrix}\right) \;.
\label{xsymp}
\end{equation}
%
Note that we do not identify the generator $X_{{\hat{\Lambda}}}$ with the symplectic matrix defining its electric-magnetic duality action%
\footnote{%
we emphasize also that, as pointed out in Subsect.\ \ref{subsubsec:symplframes}, there are isometries in $\N=2$ models which do not have duality action, namely for which the matrix in \eqref{xsymp} is null (see eq.\ \eqref{qisom})
}.\par
If we consider the variation $\delta\FF^{M}$ of the field strengths under an infinitesimal duality transformation (whose action is described by \eqref{xsymp}), the imposed symplectic condition on the matrix $X_{\hat{\Lambda}}$ and the prescription that $A^{\hat{\Lambda}}_\mu$ transforms in the co-adjoint representation of the gauge group (${\bf \nv}=\text{coadj}(\Gg)$), we obtained that the structure constants of the gauge group in \eqref{gaugealg} can be identified with the diagonal blocks of the symplectic matrices $X_{\hat{\Lambda}}$:
\begin{equation}
f_{{\hat{\Gamma}}{\hat{\Sigma}}}{}^{\hat{\Lambda}}=-X_{{\hat{\Gamma}}{\hat{\Sigma}}}{}^{\hat{\Lambda}}\;,
\label{idenfx}
\end{equation}
so that the closure condition reads
\begin{equation}
\left[X_{\hat{\Lambda}},\,X_{\hat{\Sigma}}\right]~=\,-X_{{\hat{\Lambda}}{\hat{\Sigma}}}{}^{\hat{\Gamma}}\,X_{\hat{\Gamma}}\;,
\label{gaugealg2}
\end{equation}
and results in a quadratic constraint on the tensor $\left(X_{\hat{\Lambda}}\right)^{\hat{M}}{}_{\hat{N}}$. The identification \eqref{idenfx} also implies
\begin{equation}
X_{\left({\hat{\Gamma}}{\hat{\Sigma}}\right)}{}^{\hat{\Lambda}}\=0\;.
\label{lin1}
\end{equation}
The closure condition \eqref{gaugealg2} can be thus interpreted as an invariance of the gauge generators $X_{{\hat{\Lambda}}}$ under the action of $\Gg$ itself:
\begin{equation}
\delta_{{\hat{\Lambda}}}X_{{\hat{\Sigma}}}~\equiv~ \left[X_{\hat{\Lambda}},\,X_{{\hat{\Sigma}}}\right]+X_{{\hat{\Lambda}}{\hat{\Sigma}}}{}^{\hat{\Gamma}}\,X_{\hat{\Gamma}}\=0\;.
\end{equation}
\par

\paragraph{Gauge curvature and covariant derivatives.}
Once defined the gauge connection \eqref{gconnection}, we can also write its transformation properties under a local $\Gg$-transformation
${\bf g}(x)\in\Gg$:
\begin{equation}
\Omega_{\bf g} \;\;\,\longrightarrow\quad
\Omega_{\bf g}'\,=~
    {\bf g}\;\Omega_{\bf g}\,{\bf g}^{-1}+d {\bf g}\,{\bf g}^{-1}\,=\,
    g\,A^{\prime\,\hat{\Lambda}}\,X_{\hat{\Lambda}}\;.
\label{Omtrasg}
\end{equation}
Under an infinitesimal transformation of the form \;${\bf g}(x)\equiv \Id+g\,\zeta^{\hat{\Lambda}}(x)\,X_{\hat{\Lambda}}$,\, eq.\ \eqref{Omtrasg} implies for the gauge vectors the transformation property:
\begin{equation}
\delta A^{\hat{\Lambda}}_\mu\=D_\mu\zeta^{\hat{\Lambda}} ~\equiv~
    \partial_\mu\zeta^{\hat{\Lambda}}+g\,A_\mu^{\hat{\Sigma}}X_{\hat{\Sigma}\hat{\Gamma}}{}^{\hat{\Lambda}}\,\zeta^{\hat{\Gamma}}\;,
\end{equation}
where we have introduced the $\Gg$-covariant derivative $D_\mu\zeta^{\hat{\Lambda}}$ of the gauge parameter.\par
We then define the gauge curvature%
\footnote{%
\sloppy
here we use the following convention for the definition of the components of a form: ${\omega_{(p)}=\frac{1}{p!}\,\omega_{\mu_1\dots\mu_p}\,dx^{\mu_1}\wedge \dots dx^{\mu_p}}$
}
\begin{equation}
\mathscr{F}\=F^{\hat{\Lambda}}\,X_{\hat{\Lambda}}
    \=\frac{1}{2}\;F^{\hat{\Lambda}}_{\mu\nu}\,dx^\mu\wedge
       dx^\nu\,X_{\hat{\Lambda}}
    ~\equiv~ \frac{1}{g}\left(d\Omega_{\bf g}-\Omega_{\bf g}\wedge \Omega_{\bf g}\right)\;,
\label{calF}
\end{equation}
which, in components, reads:
\begin{equation}
F_{\mu\nu}^{{\hat{\Lambda}}} \=
    \partial_\mu A^{\hat{\Lambda}}_\nu-\partial_\nu A^{\hat{\Lambda}}_\mu +
    g\,X_{{\hat{\Gamma}}{\hat{\Sigma}}}{}^{\hat{\Lambda}}\,A^{\hat{\Gamma}}_\mu\,A^{\hat{\Sigma}}_\nu\;.
\label{defF}
\end{equation}
The gauge curvature transforms covariantly under a transformation ${\bf g}(x)\in \Gg$:
\begin{equation}
\mathscr{F} \;\rightarrow\;
\mathscr{F}'={\bf g}\,\mathscr{F}\,{\bf g}^{-1}\;,
\label{gaugecovF}
\end{equation}
and satisfies the Bianchi identity:
\begin{equation}
D\mathscr{F}\equiv
    d\mathscr{F}-\Omega_{\bf g}\wedge \mathscr{F}
    +\mathscr{F}\wedge \Omega_{\bf g}=0
\;\qRq\;
D F^{{\hat{\Lambda}}}\equiv
    dF^{{\hat{\Lambda}}}+g\,X_{{\hat{\Sigma}}{\hat{\Gamma}}}{}^{{\hat{\Lambda}}}A^{{\hat{\Sigma}}}\wedge F^{{\hat{\Lambda}}}=0\;,
\end{equation}
where we have denoted by $D F^{{\hat{\Lambda}}}$ the $\Gg$-covariant derivative acting on $F^{{\hat{\Lambda}}}$.
In the ungauged Lagrangian we will have to replace the abelian field strengths by the new $\Gg$-covariant ones:
\begin{equation}
\partial_\mu A^{\hat{\Lambda}}_\nu-\partial_\nu A^{\hat{\Lambda}}_\mu
\qlq
\partial_\mu A^{\hat{\Lambda}}_\nu-\partial_\nu A^{\hat{\Lambda}}_\mu +
    g\,X_{{\hat{\Gamma}}{\hat{\Sigma}}}{}^{\hat{\Lambda}}\,A^{\hat{\Gamma}}_\mu\,A^{\hat{\Sigma}}_\nu\;.
\label{replaceF}
\end{equation}
%
In order to obtain local invariance of the Lagrangian under $\Gg$, we replace standard derivatives by covariant ones:
\begin{equation}
\dm \qlq
D_\mu\=\dm-g\,A_\mu^{\hat{\Lambda}}\,X_{\hat{\Lambda}}\;,
\label{covder}
\end{equation}
%
the covariant derivatives satisfying the relation
\begin{equation}
D^2=-g\,\mathscr{F}=-g\,F^{\hat{\Lambda}}\,X_{\hat{\Lambda}}
\;\qRq\;
\left[D_\mu,\,D_\nu\right]=-g\,F^{\hat{\Lambda}}_{\mu\nu}\,X_{\hat{\Lambda}}\;.
\label{D2F}
\end{equation}
The covariant derivatives of the scalar fields $\phi^s$ are written using the Killing vectors $k_{\hat{\Lambda}}$ associated with the action (isometry) of the gauge generator $X_{\hat{\Lambda}}$:
\begin{equation}
\dm\phi^s \qlq
D_\mu\phi^s\=\dm\phi^s-g\,A^{\hat{\Lambda}}_\mu\,k^s_{\hat{\Lambda}}(\phi)\;,
\label{covderphi}
\end{equation}
The replacements \eqref{covder}, \eqref{covderphi} amount to the \emph{introduction of minimal couplings} for the vector fields.\par
For homogeneous scalar manifolds, the left-invariant 1-form $\Omega$ \eqref{omegapro} is redefined (pulled-back on space-time) in terms of a gauged one, obtained by covariantizing the derivative on the coset representative:
\begin{equation}
\Omega_\mu= \LL^{-1}\partial_\mu \LL
\qlq
\hat{\Omega}_\mu\equiv \LL ^{-1}D_\mu \LL\=
    \LL^{-1}\left(\partial_\mu-g\,A^{\hat{\Lambda}}_\mu\,X_{\hat{\Lambda}}\right)\LL \=\hat{\pp}_\mu+\hat{\ww}_\mu\;,
\label{hatOm}
\end{equation}
where the space-time dependence of the coset representative is defined by the scalar fields $\phi^s(x)$.
%

\paragraph{Vielbein and fermions.}
The \emph{gauged} vielbein and connection are related to the ungauged ones as follows:
\begin{equation}
\hat{\pp}_\mu={\pp}_\mu-g\,A^{\hat{\Lambda}}_\mu\,{\pp}_{\hat{\Lambda}}\;;
\qquad
\hat{\ww}_\mu=
    {\ww}_\mu-g\,A^{\hat{\Lambda}}_\mu\,{\ww}_{\hat{\Lambda}}\;,
\label{gaugedPW}
\end{equation}
the matrices ${\pp}_{\hat{\Lambda}},\,{\ww}_{\hat{\Lambda}}$ being the projections onto $\mathfrak{K}$ and $\mathfrak{H}$, respectively, of $\LL^{-1}X_{\hat{\Lambda}}\LL$:
\begin{equation}
\pp_{\hat{\Lambda}}\equiv \left.\LL^{-1}X_{\hat{\Lambda}}\LL\right\vert_{\mathfrak{K}}\;;
\qquad
{\ww}_{\hat{\Lambda}}\equiv \left.\LL^{-1}X_{\hat{\Lambda}}\LL\right\vert_{\mathfrak{H}}\;.
\label{wPproj}
\end{equation}
For non-homogeneous scalar manifolds we cannot use the construction \eqref{hatOm} (based on $\LL$), but we can still define the gauged vielbein $\hat{\pp}_\mu$ and $H$-connection $\hat{\ww}_\mu$  in terms of the Killing vectors.\par\smallskip
%
Consider now a local $\Gg$-transformation ${\bf g}(x)$ whose effect on the scalars is described by eq.\ \eqref{gLh}.
Since $D$ is $G$-covariant and using \eqref{hatOm} we find:
\begin{equation}
\hat{\Omega}_\mu(g\star \phi)=h\,\hat{\Omega}_\mu(\phi)\,h^{-1}+hdh^{-1}
\qRq
\begin{cases}
\hat{\pp}(g\star \phi)=h\,\hat{\pp}(\phi)\,h^{-1}\,,\cr
\hat{\ww}(g\star \phi)=h\,\hat{\ww}(\phi)\,h^{-1}+hdh^{-1}\,,
\end{cases}
\label{PWhattra}
\end{equation}
where $ h=h(\phi,{\bf g})$. By deriving \eqref{hatOm} we find the \emph{gauged} Maurer-Cartan equations:
\begin{equation}
d\hat{\Omega}+\hat{\Omega}\wedge \hat{\Omega}
    ~=\,-g\,\LL^{-1}\mathscr{F}\LL\;,
\end{equation}
where we have used \eqref{D2F}. Projecting the above equation onto $\mathfrak{K}$ and $\mathfrak{H}$ we find the gauged version of eqs.\ \eqref{DP}, \eqref{RW}:
\begin{subeqs} 
\begin{align}
\mathscr{D}\hat{\pp} &~\equiv~
    d\hat{\pp}+\hat{\ww}\wedge \hat{\pp}+\hat{\pp}\wedge \hat{\ww}
    ~=\,-g\,F^{\hat{\Lambda}}\,{\pp}_{\hat{\Lambda}}\;,
\label{DP2}
\\
\hat{R}(\hat{\ww}) &~\equiv~
    d\hat{\ww}+\hat{\ww}\wedge \hat{\ww}
    ~=\,-\hat{\pp}\wedge\hat{\pp}-g\,F^{\hat{\Lambda}}\,{\ww}_{\hat{\Lambda}}\;.
\label{RW2}
\end{align}
\end{subeqs}
that are manifestly $\Gg$-invariant. The gauged $\mathfrak{H}$-valued curvature 2-form can be written in terms of the curvature components \eqref{Rcompo} of the manifold as:
\begingroup
\begin{equation}
\hat{R}(\hat{\ww})\=
    \frac{1}{2}\,R_{su}\,\mathscr{D}\phi^s\wedge\mathscr{D}\phi^u
    -g\,F^{\hat{\Lambda}}\,{\ww}_{\hat{\Lambda}}\;.
\end{equation}
\endgroup
The fermion fields of the theory transform under (compensating) transformation in $H$ (see \eqref{duality2}). In the gauged formulation, this is taken into account by using the gauged $H$-connection $\hat{\ww}$ in the fermion $H$-covariant derivatives, promoting the latter to $\Gg$-covariant ones and minimally coupling the fermions to the gauge fields. The gauge-covariant derivatives on a fermion field $\xi$ are then defined as
\begin{equation}
D_\mu\xi\=\nabla_\mu\xi+\hat{\ww}_\mu\star \xi\;.
\label{Dxi2}
\end{equation}
$\nabla_\mu$ being the covariant derivative containing the Levi-Civita
connection on space-time and the $\star$ symbol denoting the action of the $\mathfrak{H}$-valued connection on $\xi$, in the corresponding $H$-representation.\par\smallskip
Summarizing, local invariance of the action under $\Gg$ requires replacing everywhere the abelian field strengths by the non abelian ones, eq.\ \eqref{replaceF}, and the ungauged vielbein $\pp_\mu$ and $H$-connection $\ww_\mu$ by the gauged ones:
\begingroup
\abovedisplayskip=6pt
\abovedisplayshortskip=0pt
\belowdisplayskip=10pt
\belowdisplayshortskip=12pt
\begin{equation}
\pp_\mu\;\rightarrow\;\hat{\pp}_\mu\;;
\qquad
\ww_\mu\;\rightarrow\;\hat{\ww}_\mu\;.
\label{replacePW}
\end{equation}
\endgroup
%

\subsubsection{The gauged Lagrangian}
Now we want to discuss how to obtain a new Lagrangian, compatible
with the new local symmetry, as a deformation of the Lagrangian of the
ungauged theory. The first steps consist in covariantizing all derivatives, according to the above discussions, and replacing the abelian field strengths by the full covariant ones. Then, further deformations of the Lagrangian are required in order to restore supersymmetry and preserve gauge invariance.

\paragraph{Topological terms.}
If the symplectic duality action \eqref{xsymp} of $X_{\hat{\Lambda}}$ has a non-vanishing off-diagonal block $X_{{\hat{\Lambda}}{\hat{\Gamma}}{\hat{\Sigma}}}$, that is if the gauge transformations include Peccei-Quinn shifts, then an infinitesimal (local) gauge transformation $\zeta^{\hat{\Lambda}}(x)\,X_{{\hat{\Lambda}}}$ would produce a variation of the Lagrangian of the form \eqref{deltaLC}:
\begin{equation}
\delta\Lagr_{\textsc{bos}}\=
\frac{g}{8}\,\zeta^{\hat{\Lambda}}(x)\,X_{{\hat{\Lambda}}{\hat{\Gamma}}{\hat{\Sigma}}}\,\veps^{\mu\nu\rho\sigma}\,
    F^{\hat{\Gamma}}_{\mu\nu}F^{\hat{\Sigma}}_{\rho\sigma}\;.
\label{deltaLX}
\end{equation}
Being $\zeta^{\hat{\Lambda}}(x)$ a local parameter, the above term is no longer a total derivative and thus the transformation is not a symmetry of the action.
In \cite{deWit:1984px} it was proven that the variation \eqref{deltaLX} can be cancelled by adding to the Lagrangian a topological term
\begin{equation}
\Lagr_\text{top}
    \=\frac{1}{3}\,g\,\veps^{\mu\nu\rho\sigma}\,X_{{\hat{\Lambda}}{\hat{\Gamma}}{\hat{\Sigma}}}\;A^{\hat{\Lambda}}_\mu\,A^{\hat{\Sigma}}_\nu\,
    \left(\partial_\rho A^{\hat{\Gamma}}_\sigma +\frac{3}{8}\,g\,X_{{\hat{\Delta}}{\hat{\Pi}}}{}^{\hat{\Gamma}}\,A^{\hat{\Delta}}_\rho\,A^{\hat{\Pi}}_\sigma\right)\;,
\label{top}
\end{equation}
provided the following condition holds:
\begin{equation}
X_{({\hat{\Lambda}}{\hat{\Gamma}}{\hat{\Sigma}})}\=0\;.
\label{xsymmetr}
\end{equation}
The condition \eqref{xsymmetr}, together with the closure constraint \eqref{gaugealg2}, is part of a set of constraints on the gauge algebra which is implied by supersymmetry. Indeed, even if the Lagrangian $\LgaugO[0]$ constructed so far is locally $\Gg$-invariant, the presence of minimal couplings explicitly breaks both supersymmetry and the duality global symmetry $G$.\par

\paragraph{Yukawa terms, fermion shift matrices.}
We obtained a certain number of steps in order to construct a Lagrangian $\LgaugO[0]$ which is locally $\Gg$-invariant starting from the ungauged one. However, the obtained $\LgaugO[0]$ is no longer invariant under supersymmetry, due to the extra contributions that arise from variation of the vector fields in the covariant derivatives.\par
Consider, for instance, the supersymmetry variation of the (gauged) Rarita-Schwinger term in the Lagrangian
\begin{equation}
\Lagr_{\textsc{rs}}\=i\,\eD\,\bar{\psi}^A_\mu\,\gamma^{\mu\nu\rho}\,D_\nu\psi_{A\,\rho}
                    \+\text{h.c.}\;,
\end{equation}
where $D_\nu$ is the gauged covariant derivative defined in eq.\ \eqref{Dxi2}. Under supersymmetry variation of $\psi_\mu$ one finds
\begin{equation}
\delta\psi_\mu\=D_\mu \epsilon\+\dots\;,
\end{equation}
$\epsilon$ being the local supersymmetry parameter and the ellipses referring to terms containing the vector field strengths.
The variation of $\Lagr_{\textsc{rs}}$ produces a term
\begin{equation}
\delta\Lagr_{\textsc{rs}}
    ~=~\dots
    \+2\,i\,\eD\,\bar{\psi}^A_\mu\,\gamma^{\mu\nu\rho}\,D_\nu D_{\rho}\epsilon_A
    \+\text{h.c.}~=
    \,-i\,g\,\eD\,\bar{\psi}^A_\mu\,\gamma^{\mu\nu\rho}\,F_{\nu\rho}^{\hat{\Lambda}}\,(\ww_{\hat{\Lambda}}\epsilon)_A
        \+\text{h.c.}
\;,\label{RSvar}
\end{equation}
where we have used the property \eqref{D2F} of the gauge covariant derivative. Similarly, we can consider the supersymmetry variation of the spin-$1/2$ fields:
\begin{equation}
\delta\lambda^{I}\=i\,\hat{\pp}_\mu^{IA}\,\gamma^\mu\epsilon_A\+\dots\,,
\end{equation}
obtaining, in the variation of the corresponding kinetic Lagrangian, $O(g)$-terms of the form:
\begin{equation}
\delta\Lagr'_\text{kin}\=
    \dots\+i\,g\,\eD\,\bar{\lambda}_{I}\,\gamma^{\mu\nu}\,F_{\mu\nu}^{\hat{\Lambda}}\,{\pp}_{\hat{\Lambda}}^{IA}\,\epsilon_A
    \+\text{h.c.}
\label{lambdatra}
\end{equation}
%
%
To cancel the above $O(g)$-terms from supersymmetry variations of $\LgaugO[0]$, and to construct a gauged Lagrangian $\Lgaug$ preserving the original supersymmetries, one can apply the general Noether method%
\footnote{%
see \cite{VanNieuwenhuizen:1981ae} for a general review
}
which consists in adding new terms to $\LgaugO[0]$ and to the supersymmetry transformation laws, iteratively in the gauge coupling constant.
In our case, the procedure converges by adding terms of order one and two in $g$, so that $\Lgaug$ can be written as
\begin{equation}
\Lgaug\=\LgaugO[0]+\Delta\LgaugO[1]+\Delta\LgaugO[2]\;.
\end{equation}
The additional $O(g)$-terms are of \emph{Yukawa type} and have the general form:
\begin{equation}
\eD^{-1}\Delta\LgaugO[1]\=
    g\left(2\,\bar{\psi}^A_\mu\,\gamma^{\mu\nu}\,\psi_\nu^B\;\mathbb{S}_{AB}
    \+i\,\bar{\lambda}^{I}\,\gamma^\mu\,\psi_{\mu A}\,\mathbb{N}_{I}{}^A
    +~\bar{\lambda}^{I}\,\lambda^{J}\,\mathbb{M}_{IJ}\right)
    \+\text{h.c.}\;\;,
\label{fmassterms}
\end{equation}
characterized by the scalar-dependent matrices $\mathbb{S}_{AB}$ and $\mathbb{N}^{I A}$ called \emph{fermion shift matrices}, and a matrix $\mathbb{M}^{IJ}$ that can be rewritten in terms of the previous mixed mass tensor $\mathbb{N}^{IA}$.\par
Finally, the $O(g^2)$-terms will consist of a scalar potential:
\begingroup%
\setlength{\abovedisplayshortskip}{4pt plus 3pt}%
\setlength{\abovedisplayskip}{6pt plus 3pt minus 4pt}
\setlength{\belowdisplayshortskip}{4pt plus 3pt}%
\setlength{\belowdisplayskip}{6pt plus 3pt minus 4pt}%
\begin{equation}
\eD^{-1}\Delta\LgaugO[2]~=\,-g^2\,V(\phi) \;,
\label{spot}
\end{equation}
\endgroup%
as we shall see below.\par

\paragraph{Fermion SUSY transformations.}
Now we have to modify the fermionic transformations, adding order--$g$ terms to the supersymmetry transformation rules of the gravitino $\psi_{\mu A}$ and of the other fermions
\begin{equation}
\begin{split}
\delta_\epsilon\psi_{\mu A}&\=
    D_\mu\epsilon_A\+i\,g\;\mathbb{S}_{AB}\;\gamma_\mu\,\epsilon^B\+\dotsc\;,
\\
\delta_\epsilon\lambda_{I}&\=g\,\mathbb{N}_{I}{}^{A}\,\epsilon_A\+\dotsc
\label{fermshifts}
\end{split}
\end{equation}
These terms depend on the same fermion shift-matrices $\mathbb{S}_{AB},\,\mathbb{N}_{I}{}^{A}$ entering the mass terms.
These matrices are composite fields belonging to some appropriate representations $\Rs_\ms{\mathbb{S}},\,\Rs_\ms{\mathbb{N}}$ of the $H$ group, such that \eqref{fmassterms} is $H$-invariant.\par

\paragraph{Scalar potential.}
As we stated above, in order to cancel the $O(g^2)$-contributions resulting from the variations \eqref{fermshifts} in \eqref{fmassterms}, we need to add an $O(g^2)$-\emph{scalar potential} $V(\phi)$. The latter is totally determined by supersymmetry as a bilinear in the shift matrices by the condition
\begin{equation}
\delta_B{}^A\,V(\phi)\= g^2\,\left(\mathbb{N}_{I}{}^{A}\,\mathbb{N}^{I}{}_{B}-12\;\mathbb{S}^{AC}\,\mathbb{S}_{BC}\right)\;,
\label{WID}
\end{equation}
where we have defined \,$\mathbb{N}^{I}{}_{A}\equiv \left(\mathbb{N}_{I}{}^{A}\right)^*$ \,and\, $\mathbb{S}^{AB}\equiv \left(\mathbb{S}_{AB}\right)^*$. The above condition is called \emph{potential Ward identity} \cite{Ferrara:1985gj,Cecotti:1984wn} and defines the scalar potential as a non-linear function of the scalar fields.\par\smallskip
We must emphasize that not for all choices of the gauge group it is possible to restore supersymmetry following the above prescriptions. There are further constraints on the Lie algebra of $\Gg$ (SUSY constraints) which need to be satisfied. The latter are linear and quadratic in the gauge generators and we shall discuss them below in a convenient formalism.

\subsubsection{$\boldsymbol{G}$-covariant formulation. Embedding tensor formalism} \label{subsubsec:gaugalgembtens}
We have seen that the gauging procedure corresponds to promoting some suitable subgroup $\Gg\subset\Gel$ to a local symmetry. This subgroup is defined selecting a subset of generators within the global symmetry algebra $\mathfrak{g}$ of $G$. All the information about the gauge algebra can be encoded in a $\Gel$-covariant object $\theta$: in terms of the latter, the gauge generators can be expressed as linear combinations of the global symmetry generators $t_\alpha$ of the subgroup $\Gel\subset G$
\begin{equation}
X_{\hat{\Lambda}}\=\theta_{\hat{\Lambda}}{}^\sigma\,t_\sigma\;; \qqquad
\theta_{\hat{\Lambda}}{}^\sigma \in {\bf \nv}\times\Adj(\Gel)\;,\qquad
\label{gentheta}
\end{equation}
with \,${\hat{\Lambda}}=1,\,\dotsc,\,\nv$\, and with \,$\sigma=1,\dotsc,\,\dim(\Gel)$.\par
The $\Gel$-invariance of the original ungauged Lagrangian $\Lagr$ is restored at the level of the gauged Lagrangian $\Lagr_\text{gaug}$ provided $\theta_{\hat{\Lambda}}{}^\sigma$ is transformed under $\Gel$ as well. However, the full global symmetry group $G$ of the field equations and Bianchi identities is still broken, since the parameters $\theta_{\hat{\Lambda}}{}^\sigma$ can be viewed as a number $n_\text{el}=\dim(\Gel)$ of electric charges, whose presence manifestly break electric-magnetic duality invariance. This means we are working in a specific symplectic frame, defined by the ungauged Lagrangian we started from%
\footnote{%
it is possible to define a procedure which is completely freed from the choice of the symplectic frame, see for instance \cite{Gallerati:2016oyo,Trigiante:2016mnt}
}%
. \par\smallskip
It is useful to give a description of the gauge algebra and its consistency constraints which does not depend on the original symplectic frame, namely which is manifestly $G$-covariant. This is done by encoding all information on the initial symplectic frame in a symplectic matrix $\E\equiv(\E_M{}^N)$ and writing the gauge generators in terms of new generators as
\begin{equation}
X_M\=(X_\Lambda,\,X^\Lambda)\;,
\end{equation}
which are at least twice as many as the $X_{\hat{\Lambda}}$:
\begin{equation}
\left(\begin{matrix}
    X_{\hat{\Lambda}} \cr 0
\end{matrix}\right)
\=\E\,\left(\begin{matrix}
    X_\Lambda\cr X^\Lambda
\end{matrix}\right)\;.
\label{EXL}
\end{equation}
This description is therefore redundant and this is the price to pay in order to have a manifestly symplectic covariant formalism. We can then rewrite the gauge connection in a symplectic fashion:
\begin{equation}
A^{\hat{\Lambda}}\,X_{\hat{\Lambda}}\= A^{\hat{\Lambda}}\,\E_{\hat{\Lambda}}{}^\Lambda\,X_\Lambda
    +A^{\hat{\Lambda}}\,\E_{{\hat{\Lambda}}\,\Lambda}\,X^\Lambda
\=\AL\,X_\Lambda+\ALd\,X^\Lambda=\mathbb{A}^M_\mu\,X_M\;,
\label{syminvmc}
\end{equation}
where we have introduced the vector fields $\AL$ and the corresponding dual ones $\ALd$, that can be regarded as components of a symplectic vector
\begin{equation}
\AM~\equiv~(\AL,\,\ALd)\;.
\end{equation}
These are clearly not independent, since they are all expressed in terms of the only electric vector fields $A^{\hat{\Lambda}}$ of our theory (those entering the vector kinetic terms):
\begin{equation}
\AL=\E_{\hat{\Lambda}}{}^\Lambda\,A^{\hat{\Lambda}}_\mu\;,\qquad
\ALd=\E_{{\hat{\Lambda}}\,\Lambda}\,A^{\hat{\Lambda}}_\mu\;.
\end{equation}
%
%

\paragraph{Embedding tensor.}
The components of the symplectic vector $X_M$ are generators in the isometry algebra $\mathfrak{g}$ and thus can be expanded in a basis $t_\alpha$ of generators of $G$:
\begin{equation}
X_M\=\Theta_M{}^\alpha\,t_\alpha\;,\qquad\qquad \alpha=1,\dotsc,\,\dim(G)\;.
\label{Thdef}
\end{equation}
The coefficients of this expansion $\Theta_M{}^\alpha$ represent an extension of the definition of $\theta$ to a $G$-covariant tensor:
\begin{equation}
\theta_\Lambda{}^\sigma \,\;\longrightarrow\;\;
\Th \equiv (\theta^{\Lambda\,\alpha},\,\theta_\Lambda{}^\alpha)\;;
\qquad\; \Th\,\in\,\Rs_\text{v*}\times\Adj(G)\;,
\label{embtens}
\end{equation}
where $\Rsvst$ acts on covariant symplectic vectors, being the representation dual to the symplectic representation $\Rsv$ of the group $G$. The $\Theta$ tensor describes the explicit embedding of the gauge group $\Gg$ into the global symmetry group $G$ and combines the full set of deformation parameters of the original ungauged Lagrangian. The advantage of this description is that it allows to recast all the consistency conditions on the choice of the gauge group into $G$-covariant (and thus independent of the symplectic frame) constraints on $\Theta$.\par
Notice that, just as the redundant set of vectors $\mathbb{A}^M_\mu$, also the components of $\Theta_M{}^\alpha$ are not independent since, by eq.\ \eqref{EXL},
\begin{equation}
\theta_{\hat{\Lambda}}{}^\alpha=\E_{\hat{\Lambda}}{}^M\,\Theta_M{}^\alpha\;,
\qquad\;
0=\E^{\hat{\Lambda}\,M}\,\Theta_M{}^\alpha\;,
\label{elET}
\end{equation}
so that
\begin{equation}
\dim(\Gg)=\rank(\theta)=\rank(\Theta)\;.
\end{equation}
The above relations \eqref{elET} imply for $\Theta_M{}^\alpha$ the following symplectic-covariant condition:
\begin{equation}
\Theta_\Lambda{}^\alpha\,\Theta^{\Lambda\,\beta}-\Theta_\Lambda{}^\beta\,\Theta^{\Lambda\,\alpha}=0
\qqLqq
\Cc^{MN}\,\Theta_M{}^\alpha\,\Theta_N{}^\beta=0 \quad .
\label{locality}
\end{equation}
On the other hand, one can show that if $\Theta_M{}^\alpha$ satisfies the above conditions, there exists a symplectic matrix $\E$ which can rotate it to an electric frame, namely such that eqs.\ \eqref{elET} are satisfied for some $\theta_{\hat{\Lambda}}{}^\alpha$.\, The above equations \eqref{locality} define the so-called \emph{locality constraint} on the embedding tensor $\Theta_M{}^\alpha$ and they clearly imply:
\begin{equation}
\dim(\Gg)=\rank(\Theta)\,\le\,\nv\;,
\end{equation}
which is the preliminary consistency condition \eqref{preliminary}.\par
The electric-magnetic duality action of $X_M$, in the generic symplectic frame defined by the matrix $\E$, is described by the tensor:
\begin{equation}
X_{MN}{}^P~\equiv~
\Theta_M{}^\alpha\;t_{\alpha\,N}{}^P\=
\E^{-1}{}_M{}^{\hat{M}}\,\E^{-1}{}_N{}^{\hat{N}}\,X_{\hat{M}\hat{N}}{}^{\hat{P}}\,\E_{\hat{P}}{}^P\;.
\label{XEhatX}
\end{equation}
For each value of the index {\footnotesize $M$}, the tensor $X_{MN}{}^P$ should generate symplectic transformations, and this implies that:
\begin{equation}
X_{MNP}~\equiv~ X_{MN}{}^Q\;\Cc_{QP}=X_{MPN}\;.
\end{equation}
The remaining linear constraints \eqref{lin1}, \eqref{xsymmetr} on the gauge algebra can be recast in terms of $X_{MN}{}^P$ in the following symplectic-covariant form:
\begin{equation}
X_{(MNP)}\=0
\;\qLq\;
\begin{cases}
2\,X_{(\Lambda\Sigma)}{}^\Gamma\=X^\Gamma{}_{\Lambda\Sigma}\;,\\
2\,X^{(\Lambda\Sigma)}{}_\Gamma\=X_\Gamma{}^{\Lambda\Sigma}\;,\\ \phantom{2\,}X_{(\Lambda\Sigma\Gamma)\,}\=0\;.
\end{cases}
\label{lconstr}
\end{equation}
Notice that the second of equations \eqref{lconstr} implies that in the electric frame, in which $X^{\hat{\Lambda}}=0$, also the upper-right block of the infinitesimal gauge generators \,$\mathscr{R_v}[X_{\hat{\Lambda}}]$ vanishes, being $X_{\hat{\Gamma}}{}^{\hat{\Lambda}\hat{\Sigma}}=0$, so that the gauge transformations are indeed in $\Gel$.\par
Finally, the closure constraints \eqref{gaugealg2} can be written, in the generic frame, in the following form:
\begin{equation}
[X_M,\,X_N]~=\,-X_{MN}{}^P\,X_P
\;\qLq\;
\Theta_M{}^\alpha\Theta_N{}^\beta f_{\alpha\beta}{}^\gamma
    +\Theta_M{}^\alpha\,t_{\alpha\,N}{}^P \Theta_P{}^\gamma\=0\;.
\label{closconstr}
\end{equation}
The above condition can be rephrased, in a $G$-covariant fashion, as the condition that the embedding tensor $\Theta_M{}^\alpha$ is invariant under the action of the gauge group it defines:
\begin{equation}
\delta_M\Theta_N{}^\alpha\=0\;.
\end{equation}
Summarizing we have found that consistency of the gauging requires the following set of linear and quadratic algebraic, $G$-covariant constraints to be satisfied by the embedding tensor:
\begin{align}
\text{\textsl{Linear constraint}}&:
&&X_{(MNP)}\=0\;,&&&&
\label{linear2}
\\[1.5em]
\text{\textsl{Quadratic constraints}}&:
&&\Cc^{MN}\,\Theta_M{}^\alpha\,\Theta_N{}^\beta\=0\;,&&&&
\label{quadratic1}
\\[\jot]
&&&[X_M,\,X_N]=-X_{MN}{}^P\,X_P\;.&&&&
\label{quadratic2}
\end{align}
The linear constraint \eqref{linear2} amount to a projection of the embedding tensor on a specific $G$-representation $\Rs_{{\!}_\Theta}$ in the decomposition of the product $\Rs_\text{v*}\times\Adj(G)$ with respect to $G$
\begin{equation}
\Rsvst\times\Adj(G) \quad \stackrel{G}{\longrightarrow} \quad
\Rs_{{\!}_{\Theta}} \+ \dots
\end{equation}
and thus can be formally written as:
\begin{equation}
\mathbb{P}_{{\!}_{\Theta\!}}\cdot\,\Theta\=\Theta\;,
\end{equation}
where $\mathbb{P}_{{\!}_\Theta}$ denotes the projection on the representation $\Rs_{{\!}_\Theta}$. For this reason \eqref{linear2} is also named \emph{representation constraint}.\par
The first quadratic constraint \eqref{quadratic1} guarantees that a symplectic matrix $\E$ exists which rotates the embedding tensor $\Theta_M{}^\alpha$ to an electric frame in which the \emph{magnetic components} $\Theta^{\hat{\Lambda}\,\alpha}$ vanish. The second one \eqref{quadratic2} is the condition that the gauge algebra close within the global symmetry one $\mathfrak{g}$ and implies that $\Theta$ is a singlet with respect to $\Gg$. Let us stress, however, that constraint \eqref{closconstr} is in general stronger than simple closure: in particular we find the non-trivial relation
\begin{equation}
X_{(MN)}{}^{P}\,X_{P}\=0
\label{closure}
\end{equation}
upon symmetrization in {\footnotesize $(MN)$} of the above \eqref{closure} -- upon which the l.h.s.\ trivially vanishes, but the r.h.s.\ does not -- which clearly goes beyond closure condition.\par
In a general theory, the three constraints \eqref{linear2}, \eqref{quadratic1} and \eqref{quadratic2} should be imposed independently. In theories where all scalar fields enter the same supermultiplets as the vector ones (as it is the case of \,$\N>2$\, or \,$\N=2$\, with no hypermultiplets), the locality constraint \eqref{quadratic1} follows from the other two%
\footnote{%
in maximal supergravity, however, the closure constraint \eqref{quadratic2} follows from \eqref{linear2} and \eqref{quadratic1} and thus, once the linear constraint is imposed, the two quadratic ones are equivalent
}%
. In particular, the locality constraint \eqref{quadratic1} is independent of the others in theories featuring scalar isometries with no duality action, namely in which the symplectic duality representation $\Rsv$ of the isometry algebra $\mathfrak{g}$ is not faithful%
\footnote{%
this is the case of the quaternionic isometries in $\N=2$ theories, see eq.\ \eqref{qisom}}%
.\par\smallskip
As we have seen above, in the second part of the gauging procedure one has to restore supersymmetry after minimal couplings have been introduced and the $\Gg$-invariant Lagrangian $\LgaugO[0]$ has been constructed. However, the supersymmetric completion of $\LgaugO[0]$ requires no more constraints on $\Gg$ (i.e.\ on $\Theta$) than the linear \eqref{linear2} and quadratic ones \eqref{quadratic1}, \eqref{quadratic2} discussed above.\par

\subsubsection{Vacua and Dualities}\label{subsubsec:vacuadual}
A vacuum of a supergravity theory preserving Lorentz invariance is a maximally symmetric solution, that is, it can exhibit Minkowski, de Sitter or anti-de Sitter space-time geometry, depending on the value of the cosmological constant $\Lambda$:
\begin{equation}
\left\{
\begin{aligned}
\;\Lambda&\=0\qquad \text{Minkowski}\,,\\
\;\Lambda&~>~0\qquad \text{de Sitter}\,,\\
\;\Lambda&~<~0\qquad \text{anti-de Sitter}\,.
\end{aligned}
\right.
\end{equation}
Due to the maximal space-time symmetry, only scalar fields are allowed to have a non-vanishing (uniform) v.e.v.\ $\phi^s_0$\,:
\begin{equation}
\big\langle\phi^s(x)\big\rangle ~\equiv~ \phi^s_0 ~\equiv~ \phi_0  \;,
\end{equation}
while the vector and fermion fields vanish on the solution. This v.e.v.\ defines a point in the moduli space which is an extremum of the scalar potential $V(\phi)$:
\begin{equation}
\left.\frac{\partial V}{\partial\phi^s}\right\vert_{\phi_0}=~0\;.
\label{dV0}
\end{equation}
The value $V(\phi_0)$ of the scalar potential on the vacuum gives the effective cosmological constant for the underlying space-time geometry:
\begin{equation}
\Lambda\=V(\phi_0)\;.
\label{LV0}
\end{equation}
The Riemann tensor has the form
\begin{equation}
R_{\mu\nu\rho\sigma}~=\,
    -\frac{\Lambda}{3}\left(g_{\mu\rho}\,g_{\nu\sigma}-g_{\mu\sigma}\,g_{\nu\rho}\right)\;,
\label{RiemL}
\end{equation}
and the Ricci tensor reads
\begin{equation}
R_{\mu\nu}~=\,-\Lambda\,g_{\mu\nu}\;.
\end{equation}
The scalar potential is expressed, for extended models, by condition \eqref{WID} and, being expressed as an $H$-invariant combination of composite fields (the fermion shifts), it is invariant under the simultaneous action of $G$ on $\Theta$ and $\phi^s$:
\begin{equation}
\forall {\bf g}\in G \;:\qquad
V({\bf g}\star\phi,\,{\bf g}\star\Theta)\=V(\phi,\,\Theta)\;.
\label{Vinvar}
\end{equation}
This means that, if $V(\phi,\,\Theta)$ has an extremum in $\phi_0$:
\begin{equation}
\left.\frac{\partial}{\partial\phi^s}V(\phi,\,\Theta)\right\vert_{\phi_0}\!=~0\;,
\end{equation}
at the same time $V(\phi,\,{\bf g}\star\Theta)$ has an extremum at ${\bf g}\star\phi_0$ with the same properties,
\begin{equation}
\forall {\bf g}\in G \;:\qquad
\left.\frac{\partial}{\partial\phi^s}V(\phi,\,{\bf g}\star\Theta)\right\vert_{{\bf g}\star\phi_0}\!=~0\;,
\end{equation}
i.e.\ same value of the potential at the extremum and its derivatives.\par
If the scalar manifold of a given gauged model is homogeneous, we can map any point $\phi_0$ to the origin $\OO$, where all scalars vanish, by the inverse of the coset representative $\LL(\phi_0)^{-1}\in G$.\, We can then map a generic vacuum $\phi_0$ of a given theory (defined by an embedding tensor $\Theta$) to the origin in a theory defined by \,$\Theta'=L(\phi_0)^{-1}\star \Theta$.\, Now, if we are looking for vacua with given properties, all quantities defining the gauged theory -- fermion shifts and mass matrices -- can be computed at the origin,
\begin{equation}
\mathbb{N}(\OO,\,\Theta)\,,\;\;\mathbb{S}(\OO,\,\Theta)\,,\;\;\mathbb{M}(\OO,\,\Theta)\;,
\end{equation}
the properties of the vacuum being translated in conditions on $\Theta$. In this way, we can search for the vacua by scanning through all possible gaugings \cite{Dibitetto:2011gm,Dall'Agata:2011aa,Gallerati:2014xra}.

\paragraph{Supersymmetric vacua.}
A vacuum of the theory $\phi_0$ is said to be supersymmetric if it preserve an amount of supersymmetry. In this case there should exist a local supersymmetry parameter $\epsilon_A(x)$ along which the supersymmetry variation of the fermions vanish, when evaluated on the solution.
This follows from the fact that, along the direction of the preserved supersymmetry, the action on the vacuum gives \,$\bar{\epsilon}\,Q\,\vert 0 \rangle=0$, and thus
\begin{equation}
\delta_\epsilon \texttt{f}(x)\=
    \langle 0\vert
    \left[\,\bar{\epsilon}\,Q\,,\;\hat{\texttt{f}}(x)\,\right]
    \vert 0\rangle \=0\;,
\end{equation}
where $\texttt{f}(x)$ is a generic fermionic field and $\hat{\texttt{f}}(x)$ the corresponding field operator and where the r.h.s.\ of the above equation depends on the v.e.v.\ $\phi_0\equiv\phi_0^s$ of the scalars and geometry of the vacuum solution%
\footnote{%
analogous conditions on SUSY variations of the bosonic fields are trivially satisfied, since the latter are expressed in terms of the fermions which vanish on the background
}%
. The above conditions can be written as
\begin{subeqs} \label{KSeqs} 
\begin{align}
\delta\psi_{\mu A}&\=
    D_\mu\eps_A+i\,g\;\mathbb{S}_{AB}\,\gamma_\mu\,\eps^B\=0\;,
\label{KSeqs1}
\\
\delta\lambda_{I}&\=g\;\mathbb{N}_{I}{}^A\,\eps_A\=0\;,
\label{KSeqs2}
\end{align}
\end{subeqs}
where the tensors $\mathbb{S}_{AB}$ and $\mathbb{N}_{I}{}^A$ are evaluated at $\phi_0$. These are the \emph{Killing spinor equations for the vacuum}: if the latter admit $\mathcal{N}'$ distinct solutions (Killing spinors), the background preserves $\mathcal{N}'\le\mathcal{N}$ of the original $\mathcal{N}$ supersymmetries of the theory.\par
If one combines the imposed integrability condition on \eqref{KSeqs1}, i.e.\
\begin{equation}
0\=\nabla_{[\mu}\delta\psi_{\nu]A}\;,
\end{equation}
with the previous Killing spinor equation and the Riemann tensor form \eqref{RiemL}, it is easily demonstrated that \emph{supersymmetric vacua} can only be Minkowski ($\Lambda = 0$) or anti-de Sitter ($\Lambda < 0$). The latter, in particular, are maximally symmetric solutions with negative cosmological constant that are very interesting from a theoretical point of view in the light of the AdS/CFT holography conjecture \cite{Maldacena:1997re}.
The construction of supersymmetric solutions of supergravity theories has been studied in \cite{Caldarelli:2003pb,Meessen:2006tu,Cacciatori:2008ek,Klemm:2010mc}.


\subsection{Black holes in gauged supergravity}
As already pointed out, the construction of black hole solutions in gauged supergravity theory is essential for phenomenologically realistic cosmological models, supporting the presence of some effective cosmological constant as well as non-trivial scalar potential and scalar mass terms.\par
From a theoretical point of view, the study of gauged black hole solution has been strongly motivated by the so-called AdS/CFT duality \cite{Maldacena:1997re}, that relates $d+1$ dimensional gravity theories in Anti de Sitter (AdS) spacetime to conformal field theories (CFT) in $d$ dimensions. In particular, the conjecture states that stable AdS solutions describe conformal critical points of a suitable gauge theory defined on the boundary of the space: it is a successful realization of the holographic principle \cite{Susskind:1994vu}, asserting that the description of the bulk AdS spacetime is encoded on its boundary on which the CFT lives.\par
In general relativity, the study of exact (neutral) static black hole solutions with scalar hair was a powerful tool for clarifying different aspects of no-hair theorems \cite{Hertog:2006rr}, the role of scalar charges for black hole thermodynamics \cite{Hertog:2004bb, Anabalon:2014fla}, and issues related to their stability \cite{Hertog:2005hm, Faulkner:2010fh}.\par
After the discovery of one-parameter family of $\SO(8)$ maximal four-dimensional supergravity theories \cite{DallAgata:2012mfj}, many progresses have been made made towards the understanding of the vacuum structure and dual field theories \cite{Borghese:2012zs,Tarrio:2013qga,Gallerati:2014xra,Borghese:2012qm,Kodama:2012hu,Guarino:2013gsa,Guarino:2015tja}. Together with the original $\SO(8)$ model \cite{deWit:1982ig}, other gauged supergravities have been extended by using dyonic embedding tensor \cite{Dall'Agata:2011aa,Dall'Agata:2012sx,Anabalon:2017yhv}, featuring a richer vacuum structure and scalar field dynamics than their original counterparts.
Several procedures have then been developed for obtaining exact regular hairy black hole solutions for a general scalar potential \cite{Henneaux:2002wm,Martinez:2004nb,Hertog:2004dr,Anabalon:2012ta,Feng:2013tza,Faedo:2015jqa,Anabalon:2017yhv} and supersymmetric black hole solutions \cite{DallAgata:2010ejj,Gnecchi:2012kb,Chimento:2015rra}.

\subsubsection{Example:\, $\bm{\N=2}$\,,\, $\bm{D=4}$\, gauged SUGRA black hole}
Let us consider an extended \,$\N=2$ supergravity theory in four dimensions, coupled to $\nv$ vector multiplets and no hypermultiplets, in the presence of Fayet-Iliopoulos (FI) terms. The model describes $\nv+1$ vector fields $A^\Lambda_\mu$,\, ({\small$\Lambda$}\;$=0,\dots,\nv$) and $\ns=\nv$ complex scalar fields $z^i$ ($i=1,\dots,\ns$).\par
The bosonic \emph{gauged} Lagrangian now reads
\begin{equation}
\frac{1}{\eD}\,\Lagr_{\textsc{bos}}~=\,
-\frac{R}{2}
\+g_{i\bar{\jmath}}\,\partial_\mu z^i\,\partial^\mu \zb^{\bar{\jmath}}
\+\frac{1}{4}\,\II_{\Lambda\Sigma}(z,\zb)\,F^\Lambda_{\mu\nu}\,F^{\Sigma\,\mu\nu}
\+\frac{1}{8\,\eD}\,\RR_{\Lambda\Sigma}(z,\zb)\,\veps^{\mu\nu\rho\sigma}\,F^\Lambda_{\mu\nu}\,F^{\Sigma}_{\rho\sigma}
\-V(z,\zb)\;,
\label{boslagrGaug}
\end{equation}
where the $\nv+1$ vector field strengths are defined as usual:
\begin{align*}
F^\Lambda_{\mu\nu}\=\partial_\mu A^\Lambda_\nu-\partial_\nu A^\Lambda_\mu\;.
\end{align*}
The $\ns$ complex scalars $z^i$
span a special K\"ahler manifold $\Ms_\textsc{sk}$ and the scalar potential $V(z,\zb)$ originates from electric-magnetic FI terms. The presence of $V(z,\zb)$ amounts to gauging a $\U(1)$-symmetry of the corresponding ungauged model (with no FI terms) and implies minimal couplings of the vector fields to the fermions only.

\paragraph{Special geometry.}
A \emph{special K\"ahler manifold} $\Ms_\textsc{sk}$ is the class of target spaces spanned by the complex scalar fields in the vector multiplets of an $\N=2$ four-dimensional supergravity.\par
The geometry of $\Ms_\textsc{sk}$ can be described in terms of an \emph{holomorphic section} $\Omega^M(z^i)$ of the characteristic bundle defined over it, which is the product of a symplectic-bundle and a holomorphic line-bundle. The components of $\Omega^M(z^i)$ are written as
\begin{equation}
\Omega^M=
\left(\begin{matrix}
\X^\Lambda \cr \Fbo_\Lambda
\end{matrix}\right)\;, \qqqquad
\Lambda=0,\,\dots,\nv\;,
\end{equation}
while the \emph{K\"ahler potential} and the \emph{K\"ahler metric} have the following general form
\begin{equation}\label{Kom}
\begin{split}
\mathcal{K}(z,\zb)&~=\,
    -\log\left[\,i\;\overbar{\Omega}^T\,\Cc\;\Omega\,\right]~=\,
    -\log\left[\,i\,\left(\overbarcal{\X}^\Lambda\,\Fbo_\Lambda-{\X}^\Lambda\,\overbar{\Fbo}_\Lambda\,\right)\right]\;,
\\[1.5\jot]
g_{i\bar{\jmath}}~&=~\partial_i \partial_{\bar{\jmath}}\mathcal{K}\;.
\end{split}
\end{equation}
A change in the coordinate patch on the scalar manifold amounts to transforming $\Omega^M(z^i)$ by a corresponding constant $\Sp\big(2(\nv+1),\mathbb{R}\big)$ matrix, besides multiplying it by a holomorphic function $e^{f(z)}$. The former transformation leaves invariant the K\"ahler potential, as it is clear from the manifestly symplectic-invariant expression \eqref{Kom}, while the latter implies a corresponding \emph{K\"ahler transformation} on the potential:
\begin{equation}
\mathcal{K}(z,\zb) \;\;\rightarrow\;\; \mathcal{K}(z,\zb)-f(z)-\bar{f}(\zb)\;.
\label{Kahltransf}
\end{equation}
The choice of $\Omega^M(z^i)$ also fixes the symplectic frame (i.e.\ the basis of the symplectic fiber space) and, consequently, the non-minimal couplings of the scalars to the vector field strengths in the Lagrangian. In the \emph{special coordinate frame}, the lower components $\Fbo_\Lambda$ of the section can be expressed as the gradient, with respect to the upper entries $\X^\Lambda$, of a characteristic \emph{prepotential function} $\F(\X^\Lambda)$:
\begin{equation}
\Fbo_\Lambda\=\frac{\partial\F}{\partial \X^\Lambda}\;,
\end{equation}
where the function $\F(\X^\Lambda)$ is required to be homogeneous of degree two. The upper components $\X^\Lambda(z^i)$ are defined modulo multiplication times a holomorphic function and, in this frame, can be used as projective coordinates to describe the manifold: in a local patch in which
$\X^0\neq 0$, we can identify the scalar fields with the ratios $z^i=\X^i/\X^0$.\par\smallskip
In general a field $\Phi(z,\zb)$ on the K\"ahler manifold is a section of a $\U(1)$-bundle of weight $p$ if it transforms under a K\"ahler transformation \eqref{Kahltransf} as
\begin{equation}
\Phi(z,\zb)\;\;\rightarrow\;\;e^{i\,p\,\Img[f]}\,\Phi(z,\zb)\;,
\end{equation}
and we can define a corresponding $\U(1)$-\emph{covariant derivative} on the bundle as
\begin{equation}
\begin{split}
\D^{{}^{[\U(1)]}}_i \Phi~\equiv\,\left(\dd_i+\frac{p}{2}\,\dd_i\mathcal{K}\right)\Phi\;,
\\[\jot]
\D^{{}^{[\U(1)]}}_{\ib}\Phi~\equiv\,\left(\dd_\ib-\frac{p}{2}\,\dd_\ib\mathcal{K}\right)\Phi\;.
\end{split}
\end{equation}
Now we introduce a \emph{covariantly holomorphic vector} $\V^M$
\begin{equation}
\V^M\=e^{\frac{\mathcal{K}}{2}}\,\Omega^M\=
    \left(\begin{matrix}
          L^\Lambda \cr M_\Lambda\end{matrix}\right)\;,
\end{equation}
which is section of the $\U(1)$-line bundle with weight $p=1$, satisfying the property:
\begin{equation}
\D_{\ib}\,\V^M\=
    \left(\partial_{\bar{\imath}}-\frac{1}{2}\,\partial_{\bar{\imath}}\mathcal{K}\right)\V^M\=0\;,
\end{equation}
and we also have
\begin{equation}
\D_i\,\V^M \=
     \left(\partial_i+\frac{1}{2}\,\partial_i\mathcal{K}\right)\,\V^M\=
     \left(\begin{matrix}f_i^\Lambda\cr
           h_{i\Lambda}\end{matrix}\right)~\equiv~
    \Ucal_i^M \;,
\end{equation}
$\D_{i},\,\D_{\ib}$ being the above $\U(1)$-covariant derivatives (omitting the superscript). Under a K\"ahler transformation defined by a holomorphic function $f(z)$, the section transforms by a corresponding $\U(1)$-transformation:
\begin{equation}
\V^M \;\rightarrow\;\; e^{i\,\Img[f]}\,\V^M\;.
\end{equation}
From its definition and eq.\ \eqref{Kom}, we find that $\V^M$ satisfies the condition
\begin{equation}
\V^T\Cc\,\overbar{\V}\=i\;.
\end{equation}
In particular, the definition of this kind of manifold requires the section $\V^M$ to satisfy the properties
\begin{equation} \label{VMprop}
\begin{split}
\D_i\,\Ucal_j&\=i\;\mathcal{C}_{ijk}\,g^{k\bar{k}}\,\overbar{\Ucal}_{\bar{k}}\;,
\\[\jot]
\D_i\,\overbar{\Ucal}_{\jb}&\=g_{i\jb}\,\overbar{\V}\;,
\\[\jot]
\V^T\,\Cc\;\Ucal_i&\=0\;,
\\[\jot]
\Ucal_i^T\,\Cc\;\overbar{\Ucal}_\jb&~=\,-i\,g_{i\jb}\;,
\end{split}
\end{equation}
where $\mathcal{C}_{ijk}$ is a characteristic covariantly holomorphic tensor with weight $p=2$ which enters the expression of the Riemann tensor and defines the Pauli terms in the Lagrangian involving the gauginos.
The following identity is satisfied:
\begin{equation}
g^{i\bar{\jmath}}\,\Ucal_i^M\,\overbar{\Ucal}_{\jb}^N~=
    -\frac{1}{2}\,\M^{MN}-\frac{i}{2}\,\Cc^{MN}-\V^M\overbar{\V}^N\;,
\label{UMN}
\end{equation}
where, using property \eqref{CMC}, we have
\begin{equation}
\M^{MN}=-\,\Cc^{MP}\,\M_{PQ}\,\Cc^{QN}\;,
\end{equation}
with $\M_{PQ}$ defined in eq.\ \eqref{M}.

\paragraph{FI-terms and scalar potential.}
In $\N=2$ theories, the scalar manifold has the general form \eqref{SKQK}
\begin{equation}
\Mscal\=\Ms_\textsc{sk}\times\Ms_\textsc{qk}\;,
\end{equation}
the special K\"ahler submanifold $\Ms_\textsc{sk}$ parametrized by the complex scalar fields $z^i$ in the vector multiplets, and the quaternionic K\"ahler one $\Ms_\textsc{qk}$ by the real scalars in the hypermultiplets. The holonomy group $H$ of the scalar manifold splits according
to \eqref{HHH}
\begin{equation}
H \= H_\tts{R} \times H_\text{matt}\;,
\end{equation}
with $H_\tts{R}=\U(2)$ and $H_\text{matt}$ acting on the fields in the vector and hypermultiplets. At the same time, $H$ can be expressed by the product of the holonomy groups of $\Ms_\textsc{sk}$ and $\Ms_\textsc{qk}$ respectively:
\begin{equation}
H \= H^\textsc{sk} \times H^\textsc{qk}\;,
\end{equation}
with
\begin{equation}
H^\textsc{sk} \= \U(1) \times H^\textsc{sk}_\text{matt}\;,
\qqqquad
H^\textsc{qk} \= \SU(2) \times H^\textsc{qk}_\text{matt}\;.
\end{equation}
In the absence of hypermultiplets (that is the case under consideration) the $\SU(2)$ part of the R-symmetry group $H_\tts{R}$ becomes a global symmetry of the theory which can still be gauged, the gauging of this symmetry described by a (constant) embedding tensor $\Th$: the latter quantities are known as Fayet-Iliopoulos terms.\par
If we decide to gauge a $\U(1)$ inside $\SU(2)$, we can take $\Th$ to have only one non-vanishing component, $\theta_M=\Th[M][\alpha=1]$ and choose the remaining gauge algebra to be abelian with $X_{MN}^P=0$. In this case, the resulting theory is deformed with the introduction of abelian electric-magnetic FI terms defined by the above constant symplectic vector $\theta_M$, which encodes all the gauge parameters%
\footnote{%
even if we introduce both electric and magnetic gaugings to maintain duality
covariance, the duality group will always allow us to reduce to the case with only electric gaugings turned on (see \ref{subsubsec:gaugalgembtens}); this implies a correspondent rotation of the symplectic sections and the choice of a symplectic basis
}%
.\par\smallskip
The scalar potential $V(z,\zb)$ reads:
\begin{equation}
V\=\left(g^{i\bar{\jmath}}\,\Ucal_i^M\,\overbar{\Ucal}_{\bar{\jmath}}^N
         -3\,\V^M\,\overbar{\V}^N\right)\theta_M\,\theta_N\,=\,
    -\frac{1}{2}\,\theta_M\,\M^{MN}\,\theta_N-4\,\V^M\,\overbar{\V}^N\theta_M\,\theta_N\;,
\label{VFI}
\end{equation}
having used property \eqref{UMN}. It is easily verified that the above potential can be expressed in terms of a \emph{complex superpotential}
\begin{equation}
\W\=\V^M\,\theta_M\;,
\end{equation}
section of the $\U(1)$-bundle with $p=1$, as follows:
\begin{equation}
V\=g^{i\jb}\,\D_i\W\;\D_{\jb}\overbar{\W}-3\,|\W|^2\;.
\end{equation}
We can also define a \emph{real superpotential} $\mathcalboondox{W}=|\W|$ in terms of which the potential reads:
\begin{equation}
V\=4\,g^{i\bar{\jmath}}\,\partial_i\mathcalboondox{W}\,\partial_{\bar{\jmath}}\mathcalboondox{W}
    -3\,\mathcalboondox{W}^2\;.
\label{realsuper}
\end{equation}
The introduced $\theta_M$ terms transform in a symplectic representation $\Rsvst$ of the isometry group $G_\textsc{sk}$ of $\Ms_\textsc{sk}$ on contravariant vectors. These FI terms are analogous to the electric and magnetic charges, but while the latter can be considered as solitonic charges of the solution, the former are background quantities actually entering the Lagrangian. Moreover, even though they couple the fermion fields to the vectors, the FI terms do not define vector-scalar minimal couplings.

\paragraph{Equations of motion and isometries.}
The matrix $\M_{MN}$ can be used to write the couplings of the scalar fields to the vectors, in the equations of motion, in a formally symplectic covariant form: once written the symplectic vector of electric field strengths and magnetic duals
\begin{equation}
\FF^M_{\mu\nu}=~
    \left(
    \begin{matrix}
    F^\Lambda_{\mu\nu}\cr \Gdual_{\Lambda\mu\nu}
    \end{matrix}
    \right)\;,
\end{equation}
the equations of motion for the vector fields are expressed in the compact form
\begin{equation}
d\FF^M=0\;,\qquad\;
{}^*\FF^M=-\,\Cc^{MP}\M_{PN}(z,\zb)\,\FF^N\;.
\end{equation}
The scalar field equations can be written in the following form:
\begin{equation}
\nabla_\mu(\partial^\mu z^i)
\+\tilde{\Gamma}^i_{jk}\,\partial_\mu z^j \, \partial^\mu z^k \- \frac{1}{8}\,g^{i\bar{\jmath}}\;\FF^M_{\mu\nu}\;\partial_{\bar{\jmath}}\M_{MN}(z,\zb)\;\FF^{N\mu\nu}
\+g^{i\bar{\jmath}}\,\partial_{\bar{\jmath}}V
\=0 \;,
\end{equation}
where $\nabla_\mu$ is the covariant derivative, only containing the space-time Christoffel symbol and $\tilde{\Gamma}^i_{jk}$ is the connection on the K\"ahler manifold.\par\smallskip
Finally, the Einstein equations read:
\begin{equation}
R_{\mu\nu}\=
    2\,\partial_{(\mu} z^i\,\partial_{\nu)}
    \zb^{\bar{\jmath}}\,g_{i\bar{\jmath}}
    \+\frac{1}{2}\;\FF^M_{\mu\rho}\;\M_{MN}(z,\zb)\;\FF^{N}{\!}_\nu{}^\rho
    \-g_{\mu\nu}\,V\;,
\end{equation}
depending on the K\"ahler metric as well as on the spacetime metric .\par\smallskip
As pointed out earlier, the bundle structure defined on the scalar manifold allows to associate with a generic isometry transformation of the latter, a K\"ahler transformation and a constant symplectic transformation, belonging to the structure groups, acting on the symplectic section $\V^M$ and its derivatives. From the explicit form of the bosonic field equations and of the scalar potential, it is apparent that an isometry transformation of the scalar manifold is formally an on-shell symmetry of the theory, provided the corresponding symplectic transformation is made to act on the electric field strengths and their magnetic duals as well as on the FI terms:
\begin{align}
z^i \;\;\rightarrow\;\;  z^{\prime\,i}(z^j)\;\;:\quad
\begin{cases}
&\V^M(z',\zb')\,=\,e^{i\,\textrm{Im}(f)}\;(S^{-1})_N{}^M\;\V^N(z,\zb)\;,
\\[0.5ex]
&\theta_M\;\rightarrow\;\;\theta^\prime_M=S_M{}^N\,\theta_N\;,
\\[1ex]
&\FF^M\;\rightarrow\;\;\FF^{\prime M}=(S^{-1})_N{}^M\;\FF^N\;,
\end{cases}
\end{align}
where $S\in\Sp\big(2(\nv+1),\mathbb{R}\big)$. This formal invariance, however, involving a non-trivial transformation of the parameters (encoded in the FI terms) should be regarded as an equivalence between different theories.

\paragraph{Effective action.}
Let us consider static dyonic black hole configurations and assume a radial dependence for the scalar fields, $z^i=z^i(r)$. The most general metric ansatz, with spherical or hyperbolic symmetry, has the form
\begin{equation}
ds^2\=e^{2\,U(r)}\,dt^2-e^{-2\,U(r)}\left(dr^2+e^{2\,\Psi(r)}\,d\Sigma_\kappa^2\right)\;,
\label{metransS2H2}
\end{equation}
where \;$d\Sigma_\kappa^2=d\vartheta^2+f_\kappa^2(\vartheta)\,d\varphi^2$\; is the metric on the $2D$-surfaces $\Sigma_\kappa=\{\mathbb{S}^2,\,\mathbb{H}^2\}$, the sphere and the Lobachevskian plane, of constant scalar curvature \,$R=2\,\kappa$\, and
\begin{equation}
f_\kappa(\vartheta)\=\frac{1}{\sqrt{\kappa}}\,\sin(\sqrt{\kappa}\,\vartheta)\=
\left\{
\begin{aligned}
\;&\sin(\vartheta)\;,\quad &\kappa&=1\;;
\\
\;&\sinh(\vartheta)\;,\quad &\kappa&=-1\;.
\end{aligned}
\right.
\end{equation}
The above general metric ansatz differs from \eqref{metrans} because of the warp factor $\Psi(r)$.\par\smallskip
Now we can apply the formalism discussed in Sect.\ \ref{sec:bhconfig}, with appropriate adjustments to describe the new configuration.\par
The Maxwell equations are now satisfied using the following expression for $\FF^M$
\begin{equation}
\FF^M=\left(\begin{matrix}F^\Lambda\cr\Gdual_\Lambda\end{matrix}\right)
    =~e^{2(U-\Psi)}\,\Cc^{MP}\,\M_{PN}\,\Gamma^N\;dt\wedge dr
      +\Gamma^M\,f_\kappa(\vartheta)\;d\vartheta\wedge d\varphi
      \=d\mathbb{A}^M\;.
\label{FFk}
\end{equation}
The electric and magnetic charges are defined as
\begin{equation}
\begin{split}
e_\Lambda&~\equiv~\frac{1}{\text{vol}(\Sigma_\kappa)}\,\int_{\Sigma_\kappa}\Gdual_{\Lambda}\;,
\\[\jot]
m^\Lambda&~\equiv~\frac{1}{\text{vol}(\Sigma_\kappa)}\,\int_{\Sigma_\kappa}F^{\Lambda}\;,
\end{split}
\end{equation}
where \,$\text{vol}(\Sigma_\kappa)={\displaystyle \int} f_\kappa(\vartheta)\,d\vartheta\wedge d\varphi$.\; They can be arranged in the symplectic vector
\begingroup
\belowdisplayskip=12pt
\belowdisplayshortskip=12pt
\begin{equation}
\Gamma^M\=
\left(\begin{matrix}
m^\Lambda \cr
e_\Lambda
\end{matrix}\right)
\=\frac{1}{\text{vol}(\Sigma_\kappa)}\,\int_{\Sigma_\kappa} \FF^M \;.
\label{Gcharges}
\end{equation}
\endgroup
As we have seen in Subsect.\ \ref{subsec:bhsol}, we can obtain the equations of motion coming from the bosonic gauged Lagrangian \eqref{boslagrGaug}, with the metric ansatz \eqref{metransS2H2}, from a one-dimensional effective action that, apart from total derivative terms, has the form
\begin{equation}
\mathscr{S}_\text{eff}\=
    \mathlarger{\int} dr\,\Lagr_\text{eff}\=
    \mathlarger{\int} dr
    \,\bigg[\,e^{2\,\Psi}\,
        \Big(U^{\prime2}-\Psi^{\prime2}
        +g_{i\bar{\jmath}}\,z^{\prime i}\,\zb^{\prime\,\bar{\jmath}}\,
        \Big)
        -V_\text{eff}\,\bigg]\;,
\label{effactG}
\end{equation}
where the prime stands for derivative w.r.t.\ $r$ and where we can define an effective potential
\begin{equation}
V_\text{eff}\,=\,-\,e^{2(U-\Psi)}\,\VBH\,-\,e^{-2(U-\Psi)}\,V\,+\,\kappa\;,
\end{equation}
in terms of the scalar potential $V$ and the (charge-dependent) black hole potential $\VBH$. The latter can be written in the symplectically covariant form \eqref{VBH}
\begin{equation}
\VBH~=\,-\,\frac{1}{2}\,\Gamma^T\M\;\Gamma\;,
\label{VBH2}
\end{equation}
in terms of the magnetic and electric charges and scalar-dependent matrix $\M$.\par
Once given the effective action, one can make use of the Hamilton-Jacobi formalism and derive a system of first-order equations (\emph{flow equations}) for the warp factors $U(r)$, $\Psi(r)$ and scalar fields $z^i(r)$, $\zb^{\bar{\jmath}}(r)$.
\par

\paragraph{Supersymmetric black hole solutions.}
When interested in analysing supersymmetric configurations, one has to impose the vanishing of the SUSY transformations, in addition to solving the equations of motion.\par
The relevant supersymmetry variations can be written as:
\begin{subeqs} \label{susyvar} 
\begin{align}
\delta\psi_{\mu A}&\=
        D_\mu\eps_A\+i\;\T^{-}_{\mu\nu}\;\gamma^\nu\,\veps_{AB}\,\eps^B
        \+i\;\mathbb{S}_{AB}\,\gamma_\mu\,\eps^B\;,
\\[2ex]
\delta\lambda^{iA}&\=
        i\,\dm z^i\,\gamma^\mu\,\eps^A \-\frac{1}{2}\;g^{i\jb}\;{\bar{f}}^\Lambda_\jb\;\II_{\Lambda\Sigma}\;F^{-\Sigma}_{\mu\nu}\,\gamma^{\mu\nu}\,\veps^{AB}\,\eps_B
        \+W^{iAB}\,\eps_B\;,
\end{align}
\end{subeqs}
%
%
with $\gamma^{\mu\nu}=\gamma^{[\mu}\gamma^{\nu]}$ and where we have considered properties \eqref{VMprop}. The covariant derivatives are written as
\begin{equation}
D_\mu\eps_A\=\dm\eps_A+\frac14\,\spc\,\gamma_{ab}\,\eps_A
    +\frac{i}{2}\,\left(\sigma^2\right)_A{\!}^B\,\mathbb{A}_\mu^M\,\theta_M\,\eps_B+\frac{i}{2}\,\Q_\mu\,\eps_A\;,
\end{equation}
with
\begin{equation}
\Q_\mu\=\frac{i}{2}\left(\dd_{\ib}\K\,\dm\zb^\ib-\dd_{i}\K\,\dm{z}^i\right)\;,
\end{equation}
and, in the chosen parametrization, we also have
\begin{align*}
F^{\pm}_{\mu\nu}&\=\frac12\left(F_{\mu\nu}\pm\*F_{\mu\nu}\right)\:,\quad\;\;
\Gdual^{\pm}_{\mu\nu}\=\frac12\left(\Gdual_{\mu\nu}\pm\*\Gdual_{\mu\nu}\right)\:,
\\[2ex]
\T_{\mu\nu}&\=L^\Lambda\;\II_{\Lambda\Sigma}\;F^{\Sigma}_{\mu\nu}
    \=\frac{1}{2i}\,L^\Lambda\left(\NN-\overbar{\NN}\right)_{\Lambda\Sigma}\,F^{\Sigma}_{\mu\nu}
    ~=\,-\frac{i}{2}\left(M_\Sigma\,F^{\Sigma}_{\mu\nu}-L^\Lambda\,\Gdual_{\Lambda\mu\nu}\right)
    \=\frac{i}{2}\,\V^M\;\Cc_{MN}\;\FF^{N}_{\mu\nu}\;,
\\[2ex]
\T^{-}_{\mu\nu}&\=L^\Lambda\;\II_{\Lambda\Sigma}\;F^{-\Sigma}_{\mu\nu}\=\frac{i}{2}\,\V^M\;\Cc_{MN}\;\FF^{-N}_{\mu\nu}\;,
\\[2ex]
\T_{i\,\mu\nu}&\=\D_i \T_{\mu\nu}
    \=f^\Lambda_i\;\II_{\Lambda\Sigma}\;F^{\Sigma}_{\mu\nu}
    ~=\,-\frac{i}{2}\left(h_{i\Sigma}\,F^{\Sigma}_{\mu\nu}-f^\Lambda_i\,\Gdual_{\Lambda\mu\nu}\right)
    \=\frac{i}{2}\;\Ucal_i^M\;\Cc_{MN}\;\FF^{N}_{\mu\nu}\;,
\\[2ex]
\mathbb{S}_{AB}&\=\frac{i}{2}\,\left(\sigma^2\right)_A{\!}^C\;\veps_{BC}\;\theta_M\,\V^M
    \=\frac{i}{2}\,\left(\sigma^2\right)_A{\!}^C\;\veps_{BC}\;\W\;,
\\[2.5ex]
W^{i\,AB}&\=i\,\left(\sigma^2\right)_C{\!}^B\;\veps^{CA}\;\theta_M\;g^{i\jb}\;\overbar{\Ucal}^M_\jb\;,
\end{align*}
having used properties
\begin{equation}
\overbar{\NN}_{\Lambda\Sigma}\,F^{-\Sigma}=\,\Gdual^{-}_\Lambda\;, \qqquad
L^\Lambda\,\NN_{\Lambda\Sigma}\,=\,M_\Sigma\;.
\end{equation}
The kinetic matrix \eqref{NIR} \,$\NN=\RR+i\,\II$\, can be expressed as \cite{Gaillard:1981rj}
\begin{equation}
\NN_{\Lambda\Sigma}\=
    \dd_{\bar{\Lambda}}\dd_{\bar{\Sigma}}\overbarcal{\F}
    +2\,i\;\frac{\Img\left[\dd_{\Lambda}\dd_{\Gamma}\F\right]\;\Img\left[\dd_{\Sigma}\dd_{\Delta}\F\right]\;L^\Gamma\,L^\Delta}{\Img\left[\dd_{\Delta}\dd_{\Gamma}\F\right]\;L^\Delta\,L^\Gamma}\;,
\end{equation}
with \,$\dd_{\Lambda}=\dfrac{\dd}{\dd\X^\Lambda}$\,, \,$\dd_{\bar{\Lambda}}=\dfrac{\dd}{\dd\bar{\X}^\Lambda}$\,.\, We note that, in the special coordinate frame, the whole $\N=2$ Lagrangian can be written in terms of the holomorphic prepotential function $\F(\X)$ and its derivatives%
\footnote{%
we also emphasize that there are symplectic frames in which a prepotential
$\F(\X)$ does not exist
}.\par\smallskip
Just as we did for electric-magnetic charges in \eqref{Gcharges},
we define the \emph{central} and \emph{matter charges} as
\begin{equation}
\begin{split}
\mathscr{Z}&\=\frac{1}{\text{vol}(\Sigma_\kappa)}\,\int_{\Sigma_\kappa}\!\!\T
    \=\V^M\,\Cc_{MN}\,\Gamma^N
    \=L^\Lambda\,e_\Lambda - M_\Lambda\,q^\Lambda\;,
\\[1em]
\mathscr{Z}_i&\=\frac{1}{\text{vol}(\Sigma_\kappa)}\,\int_{\Sigma_\kappa}\!\!\T_i
    \=\D_{i}\mathscr{Z}
    \=f_i^\Lambda\,e_\Lambda - h_{\Lambda i}\,q^\Lambda\;.
\end{split}
\end{equation}
These are composite quantities that can be thought of as the physical charges measured on a solution at radial infinity. The black hole potential \eqref{VBH2} can be schematically rewritten in terms of the central charges as \cite{Ferrara:1997tw,Andrianopoli:2006ub}
\begin{equation}
\VBH\=\left|\D\Zch\right|-|\Zch|^2\;.
\end{equation}
From an explicit computation of the supersymmetry variations \eqref{susyvar}, we find the following relations for the warp factors
\begin{equation}
\begin{split}
U'&\=e^{U-2\Psi}\;\Real\left[e^{-i\alpha}\,\Zch\right]
    +e^{-U}\;\Img\left[e^{-i\alpha}\,\W\right]\;,
\\[1.5ex]
\Psi'&\=2\,e^{-U}\,\Img\left[e^{-i\alpha}\,\W\right]\;,
\end{split}
\end{equation}
and for the scalars
\begin{equation}
z^{\prime\,i}
    \=e^{-U}\,e^{i\alpha}\,g^{i\jb}\;\D_\jb\left(e^{2U-2\Psi}\,\overbarcal{\Zch}-i\,\overbar{\W}\right)\;,
\end{equation}
the above covariant derivative acting on objects with weight $p=-1$, and having introduced two projectors relating the spinor components as
\begin{equation}\label{eq:spinorproj}
\begin{split}
\gamma^0\,\eps_A&\=i\,e^{i\alpha}\,\veps_{AB}\,\eps^B\;,
\\[1ex]
\gamma^1\,\eps_A&\=e^{i\alpha}\,\delta_{AB}\,\eps^B\;.
\end{split}
\end{equation}
The Killing spinors must satisfy the relations
\begingroup
\belowdisplayskip=4pt
\belowdisplayshortskip=4pt%
\begin{equation}
\begin{split}
\eps_A&\=\chi_A\;e^{\,\frac12\left(U-i\ml{\int}dr\,\B\right)}\;,
\\[1ex]
\eps^A&\=i\,e^{-i\alpha}\,\veps^{AB}\,\gamma^0\,\eps_B\;,
\end{split}
\end{equation}
\endgroup
where we have
\begingroup
\abovedisplayskip=4pt
\abovedisplayshortskip=4pt
\begin{equation}
\begin{split}
\dd_r\chi_A&\=0\;,
\\[1ex]
\B&\=\Q_r+2\,e^{-U}\,\Real\left[e^{-i\alpha}\,\W\right]\;,
\end{split}
\end{equation}
\endgroup
and the following expression for the phase $\alpha$ holds:
\begin{equation}
\dd_r\alpha~=\,-\B\;.
\end{equation}
From the SUSY variations we obtain the property
\begin{equation}
\Img\left[e^{-i\alpha}\,\Zch\right]~=\,
    -e^{2\Psi-2U}\Real\left[e^{-i\alpha}\,\W\right]\;,
\end{equation}
and using also ansatz \eqref{FFk} for $\FF^M$, we find for the $\mathbb{A}^M_\mu$ components:
\begin{equation}
\begin{split}
\mathbb{A}^M_t\,\theta_M&\=2\,e^U\,\Real\left[e^{-i\alpha}\,\W\right]\;,
\\[2ex]
\mathbb{A}^M_r&\=0\;,
\\[2ex]
\mathbb{A}^M_\vartheta&\=0\;,
\\[1ex]
\mathbb{A}^M_\varphi&~=\,-\frac{\Gamma^M}{\kappa}\,\cos\left(\sqrt{\kappa}\,\vartheta\right)\;,
\end{split}
\end{equation}
together with the relation
\begin{equation}
\Gamma^M\,\theta_M\=\kappa\;.
\end{equation}
Because of the conditions we imposed in \eqref{eq:spinorproj} for the spinors, the construction will give us a $\frac14$-BPS solution.
\bigskip

\section{Conclusions}\label{sec:concl}
This work aims to give a self-consistent review of black hole properties and configurations in supergravity models.\par
First, special attention was posed on ungauged extended supergravity theories and their dualities, analysing the general form of black hole solutions for these models and providing an explicit construction in the relevant STU-model case.
Then, we studied in detail the gauging procedure involving the embedding tensor formalism, to be used to obtain gauged models starting from ungauged ones.
The gauged formulation was then applied to describe four dimensional black holes in $\N=2$ gauged theories, analysing also the relations to be satisfied by supersymmetric configurations.\par
Clearly, due to the vastness of the topic, some choices had to be made on what issues should be dealt with in more detail.\par

\vspace{1cm}
%
\section*{\normalsize Acknowledgments}
\vspace{-5pt}
I would like to thank professor Mario Trigiante for extremely helpful discussions during the preparation of this report.
I would also like to thank prof.\ Francesco Laviano, prof.\ Giovanni Ummarino and Fondazione CRT \,\includegraphics[height=\fontcharht\font`\B]{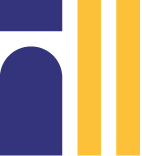}\: that partially supported these studies.


\appendix
\addtocontents{toc}{\protect\setcounter{tocdepth}{1}}
\addtocontents{toc}{\protect\addvspace{3.5pt}}%
\numberwithin{equation}{section}%
\numberwithin{figure}{section}%

\newpage

\section{Coset Geometry} \label{app:coset}

\subsection{Sigma-Model in  $D=3$} \label{subapp:sigmod}
The three-dimensional sigma-model scalar fields \;$\Phi^I\equiv\{U,\,a,\,\phi^s,\,\mathcal{Z}^M\}$\; span an homogeneous-symmetric, pseudo-Riemannian scalar manifold of the form
\begin{equation}
\Ms^{(3)}_\text{scal}\=\frac{\G}{\HHstar}\;.
\label{scalmanif}
\end{equation}
The isometry group $\G$ of the target space is the global symmetry group of the \eqref{geodaction} Lagrangian $\Lagr_{(3)}$, and $\HHstar$ is a suitable non-compact semisimple maximal subgroup of it.\par
We shall use for the scalar manifold the solvable Lie algebra parametrization, identifying the scalar fields $\Phi^I$ with parameters of a suitable solvable Lie algebra \cite{Chemissany:2010zp}. Indeed, the three-dimensional scalars $\Phi^I$ define a \emph{local} solvable parametrization, i.e.\ the corresponding physical patch $\mathscr{U}$ is isometric to a solvable Lie group generated by a solvable Lie algebra $\mathcalboondox{s}$:
\begin{equation}
\Ms^{(3)}_\text{scal} ~\supset~ \mathscr{U} ~\equiv~ e^\mathcalboondox{s}\;.
\end{equation}
The solvable Lie algebra $\mathcalboondox{s}$ is defined by the Iwasawa decomposition of the Lie algebra $\mathfrak{g}_3$ of $\G$, with respect to its maximal compact subalgebra $\mathfrak{H}_3$\,.\par
The solvable parametrization $\{\Phi^I\}$ can be expressed through the exponential map
\begin{equation}
\LL\left(\Phi^I\right)\=\exp(-a\,T_\ms{\bullet})\;\exp\left(\sqrt{2}\,\mathcal{Z}^M\,T_M\right)\;
             \exp(\phi^s\,T_s)\;\exp(2\,U H_0)\;,
\label{cosetr3}
\end{equation}
where the solvable generators $T_I=\{H_0,\;T_\ms{\bullet},\;T_s,\;T_M\}$ satisfy the commutation relations
\begin{equation}
\begin{aligned}
&[H_0,\,T_M]=\frac{1}{2}\;T_M\;;\quad&
&[H_0,\,T_s]=[T_\ms{\bullet},\,T_s]=0\;;\quad&
&[H_0,\,T_\ms{\bullet}]=T_\ms{\bullet}\;;&
\\
&[T_M,\,T_N]=\Cc_{MN}\;T_\ms{\bullet}\;;\quad&
&[T_s,\,T_M]=(T_s)^N{}_M\,T_N\;;\quad&
&[T_s,\,T_u]=-(T_{su})^{s'} T_{s'}\;,&
\label{relc1}
\end{aligned}
\end{equation}
where $(T_s)^N{}_M$ represents the symplectic representation $\mathscr{R}_{\text{s}}\left(T_s\right)$ on contravariant symplectic vectors $d\mathcal{Z}^M$.\par\smallskip
In all  $\N=2$ models with just vector multiplets, one has $\nv=\ns/2\,+\,1$, and thus the dimension of the scalar manifold in $D=3$ turns out to be $4\,\nv$\,:
\begin{equation}
\N=2   \qRq   \nv=\frac{\ns}{2}+1  \qRq
    \dim\left(\Ms^{(3)}_\text{scal}\right)=4\,\nv\;.
\end{equation}
where the manifold $\Ms^{(3)}_\text{scal}$ is a pseudo-quaternionic K\"ahler space.

\paragraph{Decompositions.}
The coset geometry is defined by the involutive \emph{pseudo-Cartan} automorphism $\zeta$ on the algebra $\mathfrak{g}_3$ of $\G$ which leaves the algebra $\halgstar$ generating $\HHstar$ invariant:
\begin{equation}
\zeta(\halgstar)\=\halgstar\;.
\end{equation}
All the formulas related to the group $\G$ and its generators, are referred to a matrix representation of $\G$ and, in particular, we shall use the fundamental one.\par
The involution $\zeta$, in the chosen representation, acts on a general matrix $X$ as:
\begin{equation}
\zeta(X)~=\,-\eta\;X^\dagger\;\eta\,,
\end{equation}
being $\eta$ an $\HHstar$-invariant metric ($\eta=\eta^\dagger,\,\,\eta^2=\Id$).\par
The pseudo-Cartan $\zeta$-involution induces a (pseudo)-Cartan decomposition of $\mathfrak{g}_3$ of the form
\begin{equation}
\mathfrak{g}_3\=\halgstar \oplus \kalgstar\;,
\label{pseudoC}
\end{equation}
where we have
\begin{equation}
\zeta\;:\qquad\;
\zeta(\halgstar)=\halgstar\;,\qquad
\zeta(\kalgstar)=-\kalgstar\;,
\end{equation}
and where the following relations hold:
\begin{equation}
[\halgstar,\;\halgstar]~\subset~\halgstar\;,\qquad [\halgstar,\;\kalgstar]~\subset~\kalgstar\;,\qquad
[\kalgstar,\;\kalgstar]~\subset~\halgstar\;.
\label{HKrels}
\end{equation}
We see that $\HHstar$ has a \emph{linear adjoint action} in the space $\kalgstar$, which is thus the carrier of an $\HHstar$-representation.\par
A general feature of $\N=2$ symmetric models is that the isotropy group $\HHstar$ has the form
\begin{equation}
\HHstar\=\SL(2,\mathbb{R})\times\G[4]'\;,
\end{equation}
and its adjoint action on $\kalgstar$ realizes the representation $({\bf 2},\,\mathscr{R}_{\text{s}})$.\par\smallskip
The decomposition \eqref{pseudoC} has to be contrasted with the ordinary \emph{Cartan decomposition} of $\mathfrak{g}_3$
\begin{equation}
\mathfrak{g}_3\=\mathfrak{H}_3\oplus\mathfrak{K}_3\;,
\label{Cartan}
\end{equation}
where the algebra $\mathfrak{g}_3$ is decomposed into its maximal compact subalgebra $\mathfrak{H}_3$, generating $\HH[3]$, and its orthogonal non-compact complement $\mathfrak{K}_3$. This decomposition is effected through the Cartan involution $\tau$, of which $\mathfrak{H}_3$ and $\mathfrak{K}_3$ represent the eigenspaces with eigenvalues $+1$ and $-1$ respectively. In the matrix representation in which we shall work, the action of $\tau$ on a matrix $X$ can be implemented as:
\begin{equation}
\tau(X)~=\,-X^\dagger\;.
\end{equation}
We shall also use the $\HHstar$-invariant symetric matrix \eqref{cm}
\begin{equation}
\M_{(3)}\left(\Phi^I\right)\=\LL\left(\Phi^I\right)\,\eta\;\LL\left(\Phi^I\right)^\dagger\;.
\end{equation}
Next we construct the left-invariant one-form and the vielbein
$\mathcalboondox{V}^\mathcal{A}={\mathcalboondox{V}_I}^\mathcal{A}\,d\phi^I$:
\begin{equation}
\LL^{-1}\,d\LL\=\mathcalboondox{V}^{\mathcal{A}}\,T_{\mathcal{A}}\=\pp+\ww\;;
\qqquad\;
\mbox{{\footnotesize$\mathcal{A}$}}\;=1,\dotsc,\,
\dim\left(\Ms^{(3)}_\text{scal}\right)\;,
\label{li1f}
\end{equation}
where $\pp=\mathcalboondox{V}^{\mathcal{A}}\,\mathbb{K}_{\mathcal{A}}$\, and \,$\ww$ are the vielbein and connection matrices, and where $\{\mathbb{K}_\mathcal{A}\}$ is a basis of $\kalgstar$ defined as
:
\begin{equation}
\mathbb{K}_\mathcal{A}\=\frac{1}{2}\,\left(
                        T_\mathcal{A}+\eta\;T_\mathcal{A}^\dagger\;\eta\right)\;,
\end{equation}
$T_\mathcal{A}=T_I$, being the solvable generators defined above.\par\smallskip
Following the prescription of \cite{Chemissany:2010zp}, the normalization of the $\HHstar$-invariant metric on the tangent space of $\G/\HHstar$ is chosen to be
\begin{equation}
g_{{\mathcal{A}\mathcal{B}}}\=
    k\;\Tr\left[\mathbb{K}_{\mathcal{A}}\,\mathbb{K}_{\mathcal{B}}\right]\,,
\end{equation}
where
\begin{equation}
k\=\frac{1}{2\;\Tr(H_0^2)}
\end{equation}
is a representation-dependent constant.\par\smallskip
The metric of the $D=3$ sigma-model has the usual form:
\begin{equation}
\begin{split}
ds^2&\=k\,\Tr\left(\pp^2\right)\=
    g_{{\mathcal{A}\mathcal{B}}}\,\pp^{\mathcal{A}}\,\pp^{\mathcal{B}}\=
\\
&\=2\,dU^2+\Gm[s][u]\,d\phi^s\,d\phi^u+\frac{1}{2}\,e^{-4U}\,\omega^2
    +e^{-2U}\,d\mathcal{Z}^T\,\M_{(4)}(\phi^s)\;d\mathcal{Z}\;,
\end{split}
\end{equation}
with \
\begin{equation}
\omega\=da+\mathcal{Z}^T\,\Cc\;d\mathcal{Z}\;.
\end{equation}

\subsection{The STU model}\label{stumodel}
The most general scalar manifold of an $\N=2$ model is described by the product of a \emph{special K\"ahler manifold} $\Ms_\textsc{sk}$, spanned by the complex scalars $z^\alpha$ in the vector multiplets, times a \emph{quaternionic K\"ahler manifold} $\Ms_\textsc{qk}$ spanned by the scalar fields $q^u$ in the hypermultiplets:
\begin{equation}
\Mscal=\Ms_\textsc{sk}\times\Ms_\textsc{qk}\;.
\end{equation}
The symplectic structure is defined only over the first factor, since only the scalars $z^\alpha$ enter the matrices $\II_{\Lambda\Sigma}$, $\RR_{\Lambda\Sigma}$.\par\smallskip
\sloppy
The STU model is an $\N=2$ supergravity coupled to three vector multiplets ${(\ns=6,\;\nv=4)}$ and where the $D=4$ scalar manifold is
\begin{equation}
\Ms^{(4)}_\text{scal}\= \frac{\G[4]}{\HH[4]}\=
    \left(\frac{\SL(2,\mathbb{R})}{\SO(2)}\right)^3
\end{equation}
\sloppy
is a complex special K\"ahler manifold, spanned by three complex scalar fields ${z^a=\text{\{S,\,T,\,U\}}}$.\par
The $D=4$ scalar metric for the STU model reads
\begin{equation}
ds^2_{(4)\,\textsc{stu}}=~\Gm[s][u]\,d\phi^s d\phi^u \=
    2\,g_{a\bar{b}}\,dz^a\,d\zb^{\bar{b}}~=\,
    -2\,\sum_{a=1}^3\frac{dz^{a}\,d\zb^{\bar{a}}}{(z^a-\zb^{\bar{a}})^2}\=
    \sum_{\mathcal{I}=1}^3\,e_i{}^\mathcal{I}\,\bar{e}_{\bar{i}}{}^\mathcal{I} \,dz^i\,d\zb^{\bar{i}}\;.
\end{equation}
We also consider the real parametrization $\phi^s=\{\epsilon_i,\,\varphi_i\}$, that is related to the complex one $z_i$ by:
\begin{equation}
\phi^s=\{\epsilon_i,\,\varphi_i\}  \;\qLq\;
z_i=\epsilon_i-\,i\,e^{\varphi_i}\;.
\end{equation}
The K\"ahler potential has the simple form:
\begin{equation}
e^{-\mathcal{K}}=~8\,e^{\varphi_1+\varphi_2+\varphi_3}\;,
\end{equation}
and, in the chosen symplectic frame (i.e.\ the special coordinate frame originating from Kaluza Klein reduction from $D=5$), the special geometry of $\Ms^{(4)}_\text{scal}$ is characterized by an holomorphic prepotential:
\begin{equation}
\mathpzc{F}(z)\=z_1\,z_2\,z_3\;.
\end{equation}
The holomorphic $\Omega^M(z)$ section of the symplectic bundle reads:
\begin{equation}
\Omega^M(z)\=\{1,\;z_1,\;z_2,\;z_3,\;-z_1\,z_2\,z_3,\;z_2\,z_3,\;z_1\,z_3,\;z_1\,z_2\}\;,
\label{omega}
\end{equation}
while the covariantly holomorphic section is given by
\begin{equation}
\V^M(z,\zb)\=e^{\frac{\mathcal{K}}{2}}\,\Omega^M(z)\;.
\end{equation}
Once defined the covariant derivative $\D_i$
\begin{equation}
\D_i \V~:=~\partial_i \V+\frac{\partial_i \mathcal{K}}{2}\,\V\;,
\end{equation}
it is possible to write the central and matter charges of a black hole solution, with quantized charges $\Gamma^M=(p^\Lambda,q_\Lambda)$, in terms of $\V^M$ and of its covariant derivative as:
\begingroup
\allowdisplaybreaks
\begin{align*}
\mathscr{Z}&\=\V^T\Cc\,\Gamma=
\\
    &\=e^{\frac{\mathcal{K}}{2}}\,(-q_0-q_1\,z_1-q_2\,z_2+p^3\,z_1\,z_2-q_3\,z_3
    +p^2\,z_1\,z_3+p^1\,z_2\,z_3-p^0\,z_1\,z_2\,z_3)\;,
\\[1.5em]
\mathscr{Z}_1&\=e_{1}{}^i\,\D_i \V^T\Cc\,\Gamma=
\\
    &~=\,-i\,e^{\frac{\mathcal{K}}{2}}\,\left(q_0+q_2\,z_2+q_3\,z_3-p^1\,z_2\,z_3
    +q_1\,\zb_1-p^3\,z_2\,\zb_1-p^2\,z_3\,\zb_1
    +p^0\,z_2\,z_3\,\zb_1\right)\;,
\\[1.5em]
\mathscr{Z}_2&\=e_{2}{}^i\,\D_i \V^T\Cc\,\Gamma=
\\
    &~=\,-i\,e^{\frac{\mathcal{K}}{2}}\,\left(q_0+q_1\,z_1+q_3\,z_3-p^2\,z_1\,z_3
    +q_2\,\zb_2-p^3\,z_1\,\zb_2-p^1\,z_3\,\zb_2
    +p^0\,z_1\,z_3\,\zb_2\right)\;,
\\[1.5em]
\mathscr{Z}_3&\=e_{3}{}^i\,\D_i \V^T\Cc\,\Gamma=
\\
    &~=\,-i\,e^{\frac{\mathcal{K}}{2}}\,\left(q_0+q_1\,z_1+q_2\,z_2-p^3\,z_1\,z_2
    +q_3\,\zb_3-p^2\,z_1\,\zb_3-p^1\,z_2\,\zb_3
    +p^0\,z_1\,z_2\,\zb_3\right)\;.
\end{align*}
\endgroup
The explicit form of the quartic invariant for the STU model is:
\begin{equation}
\begin{split}
I_4(p,q)\=&-(p^0)^2\,q_0^2-2\,q_0\left(-2\,p^1\,p^2\,p^3+p^0\,q_3\,p^3
    +p^0\,p^1\,q_1+p^0\,p^2\,q_2\right)-(p^1)^2\,q_1^2-
\\
    &-\left(p^2\,q_2-p^3\,q_3\right)^2+2\,q_1
    \left(p^1\,p^3\,q_3+q_2\,\left(p^1\,p^2-2\,p^0\,q_3\right)\right)\;.
\end{split}
\end{equation}
Upon timelike reduction to $D=3$, the scalar manifold has the form
\begin{equation}
\Ms^{(3)}_\text{scal}\=\frac{\G}{\HHstar}
    \=\frac{\SO(4,4)}{\SO(2,2)\times\SO(2,2)}\;.
\end{equation}
The generators of $\mathfrak{g}_3=\mathfrak{so}(4,4)$ can be written in terms of Cartan generators $H_\alpha$ and shift generators $E_{\pm\alpha}$
in the fundamental representation, with the usual normalization convention:
\begin{equation}
[H_\alpha,\,E_{\pm \alpha}]=\pm 2\,E_{\pm \alpha}\;;\qquad
[E_\alpha,\,E_{-\alpha}]=H_\alpha\;,
\end{equation}
where
\begin{equation}
E_{-\alpha}\=E_\alpha^\dagger\=E_\alpha^T\;.
\end{equation}
The positive roots of the algebra $\mathfrak{g}_3$ split into:
\begin{enumerate}[label=-]
\item{the root $\beta_0$ of the Ehlers subalgebra $\mathfrak{sl}(2,\mathbb{R})_E$, commuting with the algebra $\mathfrak{g}_4$ of $\G[4]$ \;($\mathfrak{g}_4\subset\mathfrak{g}_3$)\,;}
\item{the roots $\alpha_i$ of $\mathfrak{g}_4$ \;($i=1,2,3$)\,;}
\item{the roots $\gamma_M$ \;({\footnotesize$M$}\;$=1,\dotsc,8$)\,.}
\end{enumerate}
The special coordinate parametrization of \,$\Ms^{(4)}_\text{scal}$ corresponds to a solvable parametrization of the manifold in which the real coordinates $\phi^s=\{\epsilon_i,\,\varphi_i\}$ are parameters of a solvable Lie algebra generated by
\begin{equation}
T_s\=\{E_{\alpha_i},\;H_{\alpha_i}/2\}\;.
\end{equation}
The coset representative $\LL_{(4)}$ is an element of the corresponding solvable group \cite{Andrianopoli:1996bq,Andrianopoli:1996zg}, defined by the following exponentialization prescription:
\begin{equation}
\LL_{(4)}(\phi^s)\=\exp(\phi^s\,T_s)
    \=\prod_{i=1}^3\,e^{\epsilon_i E_{\alpha_i}}
    \;e^{\varphi_i\frac{H_{\alpha_i}}{2}}\;.
\label{L4}
\end{equation}
The solvable (or Borel) $\mathfrak{h_\textsc{b}}$ subalgebra of $\mathfrak{g}_3$ has the form:
\begin{equation}
\mathfrak{h_\textsc{b}}=\Span\left(T_\mathcal{A}\right)\;,\qquad T_\mathcal{A}=\{H_0,\,T_\ms{\bullet},\,T_s,\,T_M\}
\end{equation}
is used to define the parametrization of $\Ms^{(3)}_{scal}$ in terms of the $D=3$ scalars $\Phi^I$, through the coset representative \eqref{cosetr3}. This subalgebra can be defined through the identifications
\begin{equation}
H_0\=\frac{H_{\beta_0}}{2}\;;\qquad
T_\ms{\bullet}=E_{\beta_0}\;;\qquad\;
T_M=E_{\gamma_M}\;.
\end{equation}
The symplectic representation of $\{T_s\}$, in the duality representation $\mathscr{R}_{\text{s}}={\bf (2,2,2)}$ of $\G[4]$, is defined through their adjoint action on $T_M$:
\begin{equation}
[T_s,\,T_M]~=\,-(T_s)_M{}^N\;T_N\;.
\end{equation}
In order to reproduce the form of the $(T_s)_M{}^N$ in the chosen special coordinate frame \eqref{omega}, the generators $T_M$ corresponding to the roots $\gamma_M$ have to be ordered according to \eqref{gammaord}. In this basis, the symplectic representation of  $\LL_{(4)}$ defined in \eqref{L4} allows to define the matrix $\M_{(4)}$:
\begin{equation}
\M_{(4)\,MN}=-\sum_{P=1}^8\left(\LL_{(4)}\right)_M{}^P\,\left(\LL_{(4)}\right)_N{}^P\;.
\end{equation}
We give, for the sake of completeness, the matrix form of $\phi^s\,T_s$ in the symplectic representation $\mathscr{R}_{\text{s}}$\,:
\begin{equation}
\phi^s\,T_s\=\sum_{i=1}^3\epsilon_i\,E_{\alpha_i}+\varphi_i\,\frac{H_{\alpha_i}}{2}\=
    \left(
    \begin{matrix}
    A & B \cr
    \Zero & -A^T
    \end{matrix}
    \right)\;,
\end{equation}
with
\begin{equation}
\begin{aligned}
A&\=
    \left(
    \begin{array}{cccc}
    \frac{\varphi _1}{2}+\frac{\varphi _2}{2}+\frac{\varphi _3}{2} & -\epsilon _1 & -\epsilon _2 & -\epsilon _3 \\
    0 & -\frac{\varphi _1}{2}+\frac{\varphi _2}{2}+\frac{\varphi _3}{2} & 0&0 \\
    0 & 0 & \frac{\varphi _1}{2}-\frac{\varphi _2}{2}+\frac{\varphi _3}{2} &0 \\
    0 & 0 & 0 & \frac{\varphi _1}{2}+\frac{\varphi_2}{2}-\frac{\varphi_3}{2}
    \end{array}
    \right)\;,
\\
B&\=
    \left(
    \begin{array}{cccc}
    0 & 0 & 0 & 0 \\
    0 & 0 & -\epsilon _3 & -\epsilon _2 \\
    0 & -\epsilon _3 & 0 & -\epsilon _1 \\
    0 & -\epsilon _2 & -\epsilon _1 & 0
    \end{array}
    \right)\;.
\end{aligned}
\end{equation}
The pseudo-Cartan involution $\zeta$ determines the decomposition of $\mathfrak{g}_3$ into $\halgstar$ and $\kalgstar$, and is defined by the matrix \,$\eta=(-1)^{2 H_0}$\,.

\newpage

\section{Dimensional Reduction} \label{app:dimred}
The bosonic Lagrangian \eqref{boslagr} can be rewritten as:
\begin{equation}
\frac{1}{\eD}\,\Lagr_{(4)} ~=
\-\frac{R}{2}
\+\frac{1}{2}\,g^{\mu\nu}\langle J_{\mu},\,J_{\nu}\rangle
\+\frac{1}{4}\,\II_{\Lambda\Sigma}(\phi)\,F^\Lambda_{\mu\nu}\,F^{\Sigma\,\mu\nu}
\+\frac{1}{8\,\eD}\,\RR_{\Lambda\Sigma}(\phi)\,\veps^{\mu\nu\rho\sigma}\,F^\Lambda_{\mu\nu}\,F^{\Sigma}_{\rho\sigma}\;,
\label{lagr4D}
\end{equation}
having introduced the currents
\begin{equation}
J_{\mu}\=\frac12\,\M^{-1}\,\dm\M\;.
\end{equation}
The above Lagrangian describes a field theory over a 4D space-time manifold $\Sigma_4$ with coordinates $x^\mu$ and metric $\gmetr(x)$. The scalar fields have values in a target space $\Ms^{(4)}_\text{scal}$ with coordinates $\phi^s$ and metric $\Gm[s][u](\phi)$. The solutions of the scalar equations are maps from $\Sigma_4$ to $\Ms^{(4)}_\text{scal}$.\par
In particular, we considered in Section \ref{sec:constrbhsol} the case where $\Ms^{(4)}_\text{scal}$ is a non-compact homogeneous Riemannian symmetric space of the form:
\begin{equation}
\Ms^{(4)}_\text{scal}\=\frac{\G[4]}{\HH[4]}\,,
\end{equation}
$\G[4]$ being the isometry group and $\HH[4]$ its maximal compact subgroup.\par
Following the prescription of \cite{Breitenlohner:1987dg}, we shall consider only stationary (or stationary-axisymmetric) field configurations. For the latter, it is possible to reformulate the four-dimensional theory in terms of a 3D euclidean description, in analogy with the dimensional reduction technique for Kaluza-Klein theories.

\subsection{Reduction from 4 to 3 dimensions} \label{subapp:4to3}
For a field configuration allowing a Killing vector field $\xi$, we can choose a gauge such that the Lie derivative of the vector potentials $A_\mu^\Lambda$ vanishes and choose adapted coordinates such that the isometry is just a translation (e.g.\ $\xi=\dd_t$). The fields of the theory will then depend only on the remaining three coordinates $x_i$ $(i=1,2,3)$ parameterizing the orbit space $\Sigma_3$ of the action of $\xi$. In these coordinates, $\xi$ has the form $\xi = (\Upsilon,\,\Upsilon\,\omega_i)$ and the metric $\gmetr$ can be decomposed as:
\begin{equation}
\gmetr\=
    \left(
    \begin{matrix}
    \Upsilon & \Upsilon\,\omega_j \cr
    \Upsilon\,\omega_i \quad & \quad -\Upsilon^{-1}\,g^{(3)}_{ij}+\Upsilon\,\omega_i\,\omega_j \quad
    \end{matrix}\right)\;,
\end{equation}
only requiring $\Upsilon\neq0$. \, The scaled metric $g^{(3)}_{ij}$ is referred to the reduced 3D space $\Sigma_3$. In a similar way, we decompose the vector fields as:
\begin{equation}
A_\mu^\Lambda\=\left(A_0^\Lambda,\;A_0^\Lambda\,\omega_i+A_i^\Lambda\right)\=
   \left(\Z^\Lambda,\;\Z^\Lambda\,\omega_i+A_i^\Lambda\right)\;,
\end{equation}
into pieces parallel and perpendicular to $\xi$\,.\par
The Lagrangian \eqref{lagr4D} can be now rewritten (apart from surface terms) as:
\begin{equation}
\begin{split}
\frac{1}{\eD^{(3)}}\,\tilde{\Lagr}\,=\,
    &+\,\frac{R^{(3)}}{2}\-\frac{1}{2}\,g^{(3)\,ij}\langle \hat{J}_{i},\,\hat{J}_{j}\rangle
    \-\frac{1}{2\Upsilon}\,\II_{\Lambda\Sigma}(\phi)\,\dd_i\Z^\Lambda\,\dd^i\Z^\Sigma \-\frac{1}{4\Upsilon^2}\,\dd_i\Upsilon\,\dd^i\Upsilon\+
\\
    &+\,\frac{\Upsilon^2}{8}\,\omega_{ij}\,\omega^{ij}
    \+\frac{\Upsilon}{4}\,\II_{\Lambda\Sigma}(\phi)\,\left(F_{ij}^\Lambda\+\omega_{ij}\Z^\Lambda\right)\left(F^{\Sigma\,ij}+\omega^{ij}\Z^\Sigma\right)\+
\\
    &+\,\frac{1}{2\,\eD^{(3)}}\,\RR_{\Lambda\Sigma}(\phi)\,\veps^{ijk}\,\left(F_{ij}^\Lambda+\omega_{ij}\Z^\Lambda\right)\,\dd_k\Z^\Sigma\;,
\label{lagrred}
\end{split}
\end{equation}
where $R^{(3)}$ is the scalar curvature for the three-dimensional metric $g_{ij}^{(3)}$ and with
\begin{equation}
\omega_{ij}\,=\,\dd_i\omega_j-\dd_j\omega_i\;,\qqquad
F_{ij}^\Lambda\,=\,\dd_i A_j^\Lambda-\dd_j A_i^\Lambda\;.
\end{equation}
If the original field configuration was a solution of the four-dimensional field equations, then the set $\{g_{ij}^{(3)},\,\Upsilon,\,\omega_i,\,A_0^\Lambda,\,A_i^\Lambda,\,\phi^s\}$ is a solution of the three dimensional field equations derived from $\tilde{\Lagr}$ and viceversa. The field equations for the 3D vector fields $A_i^\Lambda$ and $\omega_i$ are (omitting symplectic indices)
\begin{equation}
\begin{split}
\nabla_i\left(\Upsilon\,\II(\phi)\,\left(F^{ij}+\omega^{ij}\Z\right)+\frac{1}{\eD^{(3)}}\,\RR(\phi)\,\veps^{ijk}\,\dd_k\Z\right)\=0\;,
\\
\nabla_i\left(\frac{\Upsilon^2}{2}\,\omega^{ij}+\Upsilon\,\II(\phi)\,\Z^T\left(F^{ij}+\omega^{ij}\Z\right)+\frac{1}{\eD^{(3)}}\,\RR(\phi)\,\veps^{ijk}\,\Z^T\,\dd_k\Z\right)\=0\;,
\end{split}
\end{equation}
and can be considered as Bianchi identities for the dual potentials $\Z_\Lambda$ and for the so-called \emph{twist potential} $a$.\par
Instead of using the definitions of $F_{ij}^\Lambda$ and $\omega_{ij}$, we can treat them as independent fields and add Lagrange multipliers to the Lagrangian (ensuring that they are curls)
\begin{equation}
\tilde{\Lagr}^\prime\=
    \tilde{\Lagr}+\frac{1}{2}\,\veps^{ijk}\,\Z_\Lambda\,
    \dd_i F_{jk}^\Lambda
    +\frac{1}{4}\,\veps^{ijk}\,\left(\Z_\Lambda\,\Z^\Lambda-a\right)\,\dd_i\omega_{jk}\;.
\end{equation}
The resulting field equations for $\omega_{ij}$ and $F_{ij}^\Lambda$ are
\begin{equation}
\begin{split}
\omega^{ij}&\=\frac{1}{\eD^{(3)}}\,\veps^{ijk}\,\frac{1}{\Upsilon^2}\,\varpi_k\;,
\\[\jot]
F^{\Lambda\,ij}+\omega^{ij}\Z^\Lambda&~=\,\frac{1}{\eD^{(3)}\,\Upsilon}\;\veps^{ijk}\;\II^{-1\,\Lambda\Sigma}(\phi)\,\left(\RR_{\Sigma\Pi}(\phi)\,\dd_k\Z^\Pi-\dd_k\Z_\Sigma\right)\;,
\end{split}
\end{equation}
with
\begin{equation}
\varpi_i ~=\,
    -\,\dd_ia-\left(\Z^\Lambda\,\dd_i\Z_\Lambda-\Z_\Lambda\,\dd_i\Z^\Lambda\right)\;.
\end{equation}
Inserting these expressions back into $\tilde{\Lagr}^\prime$, we obtain the Lagrangian of the three-dimensional reduced theory
\begin{equation}
\begin{split}
\frac{1}{\eD^{(3)}}\,\Lagr_{(3)} \=
    &\frac{R^{(3)}}{2}\-\frac{1}{2}\,g^{(3)\,ij}\langle \hat{J}_{i},\,\hat{J}_{j}\rangle
    \-\frac{1}{4\Upsilon^2}\,\left(\dd_i\Upsilon\,\dd^i\Upsilon+\varpi_i\varpi_j\right)
    \-\frac{1}{2\Upsilon}\,\dd_i\Z^M\,\M_{(4)MN}\,\dd^i\Z_N~\equiv
\\[\jot]
    \equiv~&\frac{R^{(3)}}{2}
    \-\frac{1}{2}\,\hat{\mathpzc{G}}_{ab}(z)\,\dd_i z^a\,\dd^i z^b\;,
\label{lagr3D}
\end{split}
\end{equation}
where $\M_{(4)MN}$ is the negative-definite matrix introduced in \eqref{M} and where the ``twist'' vector $\varpi_i$ can be rewritten in an explicit $\G[4]$ invariant form as
\begin{equation}
\varpi_i~=\,-\,\dd_i a - \Z^M\,\Cc_{MN}\,\dd_i\Z^N\;.
\end{equation}
We have obtained a non-linear $\sigma$-model with a target space $\Ms^{(3)}_\text{scal}$ parameterized by $\Phi=\{\phi,\,\Upsilon,\,\Z,\,a\}$, coupled to (three-dimensional) gravity.\par\smallskip
For a space-like Killing vector ($\Upsilon<0$) the metric on $\Ms^{(3)}_\text{scal}$ is positive definite, while for a time-like Killing vector ($\Upsilon>0$, \emph{stationary solutions}) the metric is indefinite with $2\,\nv$ negative terms due to the fields $\Z$ originating from the $\nv$ vector fields in the four-dimensional theory.

\paragraph{Invariance group and target space.}
The set of all the transformations leaving invariant the metric (on the target space $\Ms^{(3)}_\text{scal}$) and the field equations (from Lagrangian \eqref{lagr3D}) form a non-compact Lie group $\G$. The target space can be either a Riemannian symmetric space ($\Upsilon>0$ case)
\begin{equation}
\Ms^{(3)}_\text{scal}\=\frac{\G}{\HH[3]}\;, \qqquad (\Upsilon>0)
\end{equation}
where $\HH[3]$ is the maximal compact subgroup of $\G$, or a pseudo-Riemannian symmetric space ($\Upsilon<0$ stationary case) of the form
\begin{equation}
\Ms^{(3)}_\text{scal}\=\frac{\G}{\HH[3]^*}\;,\qqquad (\Upsilon<0)
\end{equation}
where $\HH[3]^*$ is a non-compact real form of $\HH[3]$.

\subsection{Stationary solutions with pseudo-Riemannian symmetric space}
If we consider the stationary case, we find a pseudo-Riemannian symmetric target space $\Ms^{(3)}_\text{scal}=\G/\HH[3]^*$. It is possible to introduce the hermitian, $\HH[3]^*$-invariant matrix $\M_{(3)}$ which, in a chosen matrix representation, reads:
\begin{equation}
\M_{(3)} ~\equiv~ \M_{(3)}(\Phi) ~\equiv~ \LL\,\eta\,\LL^\dagger \=
\M_{(3)}^\dagger\;,
\end{equation}
defined from the coset representative $\LL(\phi)$, and where $\eta$ is a suitable $\HH[3]^*$-invariant metric (see App.\ \ref{app:coset}).\par
The reduced Lagrangian \eqref{lagr3D} can be rewritten
\begin{equation}
\begin{split}
\frac{1}{\eD^{(3)}}\,\Lagr_{(3)} \=
    &\frac{R^{(3)}}{2}\-\frac{1}{2}\,\langle \hat{J}_{i},\,\hat{J}^{j}\rangle~\equiv~
\\[\jot]
    \equiv~&\frac{R^{(3)}}{2}\-\frac{\hat{\kappa}}{8}\,g^{(3)\,ij}\;
    \Tr\left({\M_{(3)}}^{-1}\dd_i\M_{(3)}\;{\M_{(3)}}^{-1}\dd_j\M_{(3)}\right)\;,
\end{split}
\end{equation}
$\hat{\kappa}$ being a constant depending on the considered representation and on the specific $\sigma$-model, and $\hat{J}_i\equiv\hat{J}^{(3)}_{i}$ being the currents
\begin{equation}
\hat{J}_i\=\frac12\,\M_{(3)}^{-1}\,\dd_i\M_{(3)}\;.
\end{equation}
The field equations for for the above Lagrangian can be written in the compact form
\begin{equation}
\begin{split}
R^{(3)}_{ij}&\=\langle \hat{J}_{i},\,\hat{J}_{j}\rangle\;,
\\[\jot]
\nabla^i \hat{J}_i&\=0\;,
\end{split}
\end{equation}
where not all the conserved currents $\hat{J}$ and not all the field equations $\nabla \hat{J}=0$ are independent.


\newpage

\hypersetup{linkcolor=blue}
\phantomsection 
\addtocontents{toc}{\protect\addvspace{4.5pt}}
\addcontentsline{toc}{section}{References} 
\bibliographystyle{mybibstyle}
\bibliography{bibliografia_old} 

\end{document}